\documentclass{jfm}

\usepackage{subfiles}
\usepackage{tabularx}
\usepackage{graphicx}
\usepackage{epstopdf,epsfig,subfloat,subfig,floatrow}
\usepackage{newtxtext}
\usepackage{newtxmath}
\usepackage{natbib}
\usepackage{amsfonts}
\usepackage{amsmath}
\usepackage{amssymb,bm,mathtools}
\usepackage{color}
\usepackage{hyperref}
\usepackage{multirow,makecell,array,booktabs}
\usepackage{soul,cancel}
\usepackage{ulem}
\usepackage{booktabs}            % 更美观的横线
\usepackage[flushleft]{threeparttable} % threeparttable 宏包
\usepackage{siunitx}             % 如果你想对齐小数点

\usepackage{lineno}

\usepackage{changes}
\usepackage{color}
\usepackage{ulem}
\usepackage{cancel}

\usepackage{soul}
\soulregister{\ref}7 % 为 ref 添加删除线
\soulregister{\citealt}7 % 为 citealt 添加删除线
\soulregister{\cite}7 % 为 citealt 添加删除线
\soulregister{\citep}7 % 为 citealt 添加删除线
\soulregister{\citet}7 % 为 citealt 添加删除线

\hypersetup{
    colorlinks = true,
    urlcolor   = blue,
    citecolor  = blue,
}

\newcommand{\RomanNumeralCaps}[1]

\allowdisplaybreaks[4]

\definecolor{orange}{RGB}{255,100,0}

\title{Model of incompressible turbulent flows via a kinetic theory
}

\author{Ziyang Xin\aff{1}, 
        Zhaoli Guo\aff{1, 2, \corresp{\email{zlguo@hust.edu.cn}}}
        \and Hudong Chen \aff{3, \corresp{\email{hdc25@zju.edu.cn}}}
        }

\affiliation{
  \aff{1} 
    State Key Laboratory of Coal Combustion, School of Energy and Power Engineering, Huazhong University of Science and Technology, Wuhan, 430074, China\\
  \aff{2} 
    Institute of Interdisciplinary Research for Mathematics and Applied Science, Huazhong University of Science and Technology, Wuhan 430074, China\\
  \aff{3} 
    College of Energy Engineering, Zhejiang University, Hangzhou 310027, China\\
  }

\begin{document}
\maketitle

%\linenumbers 

\begin{abstract}
Kinetic theory offers a promising alternative to conventional turbulence modelling by providing a mesoscopic perspective that naturally captures non-equilibrium physics such as non-Newtonian effects. In this work, we present an extension and theoretical analysis of the kinetic model for incompressible turbulent flows developed by Chen et al. (Atmos. 14(7), 1109, 2023), constructed for unbounded flows. The first extension is to reselect a relaxation time such that the turbulent transport coefficients are obtained consistently and better align with well-established turbulence theory. The Chapman-Enskog (CE) analysis of the kinetic model reproduces the linear eddy viscosity and gradient diffusion models for Reynolds stress and turbulent kinetic energy flux at the first order, and yields nonlinear eddy viscosity and closure models at the second order. 
Particularly, a previously unreported CE solution for turbulent kinetic energy flux is obtained.
The second extension is to enable the model for wall-bounded turbulent flows with preserved near-wall asymptotic behaviours. This involves developing a low-Reynolds-number model incorporating wall damping effects and viscous diffusion, with boundary conditions enabling both viscous sublayer resolution and wall function application.
Comprehensive validation against experimental and DNS data for turbulent Couette flow demonstrates excellent agreement in predicting mean velocity profiles, skin friction coefficients, and Reynolds shear stress distributions, although the near-wall normal stress anisotropy is underestimated. The results show that averaged turbulent flow behaves similarly to rarefied gas flow at finite Knudsen number, capturing non-Newtonian effects beyond linear eddy viscosity models. This kinetic model provides a physics-based foundation for turbulence modelling with reduced empirical dependence.
\end{abstract}

\section{Introduction}
Incompressible turbulent flows are ubiquitous in nature and engineering, occurring in systems ranging from atmospheric and oceanic flows to industrial processes, yet their inherent complexity continues to challenge predictive modelling efforts~\citep{pope2001turbulent}. Although advances in high-performance computing have greatly expanded the scope of direct numerical simulation (DNS) of the Navier–Stokes (NS) equations~\citep{moin1998direct}, the computational cost of fully resolving turbulent flows remains prohibitive, particularly at high Reynolds numbers and in complex geometries. Consequently, considerable effort has been devoted to turbulence modelling based on statistically averaged equations, aiming to capture key features of turbulent flows such as the mean velocity profile and the root mean square velocity fluctuations.

Among statistically averaged equations, the Reynolds averaged Navier–Stokes (RANS) equations~\citep{reynolds1895iv}, obtained by time averaging the incompressible NS equations, remain the most widely used framework in turbulence modelling. A central challenge in applying the RANS equations is to devise closure models that approximate the Reynolds stress and scalar flux introduced by the averaging process. One of the earliest closure models, the Prandtl mixing length hypothesis introduces the concept of eddy viscosity by drawing an analogy between turbulent mixing and molecular momentum transport~\citep{Prandtl1925}. To this day, eddy viscosity concepts remain a cornerstone of turbulence theory and engineering modelling. The linear eddy viscosity closure for Reynolds stress with gradient diffusion closure for scalar flux~\citep{launder1983numerical,spalart1992one,menter1997eddy,wilcox2008formulation} and nonlinear extensions~\citep{pope1975more,speziale1987nonlinear,gatski1993explicit,craft1996development,wallin2000explicit} that incorporate strain and rotation tensor invariants, have achieved success across different flow types, yet they typically rely on $ad \ hoc$ parameters that require empirical specification~\citep{durbin2018some}. The pursuit of more sophisticated closure strategies has resulted in Reynolds stress transport models that directly solve the governing equations for individual Reynolds stress tensor components, providing the ability to capture secondary flow structures and the lag phenomena between a rapidly changing applied strain rate
and the response~\citep{launder1975progress,yakhot1992development,hamlington2008reynolds}. However, these higher-order closure models introduce challenges related to the uncertain physical foundations of the required higher-order turbulence correlations and the specification of appropriate boundary conditions.

Although eddy viscosity models have achieved considerable success in RANS frameworks, they are fundamentally limited by the assumption of scale separation between the mean and fluctuating flow fields~\citep{monin2013statistical}. This limitation can be naturally addressed by seeking a description in terms of kinetic theory for gas molecules, as kinetic equations do not rely on such scale separation assumptions~\citep{cercignani1988boltzmann}. Building on this kinetic perspective, \cite{chen2003extended} introduced an extended Boltzmann kinetic framework, in which turbulent fluctuations are treated analogously to molecular thermal fluctuations. Their formulation circumvents the scale separation assumption inherent in classical eddy viscosity models, naturally recovering Reynolds stress and higher-order (nonlinear and non-Newtonian) corrections through Chapman–Enskog (CE) expansion~\citep{chapman1990mathematical}, with all model coefficients consistently determined from kinetic theory. Furthermore, by analogy, a Fokker-Planck equation explicitly tailored for turbulence modelling was developed by~\cite{heinz2007unified} and~\cite{luan2025constructing}. Using the CE expansion, \cite{luan2025constructing} derived a nonlinear eddy viscosity model from this Fokker–Planck formulation, which closely aligns with the results of the Bhatnagar-Gross-Krook (BGK) model of \cite{chen2004expanded}, differing primarily in the omission of the molecular viscous term. Encouragingly, their results demonstrate that the resulting quadratic eddy viscosity model outperforms the existing linear eddy viscosity model and achieves comparable accuracy to cubic eddy viscosity models for two-dimensional flows, without resorting to $ad \ hoc$ parameters in the Reynolds stress constitutive relation~\citep{luan2025constructing}. However, no theoretical analysis or numerical simulation has yet been reported that solves these kinetic models directly. A key drawback of these kinetic models is that they impose turbulence by analogy, replacing molecular thermal fluctuations with turbulent fluctuations, without deriving a description of the turbulence dynamics from the first principles.

Another strategy is to develop kinetic models for turbulence directly from the Boltzmann equation by systematically eliminating small-scale fluctuations through successive averaging procedures.~\cite{chen1999analysis} applied renormalization group techniques to the Boltzmann-BGK equation~\citep{bhatnagar1954model} to develop a subgrid-scale viscosity model (SGS). Subsequently,~\cite{ansumali2004kinetic} constructed a turbulence model using a mean-field approach to filter SGS effects, establishing connections with the Smagorinsky-type closures. However, a fundamental limitation arises because the averaged equilibrium distribution retains the thermodynamic temperature in the same functional role as turbulent kinetic energy. This prevents the systematic removal of molecular thermal effects through successive averaging processes, resulting in eddy viscosity formulations that exhibit unphysical dependence on molecular temperature rather than being determined solely by turbulent flow characteristics.~\cite{girimaji2007boltzmann} rigorously derives a filtered-turbulence kinetic equation from the Boltzmann--BGK framework, in which turbulence effects (Reynolds stress) enter explicitly via additional drift terms. Nevertheless, the resulting Reynolds stress contributions cannot be obtained directly from the distribution function and therefore remain unclosed; in their LES implementation, a conventional Smagorinsky-type SGS closure is employed.
%%~\cite{girimaji2007boltzmann} develops a filtered-turbulence kinetic equation rigorously from the Boltzmann-BGK equation and makes turbulence effects appear explicitly through additional drift terms. In practice, the Reynolds-stress contributions remain unclosed and must therefore be modeled, since they cannot be obtained directly from the distribution function. ~\cite{girimaji2007boltzmann} demonstrates this by adopting a conventional Smagorinsky-type SGS closure in the LES implementation.
%In practice, however, the Reynolds stress contributions remain unclosed and must be modeled; ~\cite{girimaji2007boltzmann} demonstrates this by employing a conventional Smagorinsky-type SGS closure in the LES implementation.
This limitation highlights fundamental difficulties in applying the conventional Boltzmann equation to turbulence modelling and suggests the need for alternative approaches to statistical averaging.
%~\cite{girimaji2007boltzmann} further formulated a Boltzmann kinetic equation by incorporating SGS closure to characterize filtered turbulent flows, applicable to both continuum and non-continuum effects. 
Recently,~\cite{yang2025wave} developed the wave-particle turbulence simulation method, which conjectured and numerically showed that kinetic representations may be advantageous in describing turbulent non-equilibrium physics.

Several kinetic-based approaches reformulate incompressible NS turbulence in terms of velocity distribution functions (VDFs), providing a complementary viewpoint to conventional RANS. 
\cite{lundgren1967distribution} derived an infinite hierarchy of equations for the multi-point VDF of fluid elements, starting from the incompressible NS equations. In this hierarchy, the $n$th equation contains an unknown ($n+1$)-point VDF, closely resembling the Bogoliubov-Born-Green-Kirkwood-Yvon (BBGKY)~\citep{cohen1962fundamental} hierarchy for multi-particle VDF in kinetic theory. 
%The Boltzmann equation for short-range interacting particles and the Fokker-Planck equation for long-range Coulomb interactions are derived from the BBGKY hierarchy through appropriate closure approximations. 
In subsequent work, \cite{lundgren1969model} attempted to close the system at the one-point level by employing a relaxation model identical in form to the BGK model. The resulting equation was solved for several idealized problems without solid boundaries, and good agreement with available experiments was found. 
\cite{srinivasan1977turbulent} extended the work of~\cite{lundgren1969model} to turbulent flows bounded by solid walls, developing two types of boundary conditions based on the wall function method and employing the discrete ordinate method for numerical solution. However, due to the restriction to regions outside the viscous sublayer, the near-wall Reynolds shear stress was overpredicted, leading to an incorrect shear-stress profile.
Lundgren's model equation provides a very good description of turbulence for the flow situations considered and offers an analytical tool for further study of more complex turbulent flows; yet to our knowledge, this approach has not been pursued in subsequent turbulence research. \cite{chen2023average,chen2024average} take a different approach but arrive at a similar BGK-type model for the VDF. Their analysis is developed from the idea of Klimontovich-type kinetic equation~\citep{klimontovich1969statistical} for fluid elements, which exactly satisfies the incompressible NS equations. Ensemble-averaging the Klimontovich-type equation yields a one-point kinetic equation consistent with \cite{lundgren1967distribution} and \cite{pope1983lagrangian}, in which the collision term remains unclosed and needs to be modelled. \cite{chen2023average} adopt a BGK-type approximation as a closure for this term, while avoiding conflating turbulent fluctuations with the thermodynamic temperature. Unlike the model of \cite{lundgren1969model}, the dissipation rate is incorporated into the collision term rather than added through a source term, providing a more physically realistic representation of dissipation in eddy collisions.
Overall, kinetic-based formulations provide a systematic kinetic framework for turbulence modelling, offering an alternative viewpoint to purely heuristic analogies and facilitating a connection between fluctuation statistics and macroscopic effects.
Although the kinetic model developed by \cite{chen2023average} provides a rigorous theoretical foundation for turbulence modelling, two critical limitations remain. First, the CE analysis for the Reynolds stress is based on the results of the BGK model that does not properly incorporate dissipation rate\citep{chen2004expanded}, causing the relaxation time estimation by \cite{chen2024average} to produce unrealistically high turbulent Prandtl number and higher-order transport coefficients. Second, the model was constructed for unbounded fully developed turbulent flows, and suitable near-wall treatments for wall-bounded turbulence remains to be developed.

The aim of this study is to extend, analyze, and validate the kinetic model developed by \cite{chen2023average}. Through Chapman-Enskog expansion, we derive eddy viscosity and gradient diffusion models with transport coefficients determined from kinetic theory principles, reducing reliance on  $ad \ hoc$ parameters. The theoretical results are validated through detailed comparison with experimental data and direct numerical simulations of turbulent plane Couette flow. 
The remainder of the paper is organized as follows. \S\ref{Sec: kinetic modelling} presents the kinetic model for incompressible turbulence. In \S\ref{Sec: CE}, we establish connections with traditional eddy viscosity models and compare the transport coefficients. 
\S\ref{Sec:near-wall treat} presents an extension of the model to wall-bounded turbulence. In \S\ref{Sec:results}, we validate the models through turbulent Couette flow simulations, and \S\ref{Sec:conclusion} provides concluding remarks.

\section{Kinetic model for incompressible turbulence}\label{Sec: kinetic modelling}
\subsection{Kinetic Representation of incompressible turbulence}\label{Sec: kinetic model}
Rather than relying on the ensemble-averaged Boltzmann equation originally devised for molecular gases, \cite{chen2023average} provides a comprehensive kinetic representation of incompressible turbulence at both the NS (fluid elements) level and the RANS (eddies) level. They first introduced a Klimontovich-type kinetic equation for fluid elements,
\begin{equation}\label{Eq: Klimontovich eq}
    \partial_t f + \boldsymbol{\xi}\cdot \nabla f +\boldsymbol{a}\cdot \nabla_{\boldsymbol{\xi}}f=0, \qquad \boldsymbol{a} = -\nabla p +\nu_0 \nabla^2\boldsymbol{u},
\end{equation}
where $f \equiv f(\boldsymbol x, \boldsymbol \xi, t)$ is the velocity distribution function (VDF) for fluid elements with velocity $\boldsymbol \xi \equiv \left(\xi_x, \xi_y, \xi_z\right)$ at position $\boldsymbol x \equiv \left(x, y, z\right)$ and time $t$. $\nabla$ and $\nabla_{\boldsymbol\xi}$ are the gradient operators in physical and velocity spaces, respectively. $\nu_0$ is the molecular viscosity, $p$ is the pressure determined by the incompressibility constraint, and the fluid velocity $\boldsymbol{u}$ is defined by
\begin{equation}
    \boldsymbol{u} = \int \boldsymbol{\xi} fd\boldsymbol{\xi},
\end{equation}
and $\int fd\boldsymbol{\xi}=1$ for an incompressible flow. Taking the moments of~\eqref{Eq: Klimontovich eq}, we obtain
\begin{equation}
    \begin{aligned}
        \nabla \cdot \boldsymbol{u}&=0,\\
        \partial_t \boldsymbol{u} + \left( \boldsymbol u\cdot \nabla\right)\boldsymbol{u} &= - \nabla p + \nu_0 \nabla^2 {\boldsymbol u},
    \end{aligned}
\end{equation}
which exactly reproduces the NS equations at the hydrodynamic level. The Klimontovich-type equation describes representative fluid elements that follow exactly the NS fluid streamlines. %Therefore it comes as a no surprise that its velocity moment gives rise exactly to the NS equations. 
Physically, it amounts to a Lagrangian representation of the Navier-Stokes flow.  It is in this sense that it is exact and from the first principles. 

Subsequently, applying ensemble averaging to the kinetic equation~\eqref{Eq: Klimontovich eq} provides the kinetic description of the incompressible mean turbulent field:
\begin{equation}\label{Eq:kinetic equation}
    \partial_t F + \boldsymbol \xi \cdot \nabla F + \bar{\boldsymbol a} \cdot \nabla_{\boldsymbol{\xi}} F = \mathcal{C},
\end{equation}
where $F \equiv F(\boldsymbol x, \boldsymbol \xi, t)=\langle f(\boldsymbol x, \boldsymbol \xi, t \rangle$ is an ensemble averaged VDF, with $\langle \cdot \rangle$ denoting the ensemble average. $\mathcal{C} \equiv -\nabla_{\boldsymbol{\xi}} \cdot \langle \boldsymbol{a}' f' \rangle$ is the collision operator, with $\boldsymbol{a}'$ and $f'$ denoting the fluctuating components of $\boldsymbol{a}$ and $f$, respectively. And the external force $\bar{\boldsymbol a}$ is given as
\begin{equation}\label{Eq:force term}
    \bar{\boldsymbol a} =\langle \boldsymbol{a}\rangle= - \nabla \bar p + \nu_0 \nabla^2 {\boldsymbol U},
\end{equation}
where $\nu_0$ is the molecular viscosity, $\bar p = \langle p \rangle$ and $\boldsymbol U = \langle \boldsymbol{u}\rangle$ are the mean (ensemble averaged) pressure and velocity, respectively. The mean pressure distribution is governed by the mean velocity field according to the constraint $\nabla^2 \bar{p} = -\nabla \nabla : \boldsymbol{U} \boldsymbol{U} - \nabla \nabla :\left(\langle \boldsymbol{u}' \boldsymbol{u}’ \rangle\right)$.

The collision operator $\mathcal{C}$ satisfies conservation of mass and momentum,
\begin{equation}
    \int \mathcal{C} d\boldsymbol \xi = 0, \quad \quad  \int \boldsymbol \xi \mathcal{C} d\boldsymbol \xi = 0.
\end{equation}
But the collision is not conservative with respect to energy, 
%entailing energy dissipation,
\begin{equation}\label{Eq: col K}
    \frac{1}{2} \int \left(\boldsymbol \xi-\boldsymbol U\right)^2 \mathcal{C} d\boldsymbol \xi =-\nabla\cdot\langle p' \boldsymbol{u}'\rangle+\nu_0 \nabla^2K-\epsilon,
\end{equation}
where $p'$ and $\boldsymbol{u}'$ denoting the fluctuating components of $p$ and $\boldsymbol{u}$, respectively. $\epsilon=\nu_0 \langle \nabla \boldsymbol{u}':\nabla \boldsymbol{u}' \rangle$ is the dissipation rate. The kinetic equation obtained by ensemble averaging the Klimontovich-type equation is also exact before any closure model is applied for the collision term. %Therefore, the ensemble averaged Klimontovich-type equation~\eqref{Eq:kinetic equation} can be regarded as a Boltzmann equation from the kinetic perspective, describing the transport and collision of eddy.

Mean quantities, such as the mean velocity $\boldsymbol U$, turbulent kinetic energy (TKE) $K$, Reynolds stress tensor $\boldsymbol \sigma$ and turbulent kinetic energy flux $\boldsymbol{Q}$, can be obtained by integrating the ensemble averaged VDF over the velocity space, i.e.
\begin{equation}
\begin{aligned}\label{Eq:macro quantities}
\boldsymbol{U} &= \langle \boldsymbol{u}\rangle= \int \boldsymbol{\xi} F d\boldsymbol \xi,\\
K &= \frac{1}{2} \langle \left(\boldsymbol{u}'\right)^2\rangle= \frac{1}{2} \int \left(\boldsymbol \xi-\boldsymbol U\right)^2 F d\boldsymbol \xi,\\
\boldsymbol{\sigma} &=- \langle \boldsymbol{u}'\boldsymbol{u}'\rangle= -\int (\boldsymbol{\xi}-\boldsymbol{U})(\boldsymbol{\xi}-\boldsymbol{U})  F d\boldsymbol \xi,\\
\boldsymbol{Q} &=\frac{1}{2} \langle \boldsymbol{u}'\left(\boldsymbol{u}'\right)^2\rangle= \frac{1}{2} \int (\boldsymbol{\xi}-\boldsymbol{U})(\boldsymbol{\xi}-\boldsymbol{U})^2  F d\boldsymbol \xi,
\end{aligned}   
\end{equation}
where $\int F d\boldsymbol \xi = 1$, corresponding to the normalization of the VDF for incompressible flow. More generally, arbitrary higher-order central moments of the velocity fluctuations can be obtained from the VDF as $\langle u'_{i_1} u'_{i_2} \dots u'_{i_n}\rangle=\int (\xi_{i_1}-U_{i_1})(\xi_{i_2}-U_{i_2})\dots(\xi_{i_n}-U_{i_n})Fd\boldsymbol{\xi}$.

The well-known incompressible RANS equations can be obtained by taking the moments of the kinetic equation~\eqref{Eq:kinetic equation}, 
\begin{equation}
    \begin{aligned}
        \nabla \cdot \boldsymbol{U} &= 0,\\
        \partial_t \boldsymbol{U} + \left( \boldsymbol U\cdot \nabla\right)\boldsymbol{U} &= - \nabla \bar p + \nu_0 \nabla^2 {\boldsymbol U} + \nabla \cdot \boldsymbol \sigma,\\
         \partial_t K+ \left( \boldsymbol U\cdot \nabla\right) K&= -\nabla \cdot \boldsymbol{Q}+\boldsymbol \sigma:\boldsymbol S -\nabla\cdot\langle p' \boldsymbol{u}'\rangle+\nu_0 \nabla^2K -\epsilon,
    \end{aligned}
\end{equation}
where $\boldsymbol S=\frac{1}{2}\left[\nabla \boldsymbol U + (\nabla \boldsymbol U)^T \right]$ is the rate of strain tensor. 
It can be seen that $K$, $\boldsymbol{\sigma}$, and $\boldsymbol{Q}$ are completely determined by the solution of the averaged kinetic equation without any assumptions in this kinetic representation. Moreover, in principle, transport equations for any other higher-order central moments can be derived by taking corresponding moments of the kinetic equation.

\subsection{BGK model for average turbulence dynamics}\label{Sec: bgk model}
Since the collision term lacks an explicit form in terms of the averaged variables, one needs to find an appropriate closure in the above kinetic representation. Interestingly, the asymptotic limit of the collision process $\mathcal{C}$ assumes a Gaussian form based on semi-theoretical arguments~\citep{chen2023average}, which is also consistent with experimental observations in homogeneous turbulent flows where the one-point VDF exhibits a local Gaussian form~\citep{tavoularis1981experiments,sreenivasan2021dynamics,sreenivasan2024saturation}. As a result, \cite{chen2023average} modeled the collision term $\mathcal{C}$ as a relaxation process toward the local Gaussian form with a relaxation time $\tau$, namely, a BGK form,  %specifically in the form of a BGK model,
\begin{equation}\label{Eq:BGK term}
    \mathcal{C}_{\text{BGK}}(F) = \frac{F^{eq}-F}{\tau},
\end{equation}
where $F^{eq}$ is the local equilibrium VDF, 
\begin{equation}\label{Eq: equilibrium VDF}
    F^{eq} = \left(\frac{4}{3}\pi K^{eq}\right)^{-3/2} \exp{\left[ -\frac{3\left(\boldsymbol \xi-\boldsymbol U\right)^2}{4K^{eq}}\right]}, 
\end{equation}
where $K^{eq}$ is the local equilibrium TKE. Notably, in dissipative turbulence $F^{eq}$
denotes a target function representing the asymptotic state of the BGK collision process, which is in general never achieved but serves as the direction towards which the collision process drives $F$. It excludes the non-trivial flow-induced fluctuations contained in $F$; the latter exhibits non-Gaussian properties~\citep{frisch1996turbulence} such as flatness and skewness. In the homogeneous case, the kinetic equation with the BGK model~\eqref{Eq:BGK term} admits a Gaussian solution whose variance equals the instantaneous TKE.

%denotes the local Gaussian reference distribution introduced by the BGK closure, rather than the physical terminal equilibrium state of a dissipative turbulent flow. It excludes the non-trivial flow-induced fluctuations contained in $F$; the latter exhibits non-Gaussian properties~\citep{frisch1996turbulence} such as flatness and skewness.}
%Compared with the canonical Maxwell equilibrium distribution, TKE assumes a role analogous to temperature in classical molecular thermodynamics. 
%It is worth noting that the characteristic turbulent velocity scale is often smaller than the mean velocity, i.e., $\sqrt{K} < |\boldsymbol{U}|$. Therefore, an averaged turbulent flow bears high-Mach-like similarities to compressible flows, even though it remains incompressible because pressure enforces the incompressibility constraint. 

Based on the properties of the collision term, one can obtain
\begin{equation}
 \begin{aligned}
\int F^{eq} d\boldsymbol \xi =\int F d\boldsymbol{\xi}= 1,\quad \quad \int \boldsymbol{\xi} F^{eq} d\boldsymbol \xi =\int \boldsymbol{\xi} F d\boldsymbol \xi= \boldsymbol{U}.
\end{aligned}   
\end{equation}
To simplify the analysis, \cite{chen2023average} considered statistically homogeneous, high-Reynolds-number turbulence (or equivalently, the spatially averaged TKE budget), for which the divergence-type transport terms in Eq.~\eqref{Eq: col K} do not contribute to the net balance, 
\begin{equation}\label{Eq:collision_epsilon}
    \int \frac{ \left(\boldsymbol \xi-\boldsymbol U\right)^2}{2} \frac{F^{eq}-F}{\tau}  d\boldsymbol \xi = \frac{K^{eq}-K}{\tau}=  -\epsilon,
\end{equation}
and we have the equilibrium TKE $K^{eq}=K-\tau \epsilon$  according to~\eqref{Eq:collision_epsilon}. 
In conventional RANS modeling, the pressure-diffusion term $-\nabla \cdot (p'\boldsymbol{u}')$ is rarely treated separately; instead it is commonly modeled together with the TKE flux $\boldsymbol{Q}$ using a gradient diffusion closure~\citep{launder1983numerical,wilcox1998turbulence}. In the present work, we neglect this term because it is quite small for simple flows~\citep{mansour1988reynolds,pope2001turbulent}.
Additionally, for low-Reynolds-number turbulence and/or near-wall and inhomogeneous flows, the molecular diffusion $\nu_0 \nabla^2K$ may be non-negligible and will be explicitly retained and discussed in Sec.~\ref{Sec: Low Re model}.

For the relaxation time, \cite{chen2024average} provide a theoretical estimation of $\tau = C_\tau K/\epsilon$ with $C_\tau = 6/7$ based on the fluctuation–dissipation theorem that is absent in purely hydrodynamic approaches for stationary, homogeneous and isotropic turbulent flows. It also determines the turbulent viscosity, turbulent Prandtl number, and all higher-order transport coefficients involved in $\boldsymbol \sigma$ and $\boldsymbol Q$ by an effective finite Knudsen number expansion. They obtain the turbulent viscosity $\nu_T = 2\tau K^{eq}/3 = 2 C_\tau(1-C_\tau)K^2/(3\epsilon)$ based on the CE analysis of the BGK model that does not properly incorporate dissipation of TKE~\citep{chen2004expanded}. This leads to unrealistically high turbulent Prandtl number and higher-order transport coefficients (See \S\ref{Sec: Comparison of the transport coefficients}).
Since turbulent viscosity exhibits a quadratic dependence on the coefficient $C_\tau$, we can select the another root instead of the original one:
\begin{equation}
    \tau = \frac{1}{7}\frac{K}{\epsilon}.
\end{equation}
This choice yields the same viscosity as derived by~\cite{chen2024average}, while providing more reasonable turbulent Prandtl number and improved higher-order transport coefficients.
%This coefficient is also very close to the value $1/6$ suggested by~\cite{lundgren1969model}, who assumed the relaxation time to scale as $\tau \sim L/\sqrt{K}$ where $L$ is the integral length scale of turbulence (defined through $\epsilon=K^{3/2}/L$).

In the end, the dissipation rate $\epsilon$ remains an unspecified quantity, requiring an additional scale equation to determine its evolution. The most straightforward approach is to adopt an established formulation from conventional turbulence models (e.g., the $K-\epsilon$ model) to prescribe the evolution equation for $\epsilon$. Alternatively, the conventional $K-\omega$ turbulence models can also be utilized, where the relaxation time is expressed as $\tau \sim  1/\omega$, with $\omega \sim \epsilon/K$ being the specific dissipation rate~\citep{wilcox1998turbulence}. These formulations remain incomplete from the perspective of kinetic theory, as they still depend on externally specified parameters. Therefore, a fully self-consistent kinetic theoretic description for inhomogeneous statistical turbulence would be an ultimate goal.

\section{Connection with traditional eddy viscosity models}
\label{Sec: CE}
\subsection{Chapman-Enskog analysis}
As mentioned in the Introduction, almost all traditional turbulence closures, including the eddy viscosity and gradient diffusion models, aim to approximate the Reynolds stress $\boldsymbol{\sigma}$ and the TKE flux $\boldsymbol{Q}$ by the mean variables $\boldsymbol U$ and $K$ in the RANS equations. This hydrodynamic description can be derived by looking for a normal solution to the kinetic equation~\eqref{Eq:kinetic equation} with the BGK collision term~\eqref{Eq:BGK term} by means of the CE expansion technique~\citep{chapman1990mathematical}. It should be noted that the CE expansion is primarily used to make the relationship with traditional closure models transparent.  We first use a reference velocity $U_{ref}$, reference collision time $\tau_{ref}$ and reference length $L_{ref}$ to non-dimensionalise the kinetic equation~\eqref{Eq:kinetic equation},
\begin{equation}\label{Eq:dimensionless kinetic equation}
    \frac{\partial \hat F}{\partial \hat t} + \hat {\boldsymbol \xi} \cdot \frac{\partial \hat F}{\partial \hat{\boldsymbol{x}}}+ \left(- \frac{\partial \hat {\bar p}}{\partial \hat{\boldsymbol{x}}} + \frac{1}{\text{Re}} \hat\nabla^2\hat{\boldsymbol{U}}\right) \cdot \frac{\partial \hat F}{\partial \hat{\boldsymbol{\xi}}}  =\frac{1}{\mathscr{K}} \frac{\hat F^{eq}-\hat F}{\hat \tau},
\end{equation}
where $\text{Re}=U_{ref}L_{ref}/\nu_0$ is the Reynolds number (Re), and $\mathscr{K}=U_{ref} \tau_{ref}/L_{ref}$ can be interpreted as a turbulence-based Knudsen number that quantifies the ratio of unresolved to resolved flow scales~\citep{luan2025constructing}. Notably, it vanishes when returning to the dimensional form. For simplicity, the caret indicating non-dimensional quantities will be omitted hereafter.
If $\mathscr{K}$ is regarded as a small parameter, the VDF deviations from equilibrium can be expanded in powers of $\mathscr{K}$ as
\begin{equation}\label{Eq: F Ce}
    F = F^{(0)} + \mathscr{K} F^{(1)}+\mathscr{K}^2 F^{(2)}+ \cdots ,
\end{equation}
where $F^{(0)} = F^{eq}$. Correspondingly, the temporal and spatial derivatives are also expanded formally as
\begin{equation}
    \partial_{t}=\mathscr{K} \partial_{t_1}+\mathscr{K} ^2\partial_{t_2}+\cdots, \qquad  \nabla = \mathscr{K}\nabla_1,
\end{equation}
and the force and strain rate are similarly expanded as
\begin{equation}
    \begin{aligned}
        \bar{\boldsymbol{a}} &= \mathscr{K}\bar{\boldsymbol{a}} ^{(1)}+\mathscr{K}^2\bar{\boldsymbol{a}} ^{(2)}, \quad \bar{\boldsymbol{a}}^{(1)}=-\nabla_1 \bar p, \quad \bar{\boldsymbol{a}} ^{(2)}=\nu_0\nabla_1^2\boldsymbol{U},\\
        \boldsymbol{S}&=\mathscr{K}\boldsymbol{S}^{(1)}, \quad \boldsymbol{S}^{(1)} = \frac{1}{2}\left(\nabla_1 \boldsymbol{U}+\left(\nabla_1 \boldsymbol{U}\right)^T\right).
    \end{aligned}
\end{equation}

Taking the velocity moments of the expanded VDF~\eqref{Eq: F Ce}, the expressions of $K$, $\boldsymbol \sigma$ and $\boldsymbol{Q}$ can be written as
\begin{equation}
    K = \sum_{n\geq0} \mathscr{K}^n K^{(n)}, \qquad      \boldsymbol{\sigma} = \sum_{n\geq0} \mathscr{K}^n \boldsymbol{\sigma} ^{(n)}, \qquad  \boldsymbol{Q} = \sum_{n\geq0} \mathscr{K}^n \boldsymbol{Q} ^{(n)},
\end{equation}
%\begin{equation}
%\begin{aligned}
%    K &= K^{(0)} + \mathscr{K} K^{(1)}+\mathscr{K}^2 K^{(2)}+ \cdots ,\\
%^    \boldsymbol \sigma &= \boldsymbol \sigma^{(0)} + \mathscr{K} \boldsymbol \sigma^{(1)}+\mathscr{K}^2 \boldsymbol \sigma^{(2)}+ \cdots ,\\
%    \boldsymbol Q &= \boldsymbol Q^{(0)} + \mathscr{K} \boldsymbol Q^{(1)}+\mathscr{K}^2 \boldsymbol Q^{(2)}+ \cdots.\\
%\end{aligned}
%\end{equation}
The detailed expansions and formulas at each order are provided in Appendix~\ref{Appendix: CE}. In particular, the zeroth-order approximation of TKE, Reynolds stress and TKE flux can be determined by taking moments of $F^{eq}$,
\begin{equation}
    \begin{aligned}
        K^{(0)} = K^{eq} \boldsymbol, \quad
       \boldsymbol{\sigma}^{(0)} = -\frac{2}{3}K^{eq} \boldsymbol I, \quad
       \boldsymbol{Q}^{(0)} = 0, 
    \end{aligned}
\end{equation}
where $\boldsymbol{I}$ is the unit tensor. Note that $\boldsymbol{\sigma}^{(0)}$ involves $K^{eq}$ rather than $K$, so it does not represent the isotropic portion of the Reynolds stress.

The first-order approximation can be obtained as
\begin{equation}
    \begin{aligned}
       \boldsymbol{\sigma}^{(1)} =-\frac{2}{3} K^{(1)} \boldsymbol I+ \frac{4\tau K^{eq}}{3}\boldsymbol S^{(1)}, \quad
       \boldsymbol{Q}^{(1)} = - \frac{10\tau K^{eq}}{9}\nabla_1 K^{eq}.
    \end{aligned}
\end{equation}
Truncating to first order recovers the linear eddy viscosity model for the Reynolds stress and gradient diffusion model for TKE flux,
\begin{equation}\label{Eq: 1st ce}
\begin{aligned}
    \boldsymbol \sigma\approx \boldsymbol{\sigma}^{(0)}+ \mathscr{K}\boldsymbol{\sigma}^{(1)}=-\frac{2K}{3}\boldsymbol I + 2 \nu_T \boldsymbol S,\quad
    \boldsymbol Q\approx\boldsymbol{Q}^{(0)}+ \mathscr{K}\boldsymbol{Q}^{(1)}=-\frac{\nu_T}{\text{Pr}_T} \nabla K,
\end{aligned}
\end{equation}
with $K\approx K^{(0)} + \mathscr{K} K^{(1)}$, and the turbulent viscosity $\nu_T$ and turbulent Prandtl number $\text{Pr}_T$ are given by
\begin{equation}\label{Eq: 1st ce nu_t}
    \nu_T=\frac{2\tau K^{eq}}{3}, \quad \text{Pr}_T = \frac{3K}{5K^{eq}}.
\end{equation}

The next order contributions to $\boldsymbol{\sigma}$ and $\boldsymbol{Q}$ are obtained as
\begin{equation}\label{Eq:2nd CE}
\begin{aligned}
       \boldsymbol{\sigma}^{(2)} 
        =&-\frac{2 K^{(2)}}{3} \boldsymbol{I} - \frac{4\tau}{3K^{eq}}\left(\partial_{t_1}+\boldsymbol{U}\cdot \nabla_1\right) \left[\tau  (K^{eq})^2\boldsymbol{S}^{(1)} \right]
        \\&-\frac{8 \tau^2 K^{eq}}{3} \left[\boldsymbol{S}^{(1)}\cdot \boldsymbol{S}^{(1)} - \frac{1}{3}(\boldsymbol{S}^{(1)}: \boldsymbol{S}^{(1)} )\boldsymbol{I}\right]
        +\frac{4\tau^2K^{eq}}{3}\left(\boldsymbol{S}^{(1)} \cdot \boldsymbol{\Omega}^{(1)} - \boldsymbol{\Omega}^{(1)} \cdot \boldsymbol{S}^{(1)} \right) 
        \\&-\frac{4\tau}{9}\left[\nabla_1 \left(\tau K^{eq}\nabla_1 K^{eq}\right)+\left( \nabla_1 \left(\tau K^{eq}\nabla_1 K^{eq}\right)\right)^T-\frac{2}{3}\nabla_1 \cdot \left(\tau K^{eq}\nabla_1 K^{eq}\right) \boldsymbol{I}\right] ,
        \\
        \boldsymbol{Q}^{(2)} 
        =&\frac{10\tau}{9K^{eq}}\left(\partial_{t_1}+\boldsymbol{U}\cdot \nabla_1\right) \left[\tau  (K^{eq})^2 \nabla_1 K^{eq} \right]
        +\frac{10\tau^2 K^{eq}}{9} (\nabla_1 K^{eq})\cdot \left(\nabla_1 \boldsymbol{U} \right)^T
        \\&- \frac{5\tau K^{eq}}{3} \nabla_1 \cdot \left(\frac{4\tau K^{eq}}{3} \boldsymbol{S}^{(1)}\right) + \frac{5\tau K^{eq}}{3} \boldsymbol{a}^{(1)},
\end{aligned}
\end{equation}
where $\boldsymbol{\Omega}=\frac{1}{2}\left[\nabla \boldsymbol{U} - \left( \nabla \boldsymbol{U}\right)^T\right]$ is the vorticity tensor of the mean velocity field. The second-order expansion, as shown in~\eqref{Eq:2nd CE}, provides a more accurate representation of the Reynolds stress and TKE flux. These higher-order effects, consistent with classical turbulence theory, are crucial for capturing complex turbulent behaviours such as secondary flow structures and flows subject to sudden distortions~\citep{chen2004expanded}. The second-order TKE flux term is analogous to the Burnett heat flux in rarefied flow that can describe non-Fourier heat transfer phenomena (counter-gradient heat flux)~\citep{venugopal2019non}. It could also be a potential responsible for some of the observed counter-gradient diffusion phenomena of TKE in buoyant flows~\citep{grotzbach1982direct,moeng1989evaluation,chandra2007analysis}. 

Truncating to second order, the TKE is expressed as 
\begin{equation}
    K \approx  K^{eq}+\mathscr{K}K^{(1)} +\mathscr{K}^2 K^{(2)}.
\end{equation}
Based on the assumption that $\partial_t\approx\mathscr{K}\partial_{t1}$, the Reynolds stress and the TKE flux can be approximated as
\begin{equation}\label{Eq:2nd ce q}
\begin{aligned}
    \boldsymbol \sigma\approx &\boldsymbol{\sigma}^{(0)}+ \mathscr{K}\boldsymbol{\sigma}^{(1)}+ \mathscr{K}^2\boldsymbol{\sigma}^{(2)}
    \\=&-\frac{2K}{3}\boldsymbol I + 2\nu_T\boldsymbol S- \frac{\tau}{K^{eq}}\frac{D}{D t} \left[2\nu_T K^{eq} \boldsymbol{S} \right]
    -\frac{6\nu_T^2}{K^{eq}}\left[\boldsymbol{S}\cdot \boldsymbol{S} - \frac{1}{3}(\boldsymbol{S}: \boldsymbol{S} )\boldsymbol{I}\right]
    +\frac{3\nu_T^2}{K^{eq}}\left(\boldsymbol{S} \cdot \boldsymbol{\Omega} - \boldsymbol{\Omega} \cdot \boldsymbol{S} \right) 
    \\&-\frac{2\nu_T}{3 K^{eq}}\left[\nabla \left(\tau K^{eq}\nabla K^{eq}\right)+\left( \nabla \left(\tau K^{eq}\nabla K^{eq}\right)\right)^T -\frac{2}{3}\nabla \cdot \left(\tau K^{eq}\nabla K^{eq}\right) \boldsymbol{I}\right] ,
    \\
    \boldsymbol Q \approx &\boldsymbol{Q}^{(0)}+\mathscr{K}\boldsymbol{Q}^{(1)}+\mathscr{K}^2\boldsymbol{Q}^{(2)}
    \\=&- \frac{\nu_T}{\text{Pr}_T} \nabla K +\frac{\tau}{K^{eq}} \frac{D}{D t}\left[\frac{\nu_T}{\text{Pr}_T} K^{eq}\nabla K\right]+\frac{5 \nu_T^2}{2K^{eq}} \left(\nabla K^{eq}\right)\cdot  \left(\nabla \boldsymbol{U} \right)^T-\frac{5\nu_T}{2} \nabla \cdot\left(2\nu_T \boldsymbol{S}\right)- \frac{5\nu_T}{2} \nabla \bar p,
\end{aligned}
\end{equation}
where $D/Dt \equiv \partial_{t}+\boldsymbol{U}\cdot \nabla$ is the material derivative along the mean velocity field.
The linear eddy viscosity model~\eqref{Eq: 1st ce} is in close agreement with the results obtained from traditional kinetic equations for turbulent flow (such as the Boltzmann-BGK model that does not involve TKE dissipation~\citep{chen2004expanded} and the Fokker-Planck equation~\citep{heinz2007unified,luan2025constructing}). In contrast, the nonlinear results~\eqref{Eq:2nd ce q} derived from the present model yield different material derivative of $\boldsymbol{S}$ and higher-order derivative terms of TKE. Additionally, we have derived both the gradient diffusion model and its nonlinear counterpart, which are rarely modeled in conventional turbulence models.

The material derivative in~\eqref{Eq:2nd ce q} is both interesting and revealing. 
Considering only the linear closures with respect to $\boldsymbol{S}$ and $\nabla K$ in equation~\eqref{Eq:2nd ce q}, we have
\begin{equation}
    \boldsymbol{\phi}\approx \boldsymbol \phi^{lin}-\frac{\tau}{K^{eq}}\frac{D \left(K^{eq} \boldsymbol{\phi}^{lin}\right)}{Dt} =\left(1-\frac{\tau}{K^{eq}}\frac{D K^{eq}}{D t} \right)\boldsymbol{\phi}^{lin} -\tau \frac{D \boldsymbol{\phi}^{lin}}{Dt},
\end{equation}
where $\boldsymbol{\phi} = \boldsymbol{\sigma}$ or $\boldsymbol{\phi} = \boldsymbol{Q}$, whose linear parts $\boldsymbol \phi^{lin}$ are $2\nu_T \boldsymbol{S}$ and $-\frac{\nu_T}{\text{Pr}_T}\nabla K$, respectively. First, we find that the effect of a finite relaxation time $\tau$ implies that the stress and TKE flux are not simply determined by the immediate local $\boldsymbol \phi^{lin}$, but rather emerge from the history of $\boldsymbol \phi^{lin}$ at an earlier moment and at a location further upstream in the flow, that is,
\begin{equation}
   \boldsymbol{\phi} \approx  \left(1-\frac{\tau}{K^{eq}}\frac{D K^{eq}}{Dt}\right) \boldsymbol{\phi}^{lin}\left(\boldsymbol{x}-\boldsymbol{U}\tau^*,t-\tau^*\right), \quad \tau^* =\frac{\tau}{1-\frac{\tau}{K^{eq}}\frac{D K^{eq}}{Dt}}.
\end{equation}
This memory effect could be responsible for the turbulent phenomena seen in
rapid distortion processes~\citep{yakhot1992development,hamlington2008reynolds}. Furthermore, the correction coefficient incorporates the local departure from the production-dissipation balance through the material derivative $DK^{eq}/Dt$. This non-equilibrium adjustment could address the well-documented overprediction of $K$ and $\epsilon$ by classical linear closures in scenarios involving sustained energy injection, such as homogeneous shear turbulence~\citep{yoshizawa1993nonequilibrium}. Moreover, the ``renormalized" relaxation time $\tau^*$ can also correct the unrealistically fast anisotropy decay predicted by $\tau =C_\tau  K/\epsilon$ with $C_\tau$ being a constant in homogeneous turbulence~\citep{lundgren1969model}.
Naturally, for flows characterized by slow spatial and temporal variations, these effects can be neglected, recovering the conventional eddy viscosity model.

\subsection{Comparison of the transport coefficients} \label{Sec: Comparison of the transport coefficients}
In this subsection, we conduct a comparison of the transport coefficients derived from the kinetic model under various relaxation time formulations with those obtained from the conventional $K-\epsilon$ models and quadratic nonlinear eddy viscosity models documented in the literature. For the $K-\epsilon$ model, the Reynolds stress and TKE flux are given by
\begin{equation}
    \boldsymbol{\sigma} = -\frac{2K}{3}\boldsymbol I + 2 \nu_T \boldsymbol S, \quad
    \boldsymbol Q=-\frac{\nu_T}{\text{Pr}_T} \nabla K,  
\end{equation}
with 
\begin{equation}
    \nu_T = C_{\mu}\frac{K^2}{\epsilon}.
\end{equation}

For the present choice of $\tau=K/(7\epsilon)$, we find
\begin{equation}
    C_\mu = 0.0816, \quad \quad \text{Pr}_T = 0.7;
\end{equation}
On the other hand, the choice of $\tau = 6K/(7\epsilon)$~\citep{chen2024average} gives
\begin{equation}
    C_\mu = 0.0816, \quad \quad \text{Pr}_T = 4.2;
\end{equation}
For comparison, we list the corresponding values from two most representative $K-\epsilon$ models: the standard $K-\epsilon$ model~\citep{launder1983numerical} gives
\begin{equation}
    C_\mu = 0.09, \quad \quad \text{Pr}_T = 1.0;
\end{equation}
and the renormalization group model~\citep{yakhot1992development} gives 
\begin{equation}
    C_\mu = 0.0845, \quad \quad \text{Pr}_T = 0.0719.
\end{equation}
It can be found that although the values of $C_\mu$ given by both choices of the relaxation time are close to the classical $K-\epsilon$ model, the corresponding Prandtl numbers from the two choices exhibit significant differences. In particular, the choice of $\tau = K/(7\epsilon)$ produces a turbulent Prandtl number of $0.7$, which is quantitatively consistent with the well-established $K-\epsilon$ model parameters. In contrast, the case of $\tau = 6K/(7\epsilon)$ yields a Prandtl number of $4.2$, which significantly exceeds the conventionally adopted range of $0.7\sim0.9$ in turbulence modelling~\citep{kays1994turbulent,basu2021turbulent}. It is also noted that numerous studies choose $\text{Pr}_T=1.0$ for simplicity, relying on the so-called Reynolds-analogy assumption~\citep{tennekes1972first,sutton2020atmospheric}.

Furthermore, a quantitative comparison is conducted between the Reynolds stress derived from the second-order CE expansion and the higher-order transport coefficients of some traditional quadratic nonlinear eddy viscosity models. When the nonlinear model is truncated to quadratic terms, $\boldsymbol{\sigma}$ can be expressed as
\begin{equation}
    \begin{aligned}
    \boldsymbol \sigma=&-\frac{2K}{3}\boldsymbol I + 2 C_\mu \frac{K^2}{\epsilon } \boldsymbol S-C_1 \frac{K^3}{\epsilon^2 }\left[\boldsymbol{S}\cdot \boldsymbol{S} - \frac{1}{3}(\boldsymbol{S}: \boldsymbol{S} )\boldsymbol{I}\right]-C_2 \frac{K^3}{\epsilon^2 }\left(\boldsymbol{\Omega} \cdot \boldsymbol{S} -\boldsymbol{S} \cdot \boldsymbol{\Omega}\right) \\&- C_3 \frac{K^3}{\epsilon^2 }\left[\boldsymbol{\Omega}\cdot \boldsymbol{\Omega} - \frac{1}{3}(\boldsymbol{\Omega}: \boldsymbol{\Omega} )\boldsymbol{I}\right],
    \end{aligned}
\end{equation}
where $C_1$, $C_2$ and $C_3$ are three empirical coefficients in traditional nonlinear models. Table~\ref{tab:nlevms} lists the values for these coefficients in several typical quadratic nonlinear eddy viscosity models. The nonlinear terms derived directly from the second-order CE expansion exhibit quantitative agreement with those employed in nonlinear turbulence models. The present model incorporates additional terms that simultaneously account for both the material derivative of the strain tensor and higher-order TKE gradients, which are also considered in nonlinear turbulence models. It can be seen that the choice of $\tau = 6K/(7\epsilon)$~\citep{chen2023average,chen2024average} gives higher-order transport coefficients that are larger than those of other nonlinear models, including the coefficients of the additional terms. In summary, the present kinetic model with $\tau = K/(7\epsilon)$ yields transport coefficients that are quantitatively close to those of established $K-\epsilon$ models and quadratic nonlinear eddy viscosity models, which have been obtained empirically or semi-empirically.

\begin{table*}[htbp]  % 使用table*跨双栏
\centering
\footnotesize  % 缩小字体以适应宽表格
\setlength{\extrarowheight}{6pt}  % 增加行高
\begin{tabular}{>{\centering\arraybackslash}p{2.4cm} >{\centering}p{1.5cm} >{\centering}p{1.5cm} >{\centering}p{1.5cm} >{\centering}p{1.5cm} >{\centering\arraybackslash}p{3.9cm}}
\hline  % 上横杠
Model & $C_\mu$ & $C_1$ & $C_2$ & $C_3$ & Additional terms \\
\hline  % 表头下横杠
\cite{speziale1987nonlinear} & 0.09 & $-0.054$ & 0 & 0 & $-0.054\frac{K^3}{\epsilon^2}(\dot{ \boldsymbol{S}} - \frac{1}{3} (\text{tr}{\dot{\boldsymbol{S}}}) \boldsymbol{I})$ \\[8pt]

\cite{nisizima1987turbulent} & 0.09 & $-0.274$ & 0.065 & 0.374 & $-$ \\[8pt]

\cite{rubinstein1990nonlinear} & 0.0845 & 0.230 & 0.047 & $-0.190$ & $-$ \\[8pt]

\cite{myong1990prediction} & 0.09 & 0.101 & 0.086 & 0.0178 & $W_{\alpha \beta}$ \\[8pt]

%Present model with $\tau = 6K/(7\epsilon)$ 
\cite{chen2023average, chen2024average}& 0.0816 & 0.280 & 0.140 & 0 & $-\frac{0.14}{\epsilon}\frac{D}{Dt}\left[ K^3\boldsymbol{S}/\epsilon \right] + \boldsymbol{\mathcal{Q}}$ \\[8pt]

Present model & 0.0816 & 0.047 & 0.023 & 0 & $-\frac{-0.023}{\epsilon}\frac{D}{Dt}\left[ K^3\boldsymbol{S}/\epsilon \right] + \boldsymbol{\mathcal{Q}}$ \\[8pt]
\hline  % 下横杠
\end{tabular}

\medskip
\begin{minipage}{\textwidth}
\footnotesize
%$S = (k/\varepsilon)\sqrt{2S_{ij}S_{ij}}$, \quad $\Omega = (k/\varepsilon)\sqrt{2\Omega_{ij}\Omega_{ij}}$, \quad 
$\dot{\boldsymbol{S}} = \partial_t \boldsymbol{S} + \left(\boldsymbol{U}\cdot \nabla \right)\boldsymbol{S}- \left(\nabla \boldsymbol{U}\right) \boldsymbol{S}-\boldsymbol{S}\left(\nabla \boldsymbol{U}\right)^T$, \quad 
$W_{\alpha \beta} = \frac{2}{3}\frac{\nu_0 K}{\epsilon}\left(\frac{\partial \sqrt K}{\partial x_\gamma}\right)^2(-\delta_{\alpha \beta} - \delta_{\alpha \gamma }\delta_{\beta \gamma} + 4\delta_{\alpha \eta}\delta_{\beta \eta})$,\\
$\boldsymbol{\mathcal{Q}} =-0.007\frac{K}{\epsilon}\left[\nabla \left(K^2/\epsilon\nabla K\right)+\left(\nabla \left(K^2/\epsilon\nabla K\right)\right)^T-\frac{2}{3}\nabla \cdot \left(K^2/\epsilon\nabla K\right) \boldsymbol{I}\right]$.
\end{minipage}
\caption{Comparison of the transport coefficients with typical quadratic nonlinear eddy viscosity models.}
\label{tab:nlevms}
\end{table*}

\section{Extension to wall-bounded turbulence}\label{Sec:near-wall treat}
The kinetic turbulence model developed above is valid only for fully turbulent flows, which we refer to as the ``high-Reynolds-number kinetic (HR-BGK) model". However, in the near-wall region for wall-bounded flows, the turbulence is strongly damped and the effects of molecular viscosity become paramount~\citep{hanjalic1976contribution}. Furthermore, the turbulent motion is restricted and the relaxation time $\tau$ and the turbulent Reynolds number $\text{Re}_T = K^{2}/(\epsilon \nu_0)$ tend to zero.  To address this challenge, a natural approach is to employ wall function methods from conventional turbulence modelling, which apply the HR-BGK model outside the viscous sublayer using empirical near-wall correlations. Alternatively, to enable resolution of flows within the viscous sublayer, we develop a low-Reynolds-number kinetic (LR-BGK) model that incorporates damping functions and an additional viscous diffusion term. In the following subsections, we present the LR-BGK model, and the wall boundary conditions including non-slip condition for the LR-BGK model and wall function condition for the HR-BGK model.

\subsection{Low-Reynolds-number kinetic model} \label{Sec: Low Re model}
In the near-wall region where the relaxation time $\tau$ is sufficiently small, the kinetic model asymptotically reduces to the linear
eddy viscosity~\eqref{Eq: 1st ce} and the gradient diffusion models~\eqref{Eq: 1st ce nu_t}.
Most contemporary low Reynolds number $K-\epsilon$ models retain the transport-equation structure first introduced by~\cite{jones1972prediction}; they differ mainly in the addition of (i) van Driest damping function $f_\mu$, which is small within the near-wall region, and tends to unity in bulk region, and (ii) explicit molecular diffusion term $\nu_0 \nabla^2K$ of which are essential to represent the strong viscous diffusion of TKE that dominates within the viscous near-wall region. 

First, to incorporate the van Driest damping function $f_\mu$ explicitly into the turbulent viscosity formulation, the relaxation time $\tau$ is redefined as,
\begin{equation}\label{Eq:tau with f}
    \tau = \frac{f_\tau}{7}\frac{K}{\epsilon}, \qquad f_\tau =\frac{7-\sqrt{49-24 f_\mu}}{2},
\end{equation}
where $f_\tau$ is a damping function corresponding to $f_\mu$. Note that here the physical constraint $0 \leq f_\tau \leq 1$ is considered to ensure physically meaningful values.
%To match the turbulent viscosity definition in~\eqref{Eq: 1st ce nu_t} with the classical expression incorporating the van Driest damping function $f_\mu$, we equate:
Substituting this into the definition of turbulent viscosity~\eqref{Eq: 1st ce nu_t} yields
\begin{equation}
  \nu_T=\frac{2\tau K^{eq}}{3}= C_\mu f_\mu \frac{K^2}{\epsilon}, \qquad C_\mu = 0.0816.
\end{equation}
%Then with the relationship $K^{eq}=K-\tau\epsilon$ and the expression~\eqref{Eq:tau with f}, we can obtain
%\begin{equation}
%     \frac{2f_\tau\left(1-f_\tau/7\right)}{21}=C_\mu f_\mu,
%\end{equation}
%which gives 
%\begin{equation}
%f_\tau =\frac{7-\sqrt{49-24 f_\mu}}{2}.
%\end{equation}
A number of damping functions have been proposed. For instance, the expression proposed by~\cite{jones1972prediction} depends solely on $\text{Re}_T$. Subsequent refinements by \cite{myong1990new} and \cite{nagano1990improved} introduce dependencies on the dimensionless wall coordinate ($y^+$), thus explicitly incorporating sensitivity to near-wall length scales into the turbulence model.

Secondly, the explicit molecular viscosity diffusion $\nu_0 \nabla^2K$ is incorporated by introducing an additional source term $\mathcal{S}$ into the kinetic equation
\begin{equation}\label{Eq:kinetic equation with source}
    \partial_t F + \boldsymbol \xi \cdot \nabla F + \bar{\boldsymbol a} \cdot \nabla_{\boldsymbol{\xi}} F = \mathcal{C}_{\text{BGK}}(F)+\mathcal{S},
\end{equation}
where $\mathcal{S}$ satisfies the following moment requirements,
\begin{equation}
\begin{aligned}
    &\int \mathcal{S} d\boldsymbol \xi = 0, \qquad  \int \boldsymbol \xi \mathcal{S} d\boldsymbol \xi = 0,  \qquad
    \frac{1}{2}\int  \left(\boldsymbol \xi-\boldsymbol U\right)^2\mathcal{S}d\boldsymbol \xi = \nu_0 \nabla^2K.
\end{aligned}
\end{equation}
There are many possible ways to design the specific form of the source term.  Analogous to the formulation proposed by~\cite{lundgren1969model} for the dissipation term, we can choose
\begin{equation}\label{Eq: S term F}
    \mathcal{S} = -\frac{\nu_0 \nabla^2K}{2 K^{eq}} \nabla_{\boldsymbol{\xi}} \cdot \left[(\boldsymbol{\xi}-\boldsymbol{U})F\right].
\end{equation}
%Alternatively, following the well-known Shakhov model~\citep{shakhov1968generalization} for non-equilibrium rarefied gas flows and applying the Hermite‐polynomial expansion method, we can obtain an alternative formulation,
%\begin{equation}\label{Eq: Shakhov term F}
%    \mathcal{S} = \frac{3\nu_0 \nabla^2K}{2 K^{eq}} \left(\frac{\left( \boldsymbol{\xi} - \boldsymbol{U}\right)^2}{2K^{eq}} - 1\right) F^{eq}.
%\end{equation}
%Note that the above construction is formally equivalent to replacing $F$ with its equilibrium counterpart $F^{eq}$ in~\eqref{Eq: S term F}. This approximation is justified, because the viscous diffusion of TKE is relevant only when the turbulent Reynolds number $\text{Re}_T$ (and therefore the relaxation time $\tau$) is small, a regime in which the VDF remains close to equilibrium.

%In the present study, we refer to the kinetic model \eqref{Eq:kinetic equation with source} with the relaxation time~\eqref{Eq:tau with f} as ``low Reynold number kinetic (LR-BGK) model".

\subsection{Wall boundary conditions}
 %We now address two types of wall boundary conditions, namely, no-slip  condition for the LR-BGK model and wall function condition for the HR-BGK model.
A central challenge in applying kinetic theory to turbulence is the formulation of physically consistent boundary conditions because there is no experimentally or theoretically established expression for the VDF at a solid wall. In analogy with conventional turbulence models, the wall distribution
function $F(\boldsymbol{x}_w)$ is required to satisfy the following moment constraints:
\begin{equation}\label{Eq:wall_moments}
    \begin{aligned}
        &\int \boldsymbol{\xi} F(\boldsymbol{x}_w) d \boldsymbol{\xi} = \boldsymbol{U}_w,\\
        \frac{1}{2}&\int \left( \boldsymbol{\xi} -\boldsymbol{U}(\boldsymbol{x}_w)\right)^2F(\boldsymbol{x}_w) d \boldsymbol{\xi} = K_w,\\
    \end{aligned}
\end{equation}
where $\boldsymbol{x}_w$ denotes a point on the wall interface. The prescribed wall velocity $\boldsymbol{U}_w$ and turbulent kinetic energy $K_w$ are assigned according to the turbulence model: for LR-BGK model, they reduce to strict non-slip velocity and zero TKE, whereas for HR-BGK model, they are supplied by empirical wall functions \citep{launder1983numerical}, namely the logarithmic law for velocity and the corresponding estimation of TKE. 

However, different choices of $F$ can satisfy the
integral constraints in~\eqref{Eq:wall_moments}, so that $F(\bm{x}_w)$ itself is not unique. A similar situation is encountered in rarefied gas from the kinetic theory, where the establishment of accurate solutions has led to a dedicated research field focused on modelling gas–surface interactions for boundary conditions~\citep{cercignani1971kinetic,cercignani1972scattering,lord1991some,aoki2022boundary,kosuge2025applications}. An equally specialized effort will likely be required to develop reliable wall boundary conditions in kinetic representation of the average turbulence dynamics. Here, by leveraging boundary conditions from kinetic theory, we generalize two well-known boundary conditions of rarefied gas, namely the fully diffuse Maxwell boundary condition~\citep{maxwell1879vii} and the non-equilibrium extrapolation boundary condition~\citep{zhao2002non}, to turbulent flows.

The fully diffuse Maxwell boundary condition~\citep{maxwell1879vii} is employed in the LR-BGK model to enforce the non-slip velocity and the zero TKE at the wall. Importantly, this boundary condition produces a velocity slip and a TKE jump that both diminish linearly as $\tau$ decreases. In the near-wall region, $\tau$ is sufficiently small, so the induced slip and TKE jump are negligible, and the classical non-slip and zero-TKE limits are effectively recovered. In this boundary condition, it is assumed that particles leaving the surface follow an equilibrium VDF with the wall velocity and TKE, 
\begin{equation}
    F(\boldsymbol{x}_w, \boldsymbol{\xi})= \alpha _w F^{eq}\left(\boldsymbol{\xi};\boldsymbol{U}_w,K_w\right), \qquad \boldsymbol{\xi}\cdot \boldsymbol{n}_w>0,
\end{equation}
where $\boldsymbol{n}_w$ is the unit vector normal to the wall pointing to the cell, and $\alpha _w$ is the coefficient at the wall determined by the condition of zero mass flux,
\begin{equation}
    \int_{\boldsymbol{\xi}\cdot \boldsymbol{n}_w\leq0} \left(\boldsymbol{\xi}\cdot \boldsymbol{n}_w\right)F(\boldsymbol{x}_w, \boldsymbol{\xi}) d\boldsymbol{\xi}  +     \int_{\boldsymbol{\xi}\cdot \boldsymbol{n}_w>0}\left(\boldsymbol{\xi}\cdot \boldsymbol{n}_w\right) \alpha _w F^{eq}\left(\boldsymbol{\xi};\boldsymbol{U}_w,K_w\right) d\boldsymbol{\xi} = 0,
\end{equation}
which gives
\begin{equation}
\alpha_w = -\frac{\int_{\boldsymbol{\xi}\cdot \boldsymbol{n}_w\leq0} \left(\boldsymbol{\xi}\cdot \boldsymbol{n}_w\right)F(\boldsymbol{x}_w, \boldsymbol{\xi}) d\boldsymbol{\xi} }{\int_{\boldsymbol{\xi}\cdot \boldsymbol{n}_w>0}\left(\boldsymbol{\xi}\cdot \boldsymbol{n}_w\right)F^{eq}\left(\boldsymbol{\xi};\boldsymbol{U}_w,K_w\right) d\boldsymbol{\xi}}.
\end{equation}
In practice, $K_w$ is set to a very small positive value rather than exactly zero for numerical stability.

On the other hand,  for the HR-BGK model that is applicable outside the viscous sublayer, the non-equilibrium extrapolation boundary condition~\citep{zhao2002non} in conjunction with an empirical wall function \citep{launder1983numerical} is adopted. In the near-wall region, the mean velocity follows the classical logarithmic law, and the wall function is applied only inside the log-layer ($15 \lesssim y^{+} \lesssim 200$), bypassing the viscous sublayer:
\begin{equation}\label{eq:loglaw}
\frac{U(y)}{U_{\sigma}}=
\frac{1}{\kappa}\ln \left(y^{+}\right)+B,
\end{equation}
where $U_\sigma = \sqrt{\sigma_w}$ is the friction velocity obtained from the wall shear stress $\sigma_w$ with unity density, $\kappa \approx 0.41$ is the von Kármán constant, and $B$ is an empirical constant; $y$ denotes the wall-normal distance measured from the wall into the fluid and $y^+=U_{\sigma}y/\nu_0$ is its dimensionless form. In addition, the corresponding wall value of the TKE is specified by the usual log-layer estimate, $K_w = U^2_\sigma/\sqrt{C_\mu}$.

The VDF reflected from the wall located at the interface $\boldsymbol{x}_w$ is then reconstructed by the non-equilibrium extrapolation boundary condition~\citep{zhao2002non}, 
\begin{equation}
        F(\boldsymbol{x}_w, \boldsymbol{\xi})= F^{eq}\left(\boldsymbol{\xi};\boldsymbol{U}_w,K_w\right) - \left[  F(\boldsymbol{x}_{nc}, \boldsymbol{\xi}) -  F^{eq}(\boldsymbol{x}_{nc}; \boldsymbol{U}\left(\boldsymbol{x}_{nc} \right),K\left(\boldsymbol{x}_{nc} \right)) \right], \quad \boldsymbol{\xi}\cdot \boldsymbol{n}_w>0,
\end{equation}
where $\boldsymbol{x}_{nc}$ is the
neighboring cell of $\boldsymbol{x}_w$. The equilibrium term enforces the prescribed wall function moments, while the second term extrapolates the non-equilibrium part of the VDF, ensuring compatibility with the interior solution. 
%It is worth noting that the non-equilibrium extrapolation method assumes that the deviation from equilibrium is small to permit linear extrapolation. For highly non-equilibrium turbulent flows, this method may fail to work properly.

\section{Study of turbulent plane Couette flow}\label{Sec:results}
 The turbulent plane Couette flow provides an ideal benchmark~\citep{henry1984analytical,nisizima1987turbulent,schneider1989reynolds,andersson1994modeling} to test whether the present kinetic model can accurately capture wall-bounded shear turbulence. The LR-BGK and HR-BGK models, coupled with the diffuse and non-equilibrium extrapolation boundary conditions, respectively, are employed to simulate this flow at different Reynolds numbers. The central issue in wall-bounded shear turbulence is the prediction of mean velocity profiles and the derivation of friction laws. In this study, we consider the configuration of steady turbulent Couette flow between two infinite parallel plates located at $y=\pm h$. The two plates move in opposite directions with constant velocities $\pm U_w$ in the $x$-direction, respectively. The 
 corresponding Reynolds number is defined as $\text{Re}=U_w h/\nu_0$.
% delete in R2
%\begin{figure}
%  \centerline{\includegraphics[width=0.6\textwidth]{./Figures/Fig_Couetteflow}}
%  \caption{Sketch of turbulent plane Couette flow between two infinite parallel plates separated by $2h$. The plates move in opposite directions with constant velocities $\pm U_w$ in the $x$-direction, generating a monotonic mean velocity profile $U_x(y)$.}
%\label{fig:couetteflow}
%\end{figure}
\subsection{Numerical method and parameter set-up}
As mentioned in \S\ref{Sec: bgk model}, the dissipation rate $\epsilon$ is an input parameter to determine the relaxation time $\tau$ in the present kinetic model. For the HR-BGK model, we adopt the widely used transport equation for the dissipation rate proposed by~\cite{launder1983numerical} to calculate $\epsilon$,
\begin{equation}\label{Eq: epsilon evol}
    \frac{\partial \epsilon}{\partial t} + \left( \boldsymbol{U} \cdot \nabla\right)\epsilon= \nabla \cdot \left(\frac{\nu_T}{C_\epsilon} \nabla \epsilon \right)+C_{\epsilon1}\frac{\epsilon}{K}\boldsymbol{\sigma}:\nabla \boldsymbol U - C_{\epsilon2}\frac{\epsilon^2}{K},
\end{equation}
where $\nu_T= 0.09 K^2/\epsilon$, $C_\epsilon = 1.3$, $C_{\epsilon1} = 1.44$ and $C_{\epsilon2} = 1.92$. It is worth highlighting that both the turbulent kinetic energy $K$ and the stress tensor $\boldsymbol{\sigma}$ are obtained directly from the solution of kinetic equation; therefore, equation~\eqref{Eq: epsilon evol} retains higher-order non-equilibrium information that is absent from the linear eddy viscosity model. The wall boundary condition for Couette flow is implemented using a wall function approach. In this approach, the wall dissipation rate $\epsilon_w$ is prescribed based on the standard logarithmic layer estimate as $\epsilon_w = U_\sigma ^3/(\kappa y_1)$, where $y_1$ is the wall-normal distance from the wall to the center of the first grid cell adjacent to the wall. 

Conversely, for the LR-BGK model, we employ an alternative $\epsilon$-transport equation developed by~\cite{nagano1990improved}, where the turbulent viscosity and model coefficients are modified as $\nu_T= 0.09 f_\mu K^2/\epsilon$, $C_{\epsilon 1} = 1.45$ and $C_{\epsilon2} = 1.92f_{\epsilon2}$. The damping functions $f_\mu$ and $f_{\epsilon 2}$ are utilized to account for near-wall damping, with their specific forms defined therein.
The corresponding wall boundary condition is,
\begin{equation}
    \epsilon_w = \nu_0 \left(\frac{\partial \sqrt{K}}{\partial y}\right)^2.
\end{equation}

Since the turbulent Couette flow is a quasi-one-dimensional problem, the computational cost can be reduced by introducing the following reduced distribution functions,
\begin{equation}
\left(\Phi_1, \Phi_2, \Phi_3,\Phi_4\right)^T=\int \left(1, \xi_x,  \xi^2_x,\xi^2_z\right)^T F\left(\boldsymbol{x},\boldsymbol{\xi},t\right) d \xi_xd\xi_z.
\end{equation}
These reduced VDFs carry distinct essential physics in the context of turbulent transport. $\Phi_1$ represents the local mass density distribution in velocity space, directly related to the continuity equation and mass conservation. $\Phi_2$ corresponds to the streamwise momentum distribution, governing the transport of momentum in the flow direction and thus controlling the Reynolds shear stress generation. $\Phi_3$ characterizes the streamwise TKE distribution, which is crucial for understanding the anisotropic energy transport and the development of velocity fluctuations (or Reynolds normal stress). Finally, $\Phi_4$ represents the vertical kinetic energy distribution, capturing the cross-flow energy transport mechanisms. Together, these reduced VDFs encapsulate the essential physics of momentum and energy transport in wall-bounded turbulent flows while significantly reducing the computational complexity compared to the full three-dimensional kinetic equation.

The reduced kinetic equations for the four $\Phi_k$'s are then solved using a discretized velocity method; further details can be found in Appendix~\ref{sec: dvm}. To ensure accurate approximation of macroscopic moments in velocity space across various Re, computational parameters, including the spatial mesh size and the discretization of velocity
space, must be carefully selected through convergence studies until solution variations become negligible.

In our simulations, the computational domain is set to be $[-h,h]$. For the HR-BGK model, the computational domain is uniformly discretized, with the cell size determined according to different Re. The selection criterion ensures that the first cell center point is located inside the logarithmic layer. For the LR-BGK model, the nonuniform meshes are adopted to improve the prediction by using a locally refine mesh close to the boundary. The mesh points are given by $y_j/(2h)=\left(\zeta _j+\zeta _{j+1}\right)/2$ for $0\leq j\leq N_y-1$, where $\zeta _j$ is defined by
\begin{equation}
    \zeta_j = \frac{\tanh \left(a\left(j/N_y-1/2\right)\right)}{2 \tanh\left( a/2\right)}, \quad j=0,1,\cdots,N_y,
\end{equation}
with $a$  being a dimensionless stretching parameter controlling the characteristics of the mesh distribution, and $N_y$ is the number of mesh cells. In the present study, $a$ is taken as $2.5$. We utilize $N_y=401$ cells for the LR-BGK model, ensuring that the center of the first off-wall cell lies within the viscous sublayer for all cases considered. The velocity space truncated at $[-5\sqrt{U_w},5\sqrt{U_w}]$ is discretized by the trapezoidal rule~\citep{wu2014solving} with 4000 non-uniform velocity grid points. Furthermore, the time step is set to $0.05\times 5\sqrt{U_w}/(\Delta y_{min})$ with $ \Delta y_{min}$ being the minimum cell size, which satisfies the Courant-Friedrichs-Lewy (CFL) condition. The steady-state solution is considered to be achieved when the maximum relative change in the velocity field between two consecutive iterations falls below $10^{-10}$.

\subsection{Comparison with experiments and direct numerical simulations}
The present kinetic model is validated by comparing the predictions against benchmark data from experiments and DNS solutions. The comparison focuses on key flow parameters including mean velocity profiles, dimensionless velocity distributions, skin friction coefficients, and Reynolds stress components.

\begin{figure}[t]
\centering
  \subfloat[]{\includegraphics[scale=0.45,clip=true]{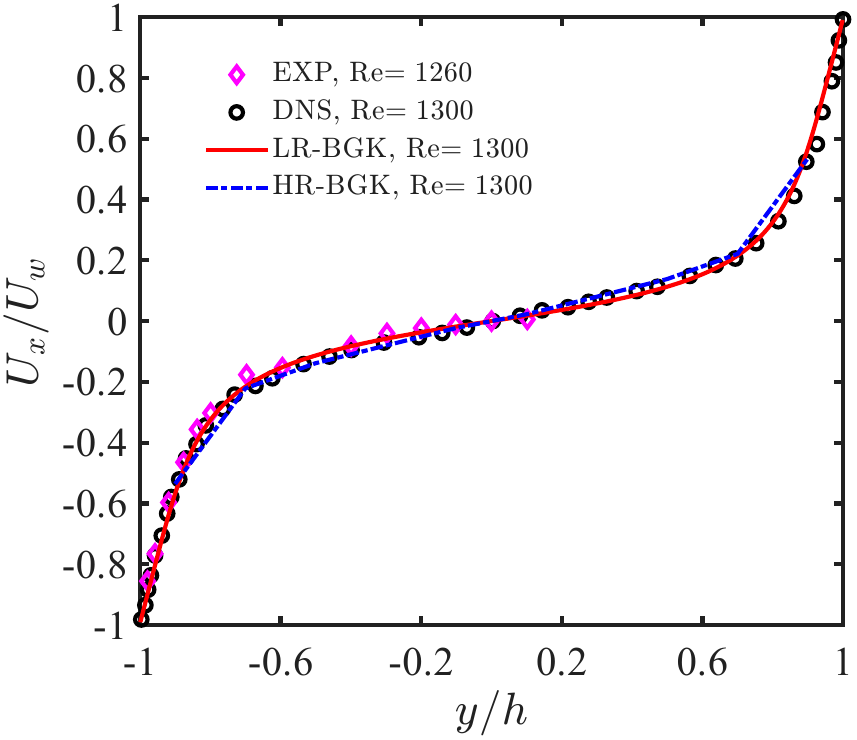}\label{fig:comparea_uxbar_Re1300}} \quad
  \subfloat[]{\includegraphics[scale=0.45,clip=true]{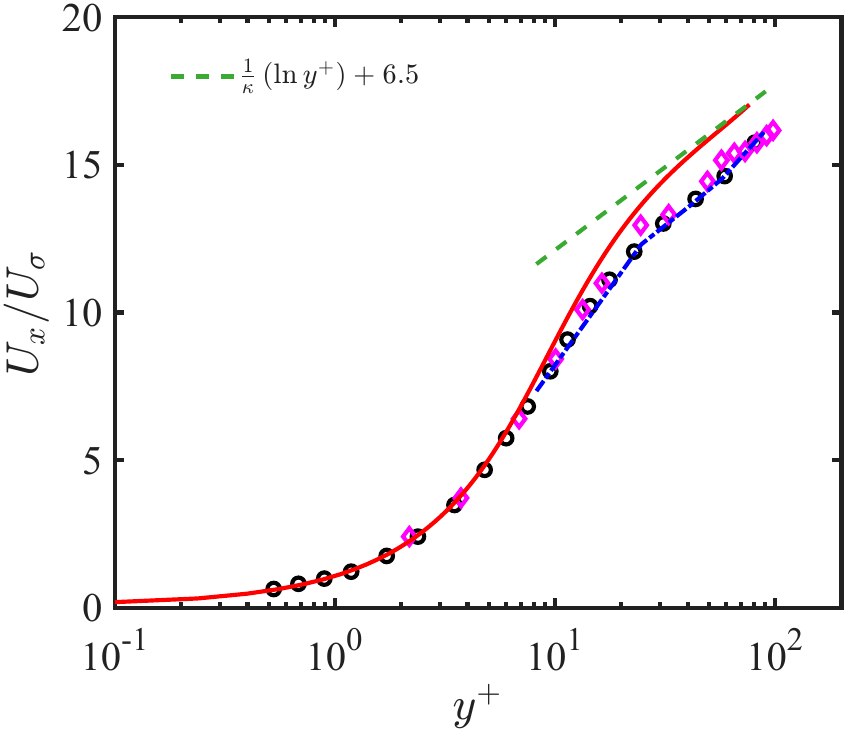}\label{fig:comparea_uxplus_Re1300}}
  \caption{Mean velocity profiles scaled in (a) outer and (b) wall units  of the turbulent Couette flow at $\text{Re} = 1300$. 
  %The solid and dash‑dot lines represent the predictions of the LR-BGK model and the HR-BGK model at $\text{Re} = 1300$, respectively.The diamonds denote the experimental data of~\cite{bech1995investigation} at $\text{Re} = 1260$, and the circles denote the DNS data of~\cite{bech1995investigation} at $\text{Re} = 1300$. The dashed line represents the classical logarithmic law profile $\left(\ln y^+\right)/0.41+6.5$. 
  }
\label{fig:comparea_ux_Re1300}
\end{figure}

\begin{figure}
\centering
  \subfloat[]{\includegraphics[scale=0.45,clip=true]{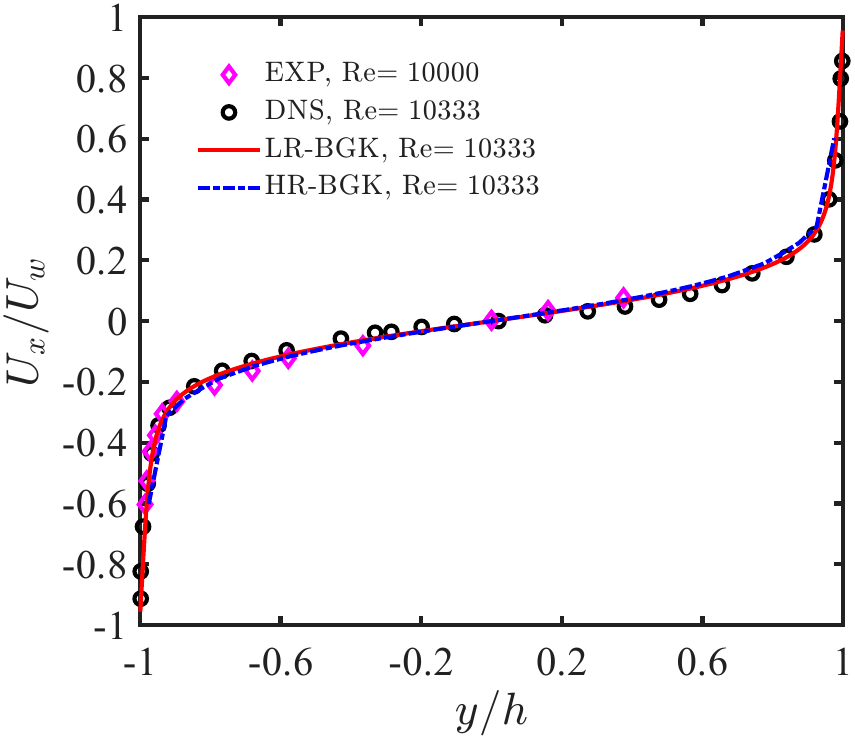}\label{fig:comparea_uxbar_Re10133}} \quad
  \subfloat[]{\includegraphics[scale=0.45,clip=true]{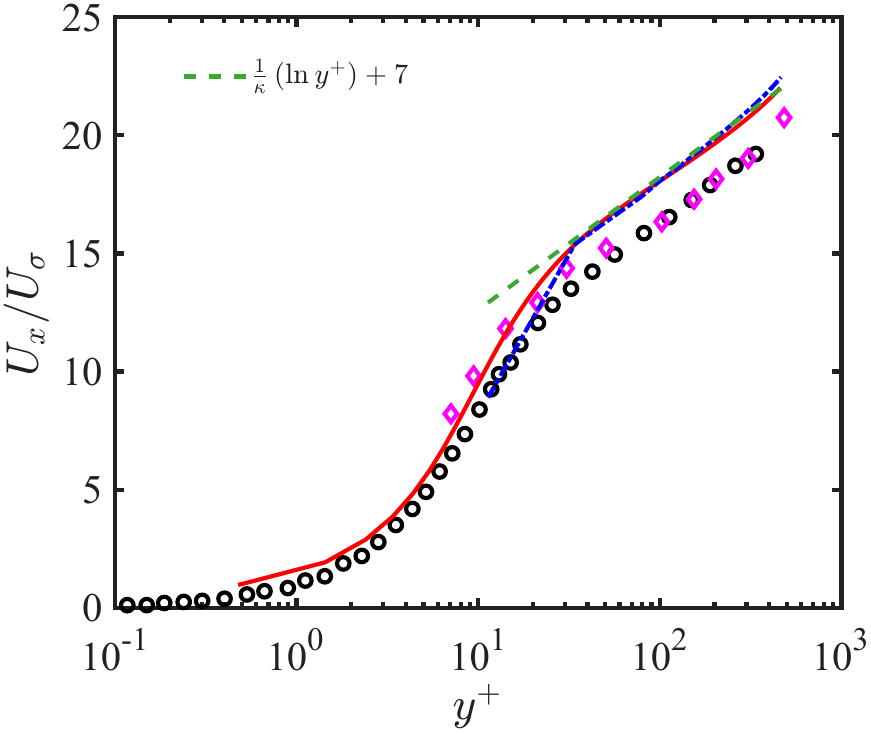}\label{fig:comparea_uxplus_Re10133}}
    \caption{Mean velocity profiles scaled in (a) outer and (b) wall units  of turbulent Couette flow at $\text{Re} = 10133$.
    %The solid and dash‑dot lines represent the predictions of the LR-BGK model and the HR-BGK model at $\text{Re} = 10133$, respectively.The diamonds denote the experimental data of~\cite{robertson1959study} at $\text{Re} = 10000$, and the circles denote the DNS data of~\cite{pirozzoli2014turbulence} at $\text{Re} = 10133$. The dashed line represents the classical logarithmic law profile $\left(\ln y^+\right)/0.41+7$.
    }
\label{fig:comparea_ux_Re10133}
\end{figure}

The mean velocity profiles predicted by the LR-BGK and HR-BGK models are compared with the experimental~\citep{bech1995investigation,robertson1959study} and DNS~\citep{bech1995investigation,pirozzoli2014turbulence} data at $\text{Re} = 1300$ and $10133$, as shown in figures~\ref{fig:comparea_ux_Re1300} and~\ref{fig:comparea_ux_Re10133}. The results are presented in both outer units (scaled by $U_w$ and $h$) and wall units (scaled by $U_\sigma$ and $y^+$) to examine the performance of the models across different flow regions. In the case of $\text{Re} = 1300$,  Figure~\ref{fig:comparea_uxbar_Re1300} demonstrates that both models accurately capture the overall velocity distribution in outer coordinates, closely matching the experimental and DNS data  of~\cite{bech1995investigation}. Specifically, in figure~\ref{fig:comparea_uxplus_Re1300}, the HR-BGK model shows excellent agreement throughout the entire flow domain, while the LR-BGK model exhibits slightly larger deviations in the logarithmic layer. This is because HR-BGK targets the outer region (bypassing the viscous sublayer), whereas LR-BGK includes stronger near-wall corrections that can slightly influence the log-layer profile.
%Specifically, the LR-BGK model shows excellent agreement throughout the entire flow domain, while the HR-BGK model exhibits slightly larger deviations in the near-wall region, which can be expected since the viscous sublayer is bypassed in the wall function boundary condition. 
At a higher Reynolds number ($\text{Re} = 10133$), figure~\ref{fig:comparea_ux_Re10133} shows that both the HR-BGK and LR-BGK models maintain their predictive accuracy, in good agreement with the experimental data of~\cite{robertson1959study} and DNS data of~\cite{pirozzoli2014turbulence}, though both models exhibit slightly larger deviations in the near-wall region. The logarithmic law profile~\eqref{eq:loglaw} is well-captured by both models with $\kappa = 0.41$, using $B = 6.5$ and $7.0$ for $\text{Re} = 1300$ and $10133$, respectively.

\begin{figure}
\centering
  {\includegraphics[scale=0.5,clip=true]{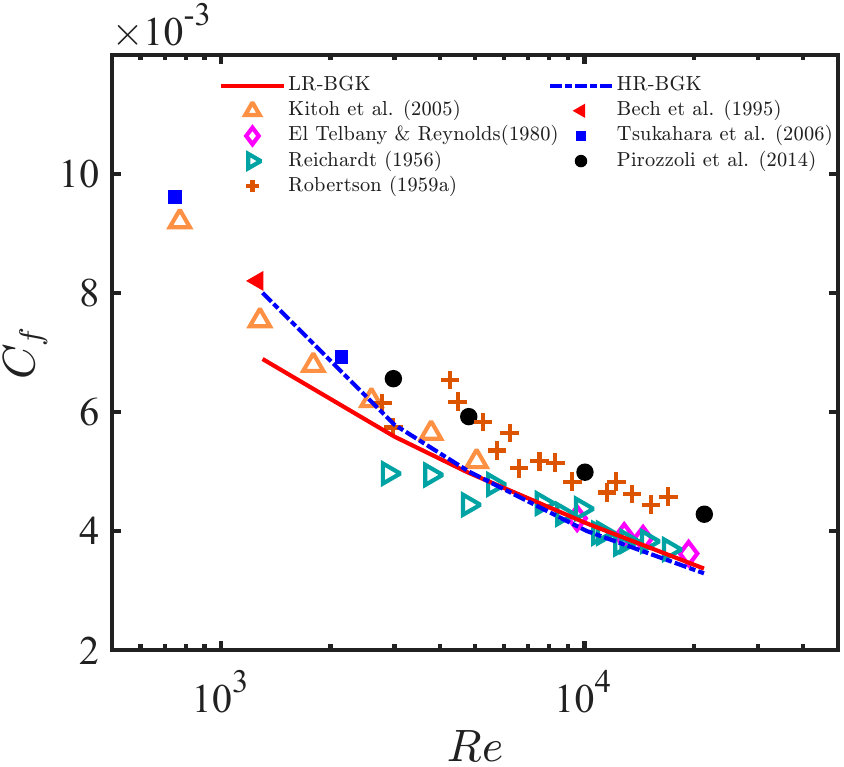}}
    \caption{Variation of skin friction coefficient with Reynolds number. 
    %The solid and dash‑dot lines represent the predictions of the LR-BGK model and the HR-BGK model, respectively. Open symbols refer to experimental data by~\cite{kitoh2005experimental} (triangles),~\cite{el1980velocity} (diamonds),~\cite{reichardt1956geschwindigkeitsverteilung} (right‑pointing triangles),~\cite{robertson1959turbulent} (pluses). Filled symbols refer to DNS data by~\cite{bech1995investigation} (left‑pointing triangles),~\cite{tsukahara2006dns} (squares),~\cite{pirozzoli2014turbulence} (circles).
    }
\label{fig:comparea_Cf}
\end{figure}

Figure~\ref{fig:comparea_Cf} presents the variation of the skin friction coefficient $C_f$ with Reynolds number, together with the comparisons with extensive experimental~\citep{kitoh2005experimental,el1980velocity,reichardt1956geschwindigkeitsverteilung,robertson1959turbulent} and DNS~\citep{bech1995investigation,tsukahara2006dns,pirozzoli2014turbulence} databases. The friction coefficient is defined as 
\begin{equation}
    C_f = \frac{\sigma_w}{\frac{1}{2}U_w^2},
\end{equation}
where $\sigma_w$ is the wall shear stress obtained directly from the VDFs. Both models demonstrate excellent agreement with the experimental data for Re between $10^3$ to $10^4$. The LR-BGK model shows minor deviations in the lower Re range, with deviations typically less than $5\%$ from the data of~\cite{kitoh2005experimental, bech1995investigation, tsukahara2006dns}. The HR-BGK model maintains consistent accuracy across the entire Re range, showing good agreement with both experimental and DNS data. 

\begin{figure}
\centering
  \subfloat[]{\includegraphics[scale=0.44,clip=true]{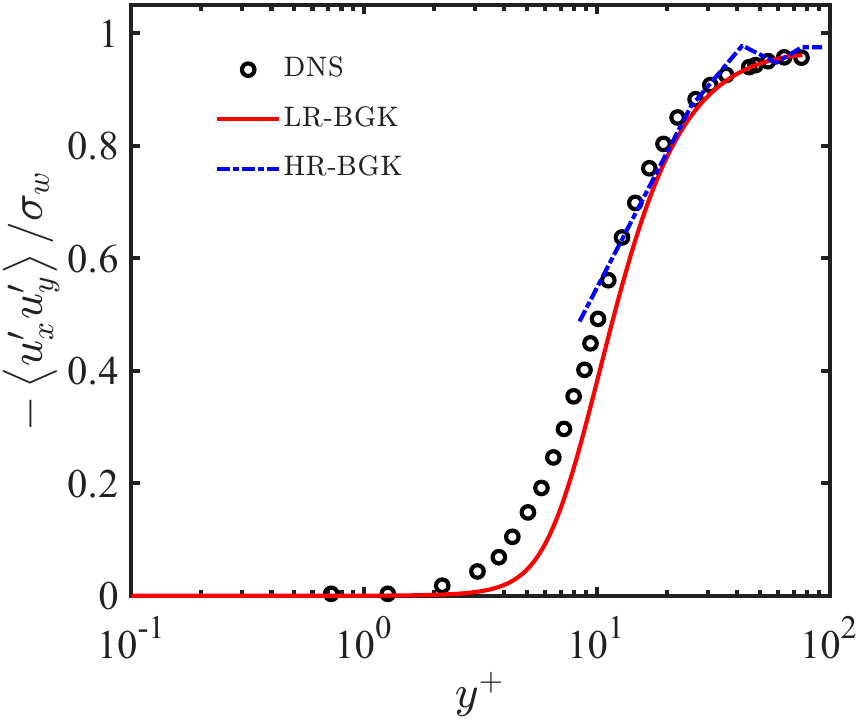}} \quad
  \subfloat[]{\includegraphics[scale=0.44,clip=true]{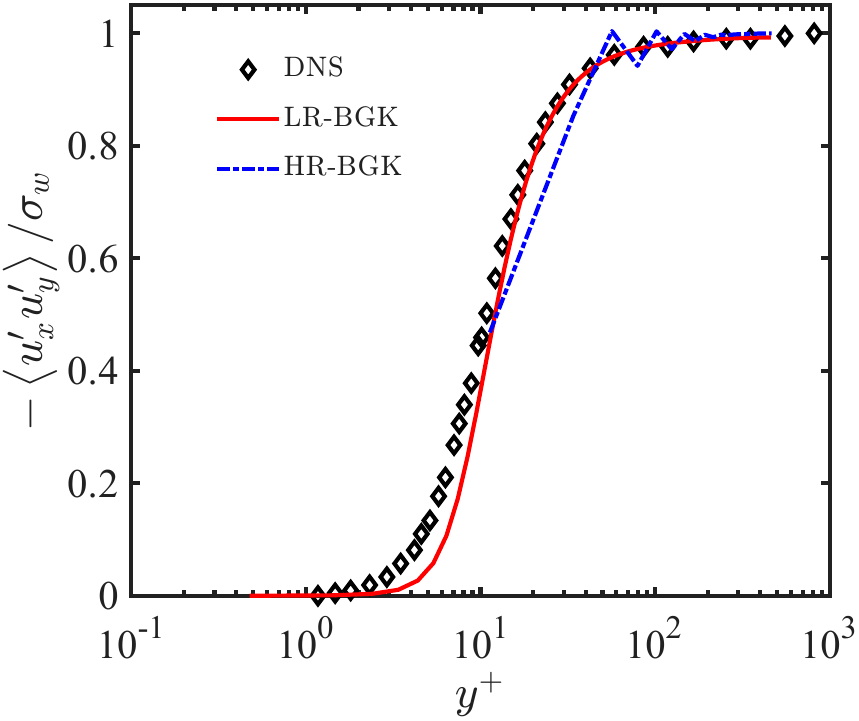}}
    \caption{Reynolds shear stress profiles scaled in wall units of turbulent Couette flow at  (a) $\text{Re} = 1300$ and (b) $\text{Re} = 10133$. 
    %The solid and dash‑dot lines represent the predictions of the LR-BGK model and the HR-BGK model, respectively. The circles and the  diamonds denote the DNS data of~\cite{bech1995investigation} and~\cite{pirozzoli2014turbulence}, respectively.
    }
\label{fig:comparea_sxy}
\end{figure}

%The Reynolds shear stress profiles provide a stringent test of the capacity of the models to capture turbulent momentum transport mechanisms. 
Figure~\ref{fig:comparea_sxy} compares the dimensionless Reynolds shear stress $ -\langle u'_x u'_y \rangle/\sigma_w$ predicted by both kinetic models with DNS data~\citep{bech1995investigation,pirozzoli2014turbulence} as Re ranges from 1300 to 10133. It is noteworthy that the HR-BGK model, due to its coarser grid resolution, exhibits minor numerical oscillations in the near-wall region. Both kinetic models exhibit excellent agreement with the DNS data, accurately capturing both the peak value and its location, particularly in the logarithmic and outer regions ($y^+ \gtrsim 15$).  However, the LR-BGK model shows some deviations in the transition region ($1 \lesssim y^+ \lesssim 15$) due to the inherent complexity of this region where viscous and turbulent effects compete. 

\begin{figure}
\centering
  \subfloat[]{\includegraphics[scale=0.44,clip=true]{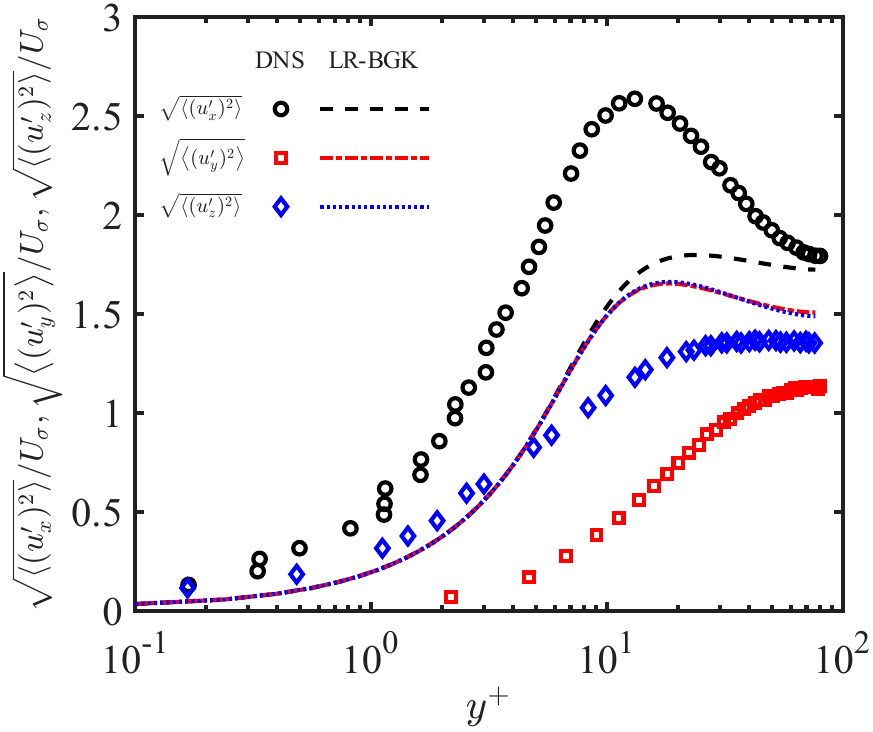}} \quad
  \subfloat[]{\includegraphics[scale=0.44,clip=true]{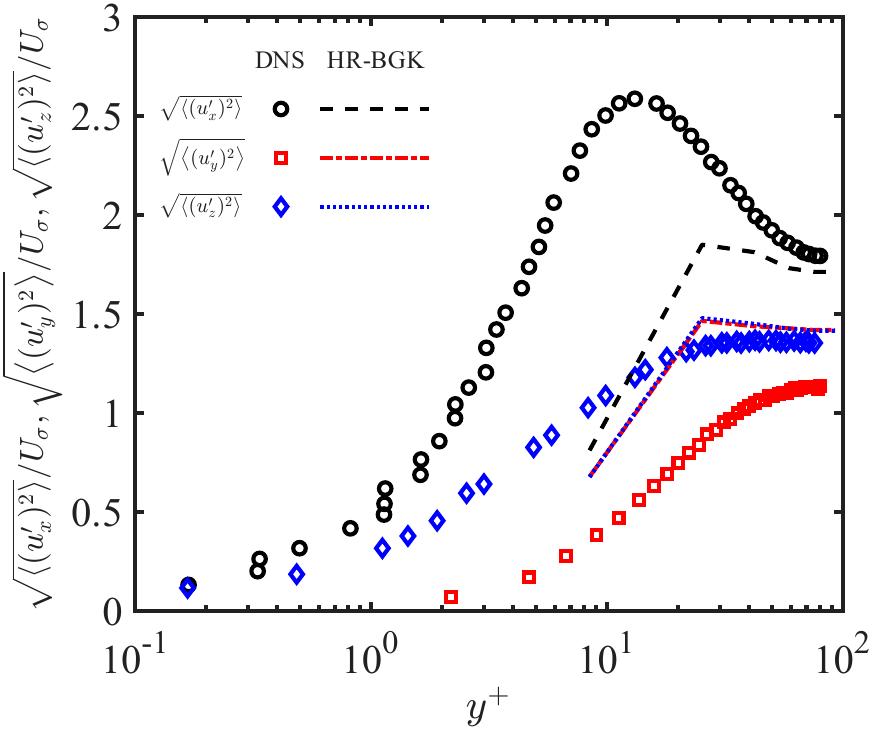}}
    \caption{Distribution of the velocity fluctuation intensities (Reynolds normal stress) scaled in wall units of turbulent Couette flow from (a) the LR-BGK model and (b) the HR-BGK model at $\text{Re} = 1300$. 
    %%The dash, dash-dot and dot lines represent $\sqrt{\left \langle (u_x')^2\right \rangle}/U_\sigma$, $\sqrt{\left \langle (u_y')^2\right \rangle}/U_\sigma$ and $\sqrt{\left \langle (u_z')^2\right \rangle}/U_\sigma$, respectively. The symbols denote the DNS data of~\cite{bech1995investigation}: $\sqrt{\left \langle (u_x')^2\right \rangle}/U_\sigma$ (circles), $\sqrt{\left \langle (u_y')^2\right \rangle}/U_\sigma$ (squares) and $\sqrt{\left \langle (u_z')^2\right \rangle}/U_\sigma$ (diamonds).
    }
\label{fig:comparea_s_Re1300}
\end{figure}

\begin{figure}
\centering
  \subfloat[]{\includegraphics[scale=0.44,clip=true]{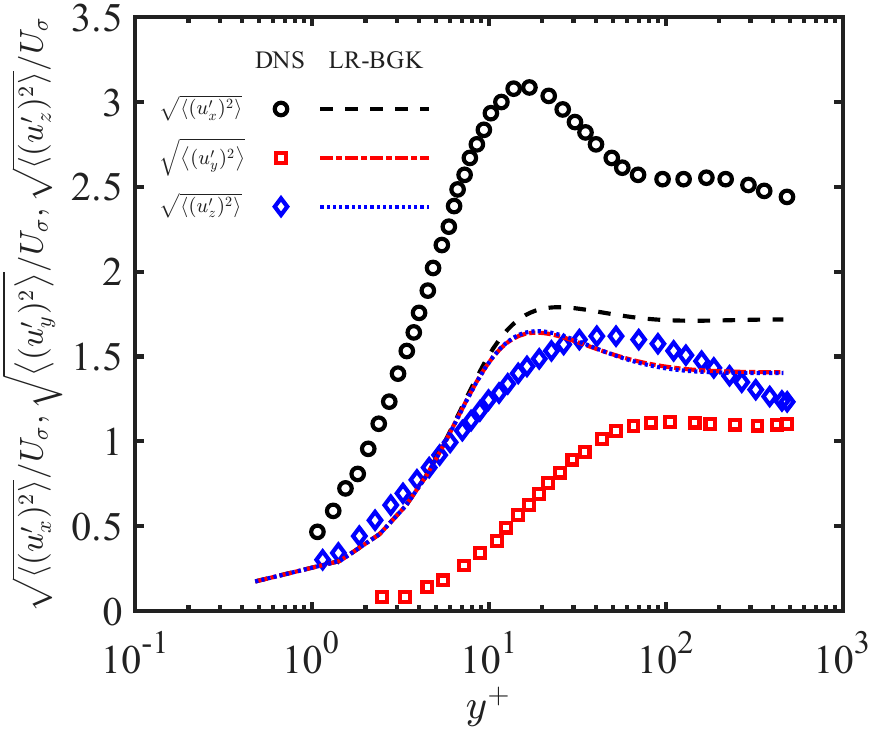}} \quad
  \subfloat[]{\includegraphics[scale=0.44,clip=true]{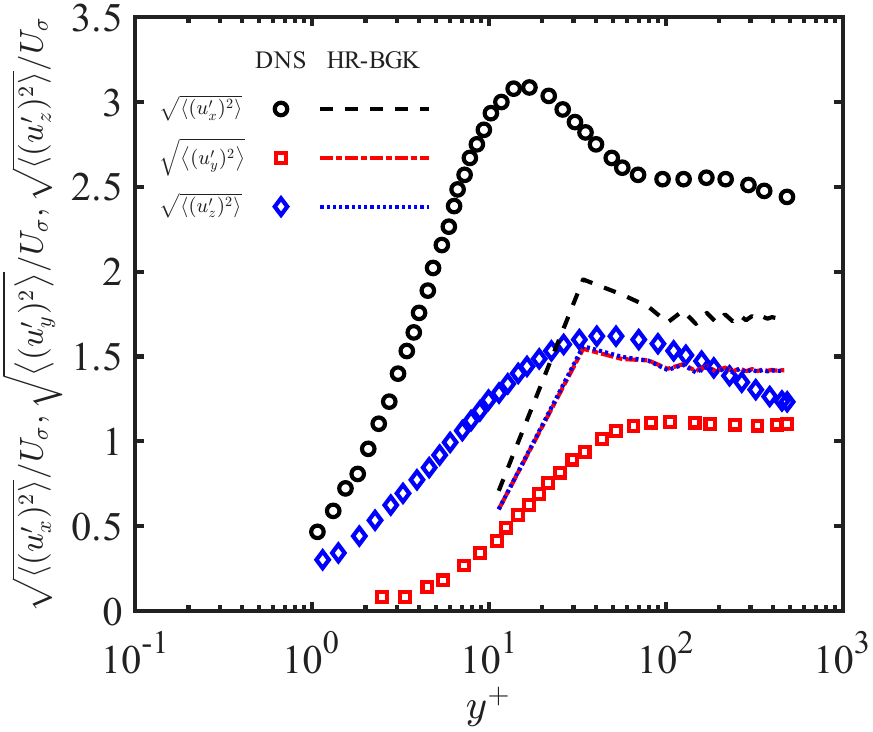}}
    \caption{Distribution of the velocity fluctuation intensities (Reynolds normal stress) scaled in wall units of turbulent Couette flow from (a) the LR-BGK model and (b) the HR-BGK model at $\text{Re} = 10133$.
    %The dash, dash-dot and dot lines represent $\sqrt{\left \langle (u_x')^2\right \rangle}/U_\sigma$, $\sqrt{\left \langle (u_y')^2\right \rangle}/U_\sigma$ and $\sqrt{\left \langle (u_z')^2\right \rangle}/U_\sigma$, respectively. The symbols denote the DNS data of~\cite{pirozzoli2014turbulence}: $\sqrt{\left \langle (u_x')^2\right \rangle}/U_\sigma$ (circles), $\sqrt{\left \langle (u_y')^2\right \rangle}/U_\sigma$ (squares) and $\sqrt{\left \langle (u_z')^2\right \rangle}/U_\sigma$ (diamonds).
    }
\label{fig:comparea_s_Re10133}
\end{figure}

Figures~\ref{fig:comparea_s_Re1300} and~\ref{fig:comparea_s_Re10133} present detailed comparisons of the velocity fluctuation intensities (normal Reynolds stress) in all three spatial directions  which can provide insight into the anisotropic nature of near-wall turbulence, which the linear eddy viscosity model~\eqref{Eq: 1st ce} cannot capture~\citep{launder1974application}. The velocity fluctuation intensities predicted by the present models show noticeable differences from the DNS data~\citep{bech1995investigation,pirozzoli2014turbulence}, with the discrepancies becoming more pronounced at higher Re. At $\text{Re} = 1300$, the dimensionless streamwise $\sqrt{\left \langle (u'_x)^2\right \rangle}$ and vertical $\sqrt{\left \langle (u'_z)^2\right \rangle}$ components exhibit relatively small deviations from DNS data near the channel centerline, while the differences become increasingly evident as Re increases to $10133$.  The LR-BGK model demonstrates superior performance in capturing the peak locations of the streamwise fluctuations, where the peak occurs around $y^+ \approx 10$. Notably, the largest discrepancies relative to DNS occur in the near-wall/buffer region (approximately $y^+\approx 1\sim10$), where the separation among the normal stress components is underpredicted; this highlights a limitation of the present formulation and motivates further improvements.
Although the magnitudes of the normal stress components are consistent with DNS, a clear separation among these components is still not reproduced by the present models. Nevertheless, they demonstrate the capability to capture non-Newtonian (nonlinear eddy viscosity) effects as mentioned in \S\ref{Sec: CE}. This non-equilibrium phenomenon, often reported in rarefied gas flows at finite Knudsen numbers, exhibits the anisotropic stress behaviour that cannot be captured by the NS equations~\citep{xu2007multiple}.
%More accurate results should be interested in exploring higher order information about the collision term $\mathcal{C}(F) \equiv -\nabla_{\boldsymbol{\xi}} \cdot \langle \boldsymbol{a}' f' \rangle$ and modelling them accurately.

\subsection{Non-equilibrium characteristics and velocity distribution functions}
\begin{figure}
\centering
  {\includegraphics[scale=0.27,clip=true]{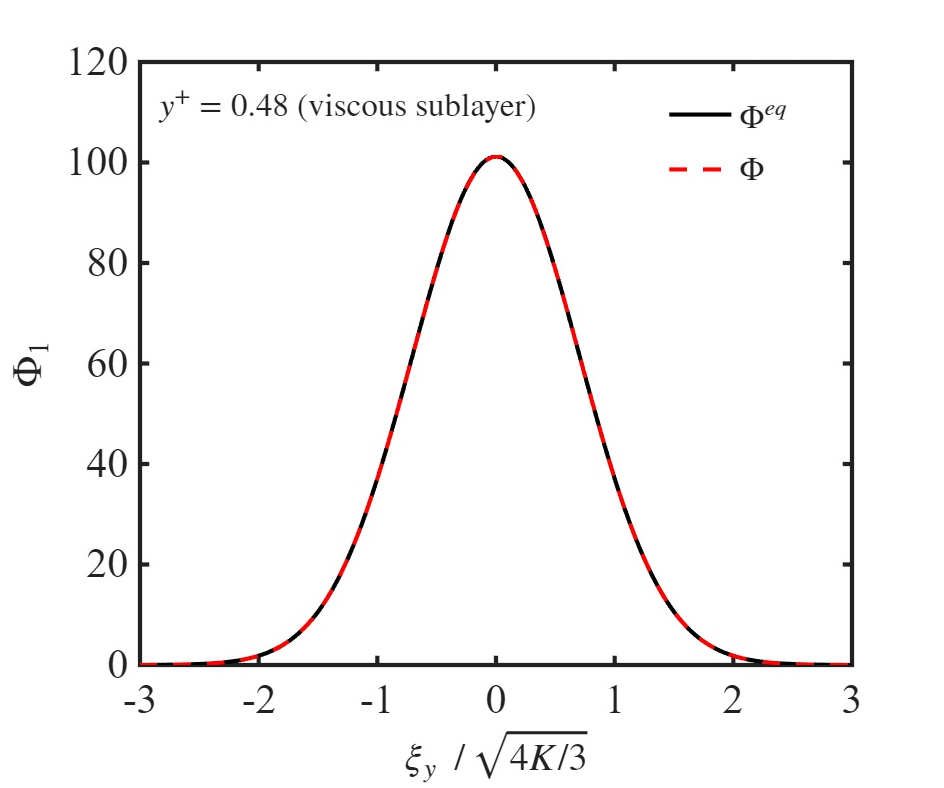}}
  {\includegraphics[scale=0.27,clip=true]{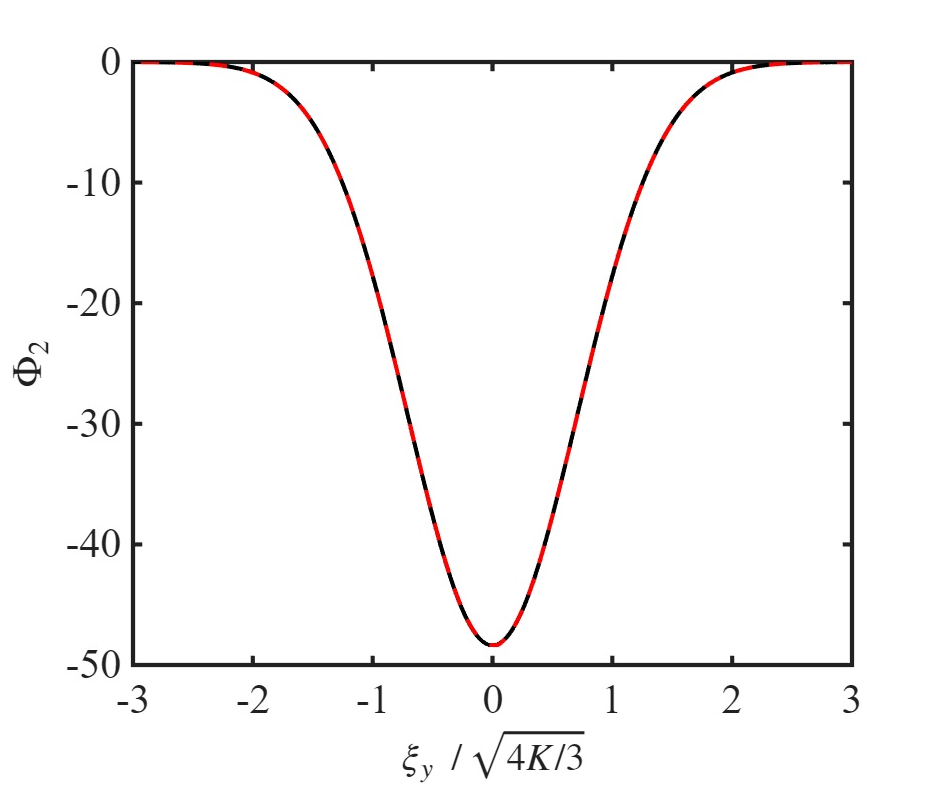}}
  {\includegraphics[scale=0.27,clip=true]{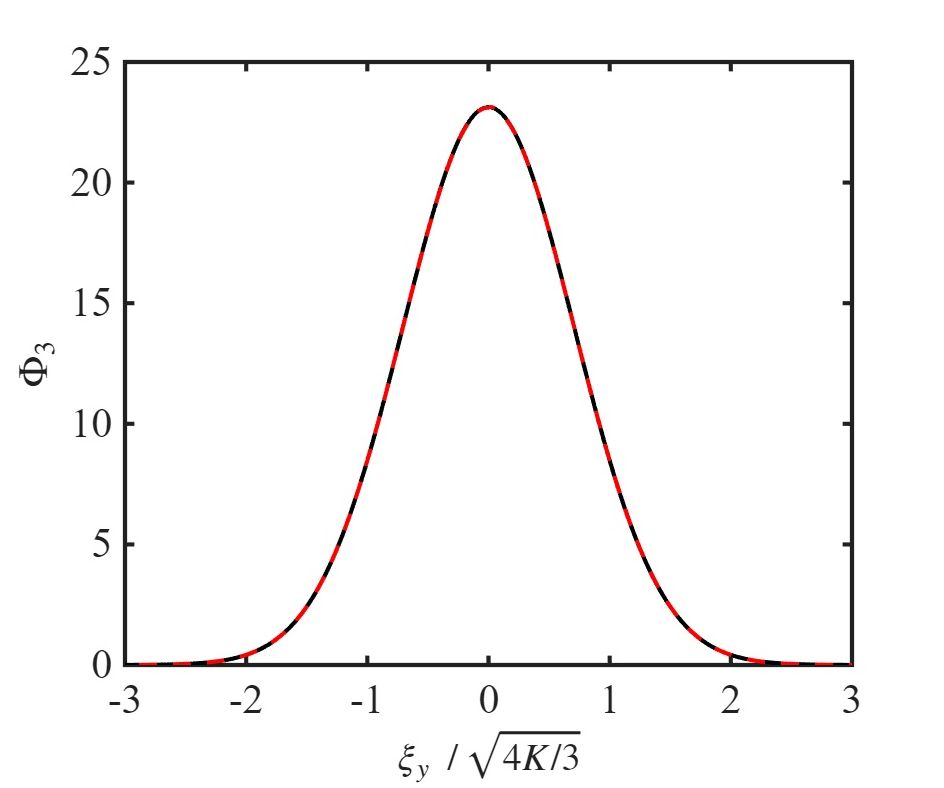}}
  \\
{\includegraphics[scale=0.27,clip=true]{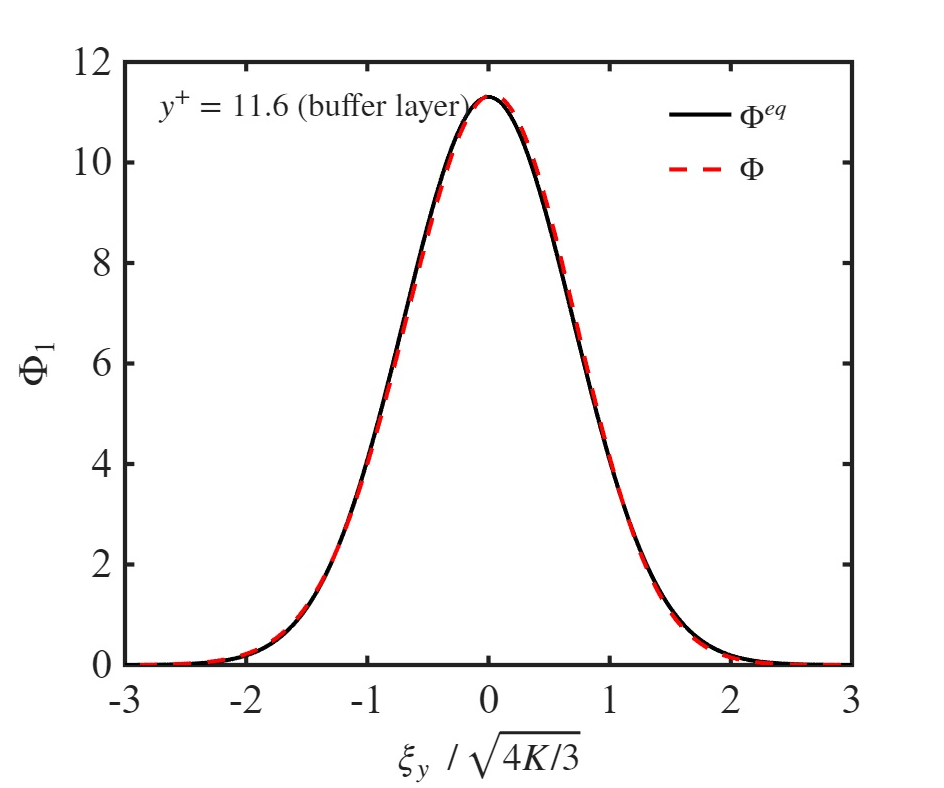}}
  {\includegraphics[scale=0.27,clip=true]{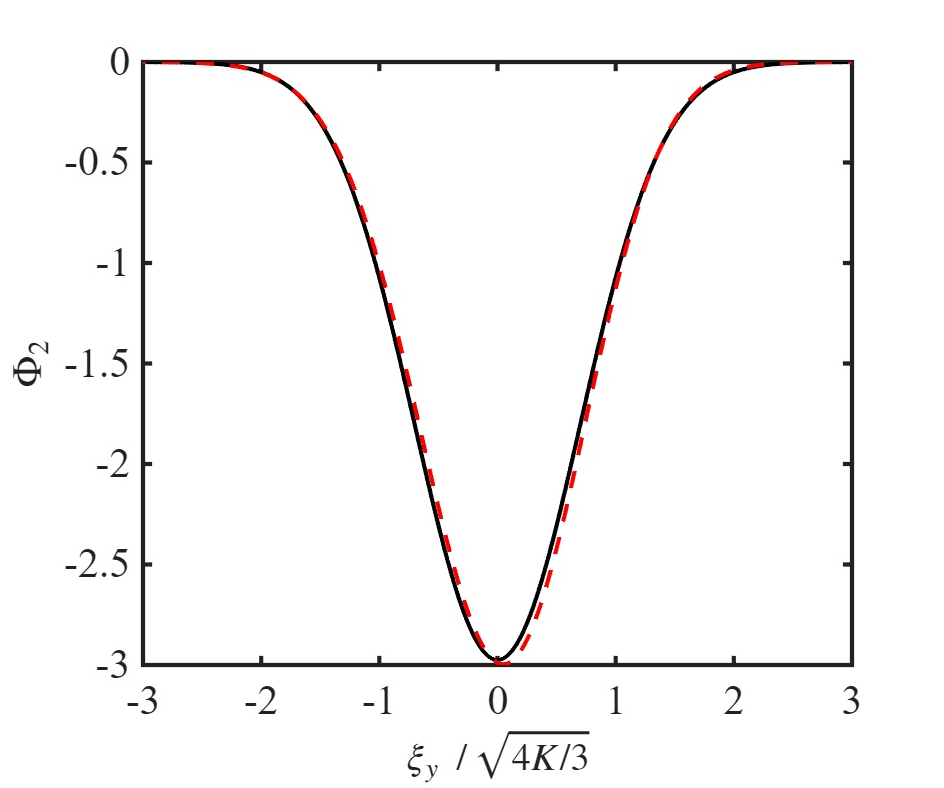}}
  {\includegraphics[scale=0.27,clip=true]{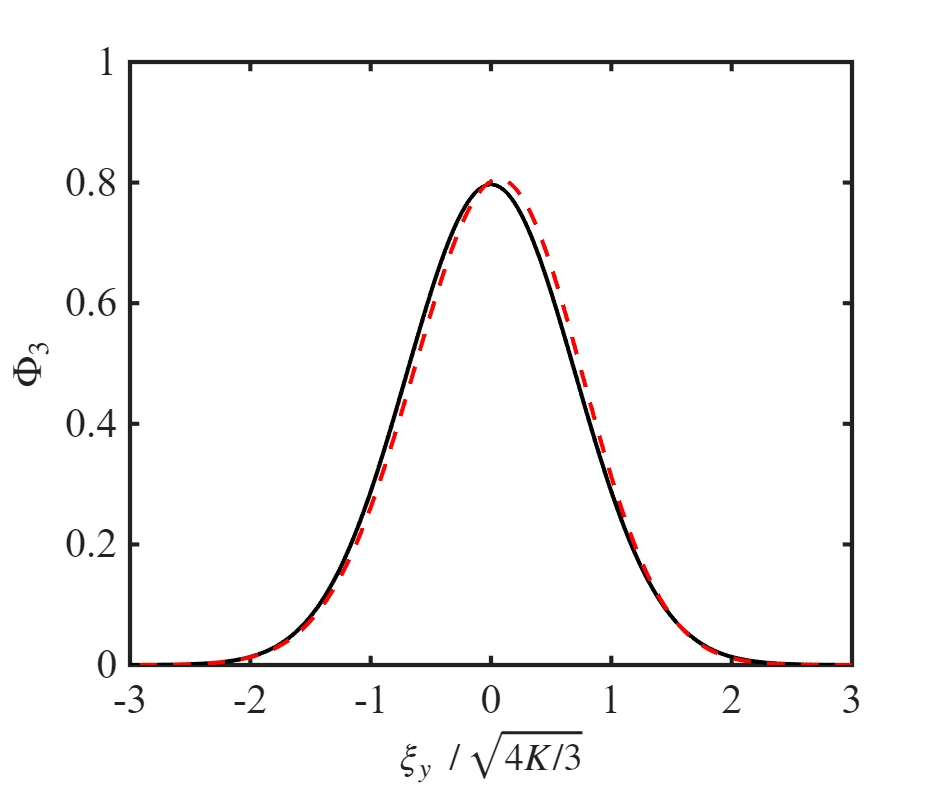}}
    \\
{\includegraphics[scale=0.27,clip=true]{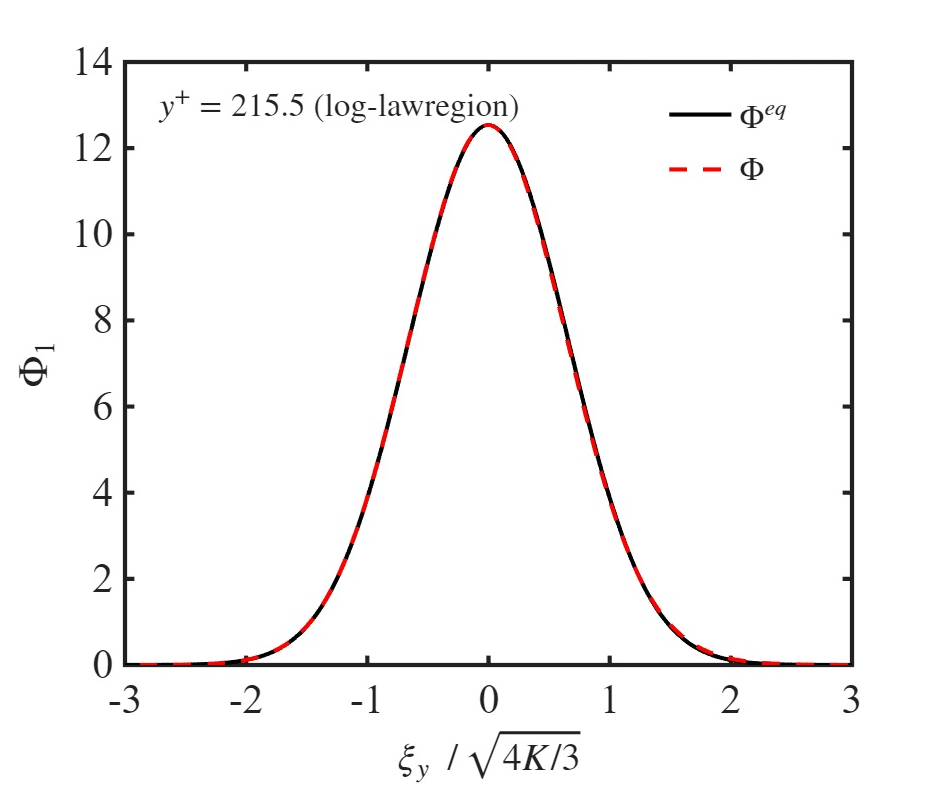}}
  {\includegraphics[scale=0.27,clip=true]{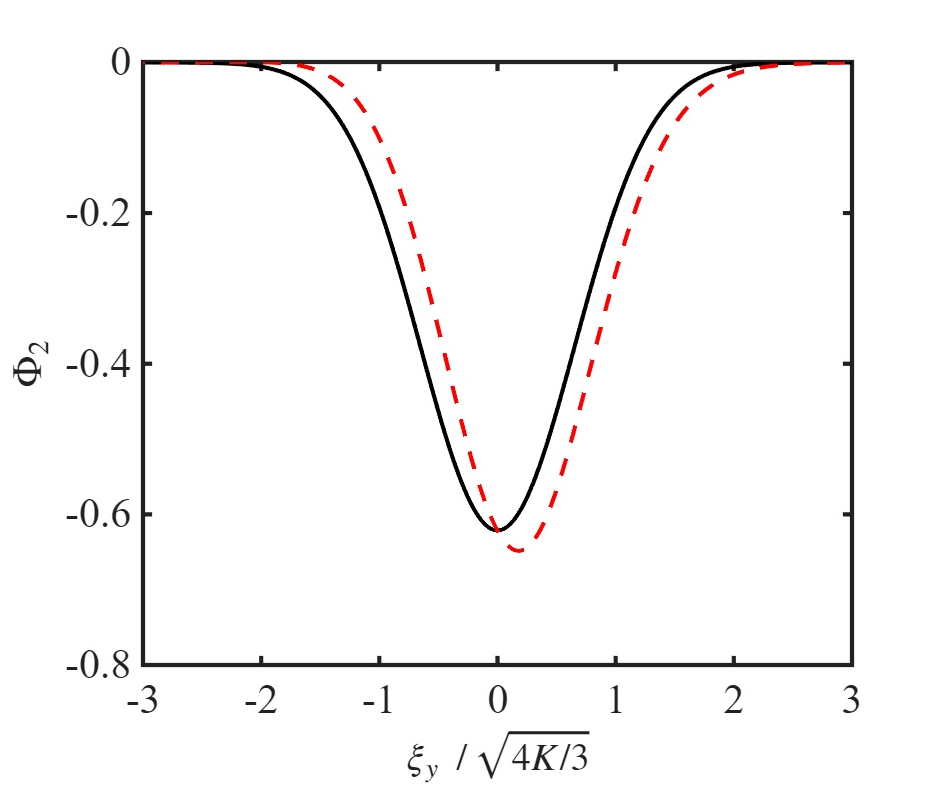}}
  {\includegraphics[scale=0.27,clip=true]{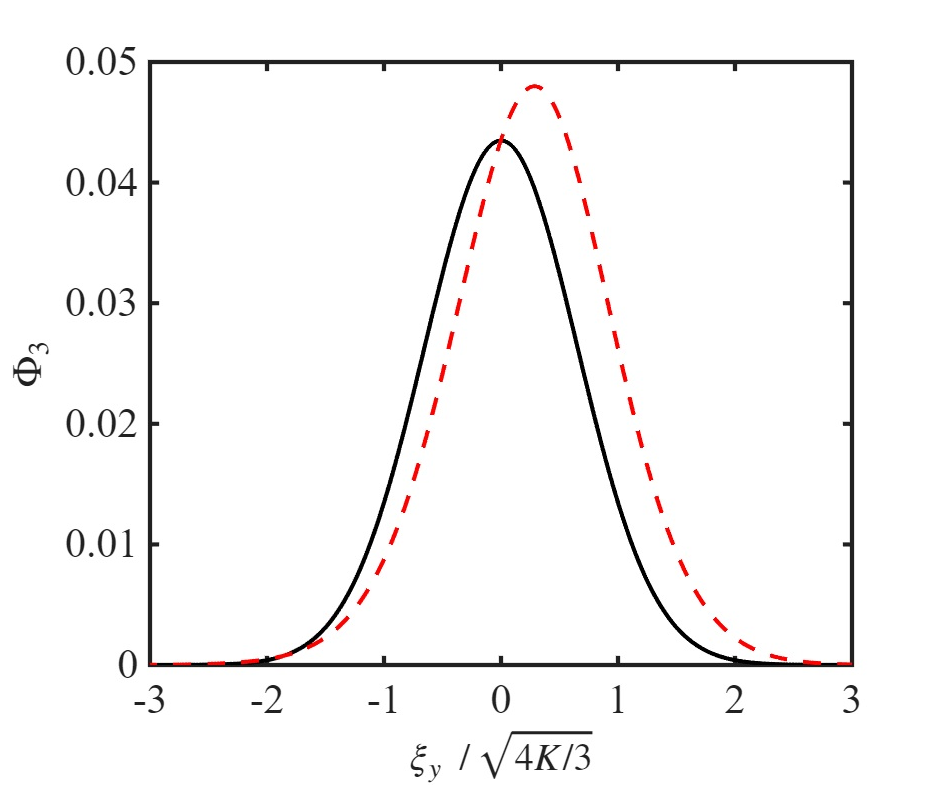}}
      \\
{\includegraphics[scale=0.27,clip=true]{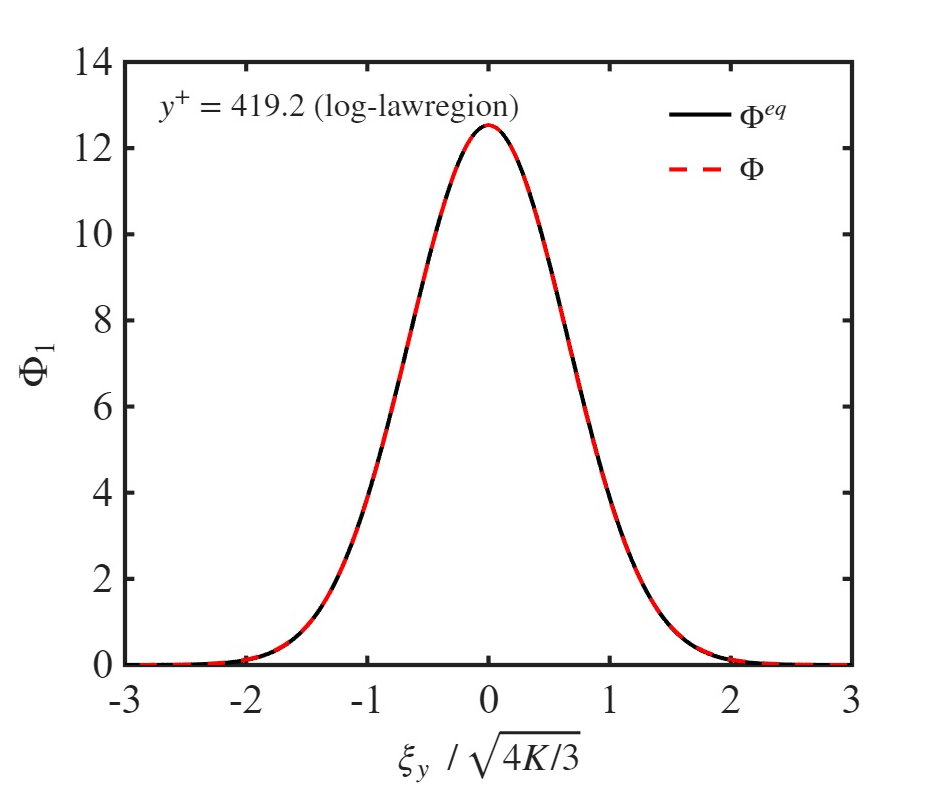}}
  {\includegraphics[scale=0.27,clip=true]{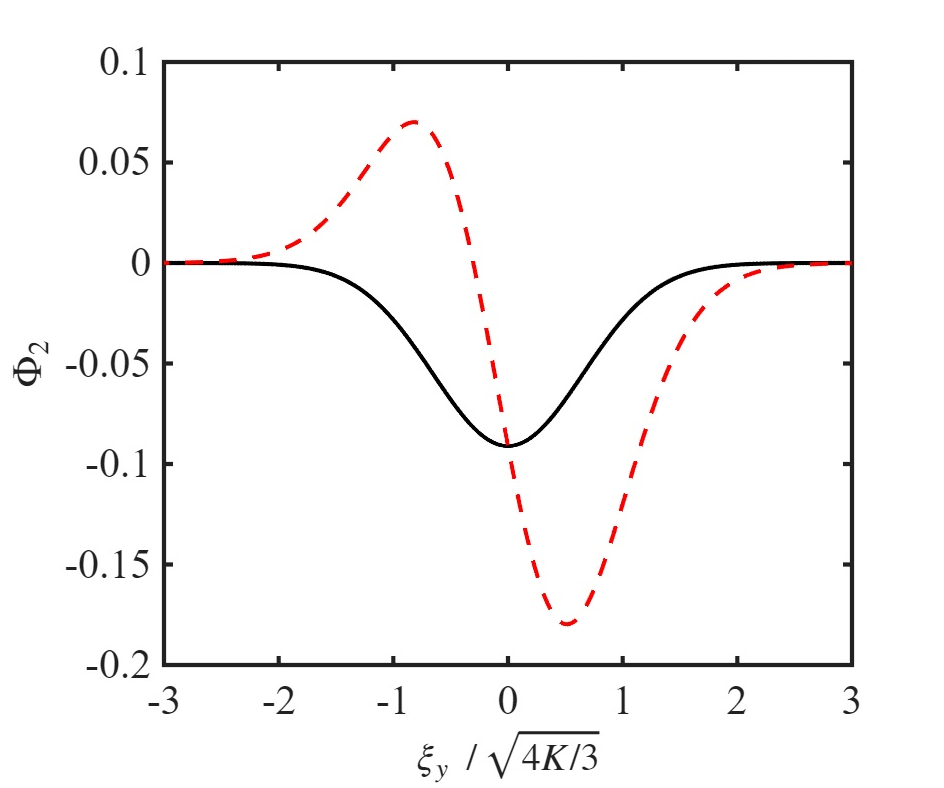}}
  {\includegraphics[scale=0.27,clip=true]{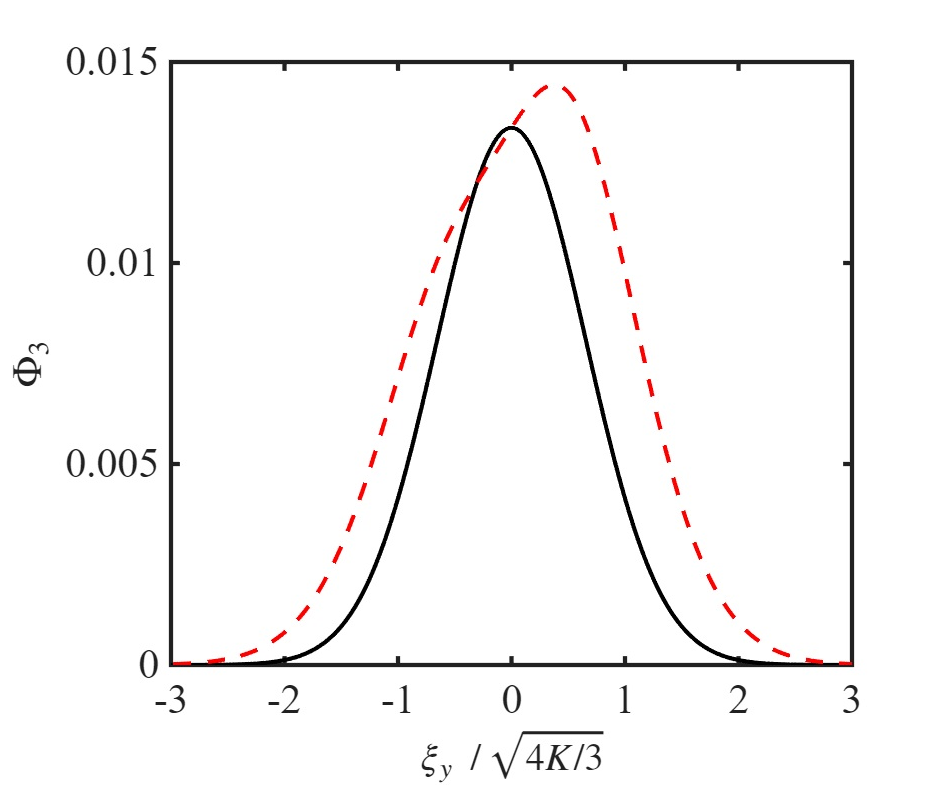}}
      \\
{\includegraphics[scale=0.27,clip=true]{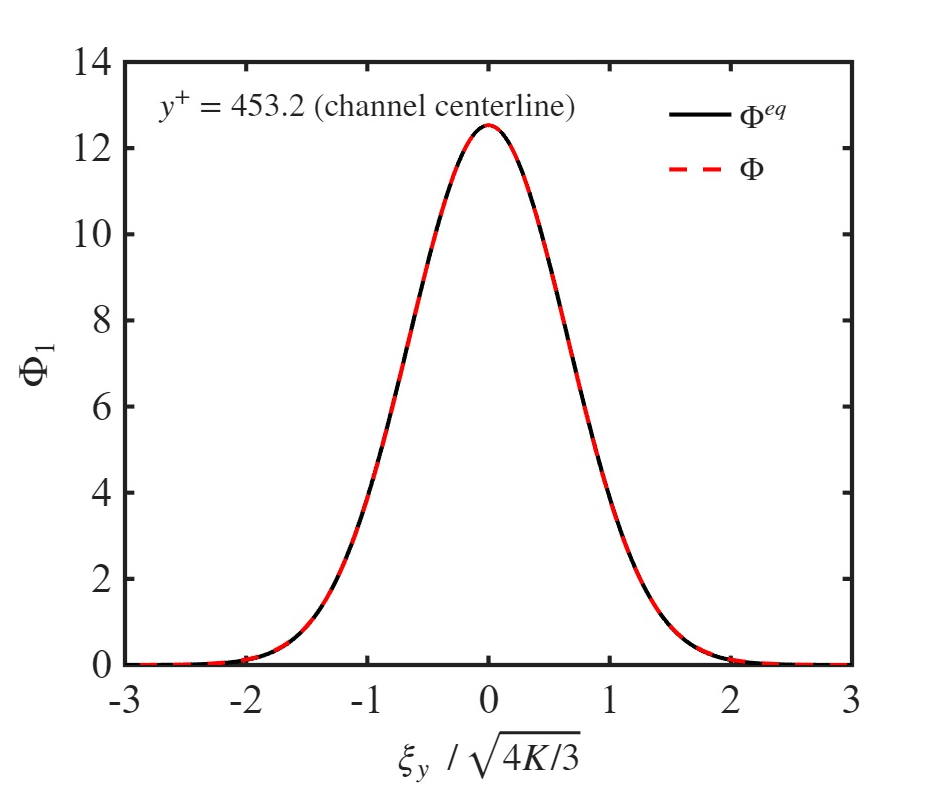}}
  {\includegraphics[scale=0.27,clip=true]{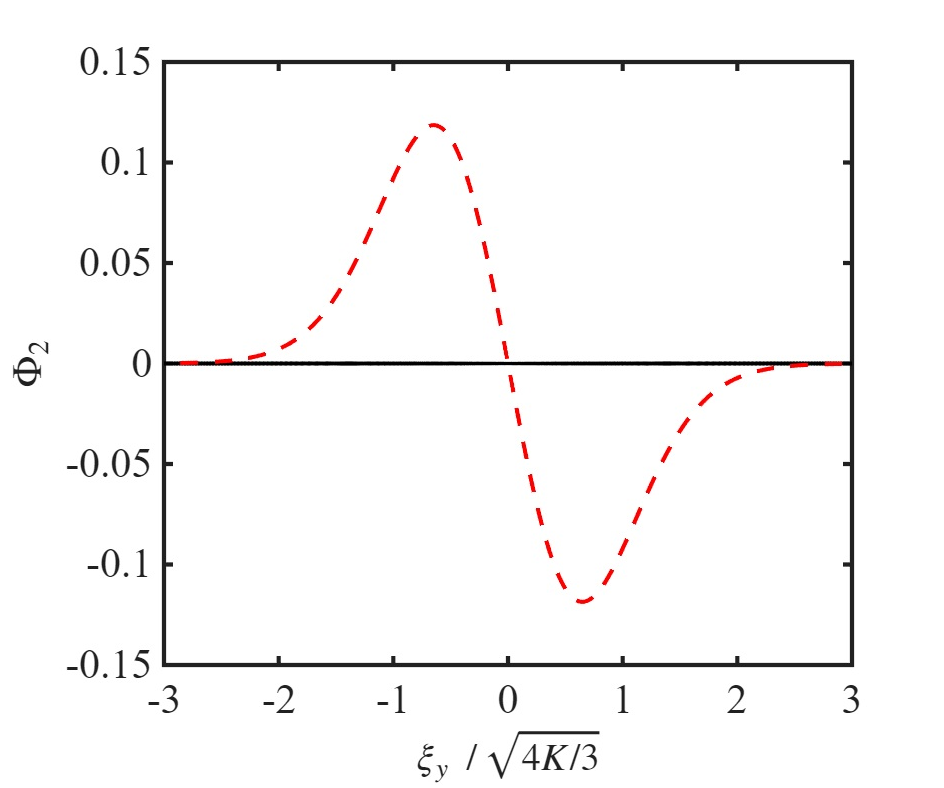}}
  {\includegraphics[scale=0.27,clip=true]{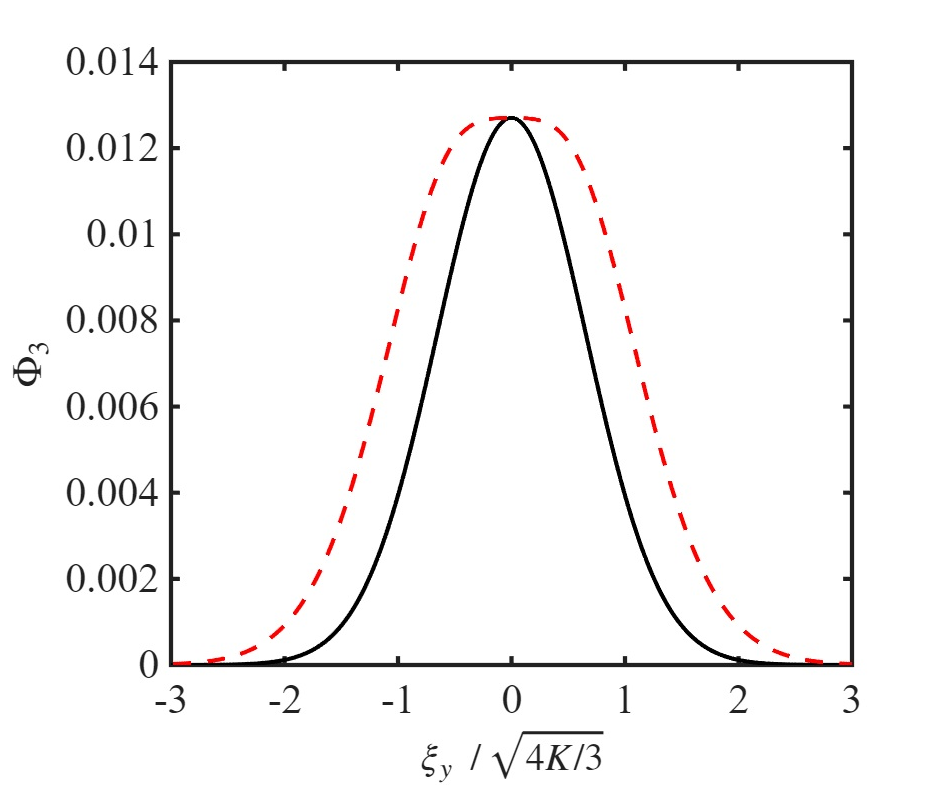}}
    \caption{The reduced velocity distribution functions $\Phi_1$ (first column), $\Phi_2$ (second column) and $\Phi_3$ (third column) from the LR-BGK model at $\text{Re} = 10133$ as a function of the microscopic velocity $\xi_y$ 
    %scaled by local parameters $\sqrt{4K/3}$ 
    at different wall-normal positions.}
\label{fig:lowRe_VDFs}
\end{figure}

The non-equilibrium and non-Newtonian characteristics predicted by the present kinetic model can be essentially explained from the VDFs. We now examine the reduced VDFs obtained from the LR-BGK model at $\text{Re}=10133$, comparing the computed VDFs with their corresponding local equilibrium distributions to quantify the degree of departure from equilibrium.
Figure~\ref{fig:lowRe_VDFs} presents the reduced velocity distribution functions $\Phi_1$, $\Phi_2$, and $\Phi_3$ as functions of microscopic velocity $\xi_y$ in various normal wall positions ranging from the viscous sublayer to the centerline of the channel, compared against the local equilibrium VDFs derived from the local mean velocity and TKE. Here $\xi_y$ in the region $\left[-3\xi_m,3\xi_m\right]$ with $\xi_m = \sqrt{4K/3}$ represents the most probable velocity scale based on the local turbulent kinetic energy.  The fourth VDF $\Phi_4$, which exhibits behaviour similar to $\Phi_1$ by remaining close to equilibrium (not shown here for brevity). 

The most striking observation is the markedly different behaviours exhibited by the various reduced VDFs. The distribution function $\Phi_1$, which represents the zeroth-order moment in the wall-normal direction, remains remarkably close to its local equilibrium state across all wall-normal positions. This near-equilibrium behaviour indicates that the mass conservation constraint is well satisfied throughout the flow domain, consistent with the incompressibility condition. In sharp contrast, $\Phi_2$ and $\Phi_3$ exhibit pronounced departures from their respective equilibrium distributions, particularly in the outer region of the flow. Here the "equilibrium" baseline refers to the local isotropic Gaussian distribution $F^{eq}$ (and its reduced counterparts $\Phi^{eq}_k$) constructed from the local mean velocity and TKE. Therefore the deviations primarily reflect shear-induced anisotropy and wall-normal inhomogeneity in the averaged turbulence dynamics. The VDF $\Phi_2$, associated with the streamwise momentum transport, shows increasing non-equilibrium effects as the distance from the wall increases. Notably, at $y^+ \approx 419$ and near the channel centerline, $\Phi_2$ exhibits a clear departure from the isotropic Gaussian baseline, characterized by an odd-in-$\xi_y$ (sign-changing) deviation. This feature is directly related to the Reynolds shear stress, since $\sigma_{xy}=-\langle u_x' u_y'\rangle$ is obtained as the moment of the reduced distribution (in our definition, $\sigma_{xy}=-\int \xi_y \Phi_2 d\xi_y+U_x U_y$ for $U_y\approx0$). The observed positive deviation for $\xi_y<0$ and negative deviation for $\xi_y>0$ contributes to the negative correlation $-\langle u_x'u_y'\rangle$, consistent with the leading-order (first-order CE) closure and also with linear eddy-viscosity modeling.
Similarly, $\Phi_3$, which is related to the streamwise kinetic energy transport, demonstrates substantial non-equilibrium characteristics with a comparable spatial evolution pattern. The VDF $\Phi_3$ shows progressively stronger departures from equilibrium as the wall-normal distance increases, with the outer region exhibiting the most pronounced non-equilibrium effects. At $y^+ \approx 419$ and near the channel centerline, 
$\Phi_3$ exhibits a clear asymmetry/shift relative to the isotropic baseline. This feature indicates a nonzero wall-normal transport of TKE flux, which corresponds to a higher-order velocity moment of the VDF and is captured at leading order by the first-order CE closure.

While the leading-order trends of these effects can be understood qualitatively within the first-order CE closure, we further quantify the degree of departure from equilibrium and assess the contributions beyond the first-order closure by examining the second-order moment of $F-F^{eq}$,
\begin{equation}\label{Eq:d sigma}
    \boldsymbol{\sigma}^{neq}=-\int \left(\boldsymbol{\xi}-\boldsymbol{U}\right)\left(\boldsymbol{\xi}-\boldsymbol{U}\right) (F-F^{eq})d\boldsymbol{\xi}.
\end{equation}
It should be noted that the zeroth and first-order moments of $F-F^{eq}$ vanish identically. In the case of plane Couette flow, equation~\eqref{Eq:2nd ce q} reduces to:
\begin{equation}\label{Eq:d sigma CE}
    \begin{aligned}
    \sigma^{CE}_{xy}&=\nu_T\frac{\partial U_x}{\partial y},\\
    \sigma^{CE}_{xx}&=-\frac{2}{3}K-\frac{2\nu_T^2}{K^{eq}}\left(\frac{\partial U_x}{\partial y}\right)^2+\frac{2 \nu_T}{3K^{eq}}\frac{\partial}{\partial y} \left(\nu_T\frac{\partial K^{eq}}{\partial y}\right),\\
    \sigma^{CE}_{yy}&=-\frac{2}{3}K+\frac{\nu_T^2}{K^{eq}}\left(\frac{\partial U_x}{\partial y}\right)^2-\frac{4\nu_T}{3K^{eq}}\frac{\partial}{\partial y} \left(\nu_T\frac{\partial K^{eq}}{\partial y}\right),\\
    \sigma^{CE}_{zz}&=-\frac{2}{3}K+\frac{\nu_T^2}{K^{eq}}\left(\frac{\partial U_x}{\partial y}\right)^2+\frac{2\nu_T}{3K^{eq}}\frac{\partial}{\partial y} \left(\nu_T\frac{\partial K^{eq}}{\partial y}\right),
    \end{aligned}
\end{equation}
where $\nu_T = 0.0816 f_\mu K^2/\epsilon$. These expressions indicate significant differences among the three normal stress components, demonstrating the capability of the present kinetic model to capture anisotropic stress behaviour. 

\begin{figure}
\centering
  \subfloat[]{\includegraphics[scale=0.44,clip=true]{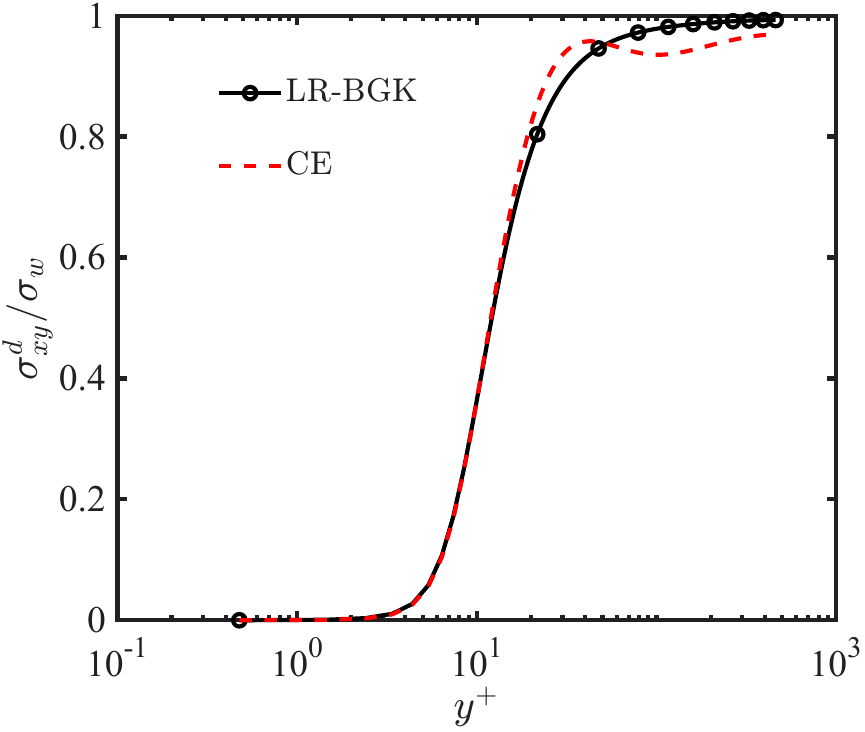}\label{fig:comparea_sxyd_Re10133}} \quad
  \subfloat[]{\includegraphics[scale=0.44,clip=true]{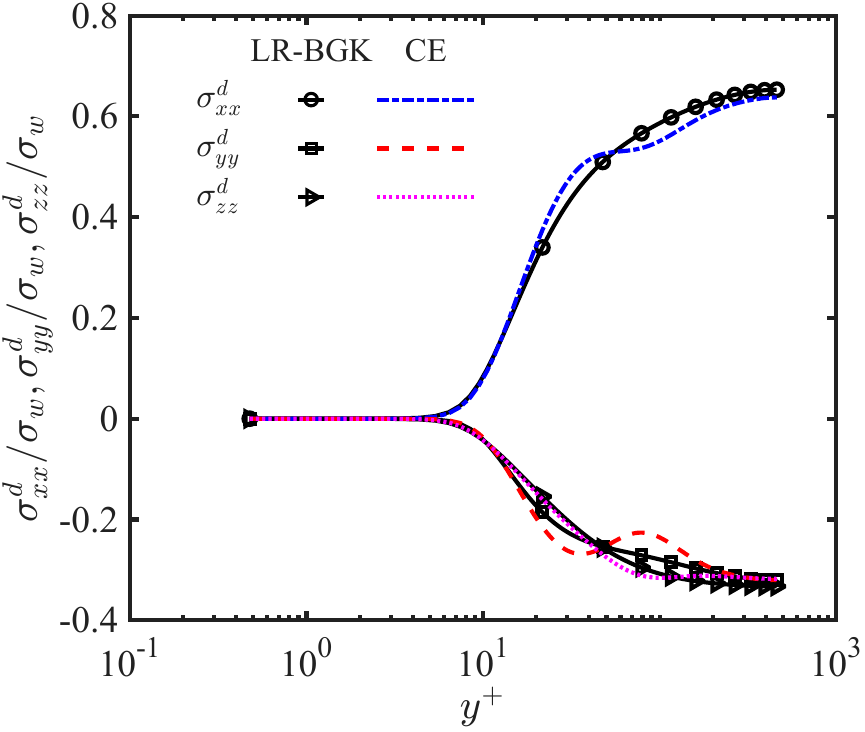}\label{fig:comparea_snd_Re10133}}
    \caption{Deviatoric Reynolds (a) shear stress component and (b) normal stress components in wall units of turbulent Couette flow at $\text{Re} = 10133$. Here, the deviatoric Reynolds stress is defined as $\boldsymbol{\sigma}^d=\boldsymbol{\sigma}-\frac{1}{3} \left(\text{tr}\boldsymbol{\sigma}\right) \boldsymbol{I}$.
    }
\label{fig:comparea_sd_Re10133}
\end{figure}

Figure~\ref{fig:comparea_sd_Re10133} presents the deviatoric part of $\boldsymbol{\sigma}^{neq}$ predicted by LR-BGK model at $\text{Re}=10133$ and compare them with the CE expansion results given by equation~\eqref{Eq:2nd ce q}.
It can be seen that the deviatoric Reynolds stress shows good agreement between the LR-BGK model and the CE expansion throughout the flow domain, validating the theoretical consistency of the kinetic approach. The minor discrepancies observed can be attributed to higher-order terms in the CE expansion that are not captured by the second-order analysis. Furthermore, the deviatoric Reynolds normal stress components better reflect the degree of departure from non-equilibrium of the VDFs mentioned above.  In the present model calculations, the deviatoric normal stress components remain close to zero in the viscous sublayer. Across the buffer region, they begin to deviate from zero, with $\sigma^{d}_{xx}$ increasing to positive values while $\sigma^{d}_{yy}$ and $\sigma^{d}_{zz}$ decrease to negative values. Their magnitudes continue to grow toward the outer layer, where the non-equilibrium signatures become more evident. This relatively weak near-wall non-equilibrium signature may also reflect a limited capability of the present formulation to represent buffer-layer non-equilibrium effects, which could contribute to the relatively large discrepancies in the normal stress predictions observed in figure~\ref{fig:comparea_s_Re10133} in the near-wall region ($1 \lesssim y^+ \lesssim 10$) compared with DNS data. A possible cause is that, especially in the near-wall region, the relaxation time $\tau$ in the present single-relaxation-time BGK closure is chosen for obtaining the desirable shear stress, as a result, the effective turbulent Knudsen number ($\sim \tau |S|$) becomes too small, so that all other Reynolds stress components are suppressed. One of the future investigations is how to adjust  $\tau$ to match shear stress without sacrificing the non-equilibrium effects for normal stress components, including generalization of the single relaxation time collision model to involve multiple relaxation times.
%A possible interpretation is that, in the near-wall region, the effective relaxation time $\tau$ in the present single-relaxation-time closure is too small relative to the characteristic time scale of anisotropy generation (e.g. $1/|S|$), so the model relaxes too rapidly toward an isotropic local equilibrium and under-represents buffer-layer non-equilibrium signatures. This observation suggests that improving the model's ability to represent anisotropic relaxation may be a useful direction for future extensions of the kinetic closure. For example, one may adopt an ellipsoidal-statistical BGK (anisotropic Gaussian) local equilibrium, or introduce a dual-relaxation-time formulation with a small number of additional parameters to regulate anisotropic relaxation.} 

%In the viscous sublayer and buffer region, the system remains essentially in equilibrium with all three normal stress components close to zero, while in the outer layer non-equilibrium effects become more pronounced as the deviations from zero increase progressively.

Taken together, the VDFs and their associated non-equilibrium moments provide a framework-specific interpretation for the ability of the present kinetic models to represent anisotropic Reynolds stress behaviour and related non-Newtonian effects. Within this interpretation, departures from local equilibrium reflect finite-relaxation effects in the kinetic closure, analogous in spirit to non-equilibrium transport in rarefied-gas kinetics.
%These non-equilibrium characteristics, as revealed through both the VDFs and their non-equilibrium moments, provide a fundamental explanation for the superior performance of the kinetic models in capturing the anisotropic Reynolds stress behaviour and the associated non-Newtonian effects. The departure from local equilibrium enables the kinetic approach to naturally account for the complex momentum and energy transport mechanisms that arise from the finite relaxation times of turbulent eddies, analogous to the finite collision times in rarefied gas flows. This non-equilibrium  characteristic is particularly important in wall-bounded flows, where the presence of the wall creates strong gradients and anisotropic conditions that drive the system away from local Gaussian equilibrium.

%\begin{figure}
%\centering
%  \subfloat{\includegraphics[scale=0.5,clip=true]{./Figures/FIG_sxyd_Re10133_lowRe.pdf}}
%  \caption{ Deviatoric Reynolds shear stress component in wall units of turbulent Couette flow at $\text{Re} = 10133$. The solid lines with open circles and dash-dot lines represent the predictions of the LR-BGK model~\eqref{Eq:d sigma} and the CE expansion results~\eqref{Eq:d sigma CE}, respectively. Here, the deviatoric Reynolds stress is defined as $\boldsymbol{\sigma}^d=\boldsymbol{\sigma}-\frac{1}{3} \left(\text{tr}\boldsymbol{\sigma}\right) \boldsymbol{I}$.}
%\label{fig:comparea_sxyd_Re10133}
%\end{figure}

%\begin{figure}
%\centering
%  \subfloat{\includegraphics[scale=0.5,clip=true]{./Figures/FIG_sd_Re10133_lowRe.pdf}}
%    \caption{Deviatoric Reynolds normal stress components in wall units of turbulent Couette flow at $\text{Re} = 10133$. The solid lines with circles, squares, and triangles represent ${\sigma}_{xx}^d$, ${\sigma}_{yy}^d$, and ${\sigma}_{zz}^d$ predicted by the LR-BGK model~\eqref{Eq:d sigma}, respectively. The dash-dot, dash and dot lines represent ${\sigma}_{xx}^d$, ${\sigma}_{yy}^d$, and ${\sigma}_{zz}^d$ predicted by the CE expansion results~\eqref{Eq:d sigma CE}, respectively.
%    Here, the deviatoric Reynolds stress is defined as $\boldsymbol{\sigma}^d=\boldsymbol{\sigma}-\frac{1}{3} \left(\text{tr}\boldsymbol{\sigma}\right) \boldsymbol{I}$.}
%\label{fig:comparea_sd_Re10133}
%\end{figure}

\section{Conclusions} \label{Sec:conclusion}
In this work, we extended and analyzed the kinetic theoretic model for turbulence derived from the first principles of incompressible Navier-Stokes equation by~\cite{chen2023average} and demonstrated improved applicability and physical consistency. By redefining the relaxation time as $\tau=K/(7\epsilon)$, the Chapman–Enskog expansion of the kinetic model yielded linear and nonlinear eddy viscosity and gradient diffusion closures with transport coefficients ($C_\mu=0.0816, \text{Pr}_T=0.7$) that are in quantitative agreement with those of conventional turbulence models. The analysis of our kinetic model reveals that its second-order expansion generates various material derivatives and higher-order derivatives of TKE.
%Unlike previous results for the Boltzmann-BGK equation without TKE dissipation~\citep{chen2004expanded} or for the Fokker-Planck equation~\citep{heinz2007unified,luan2025constructing}, the analysis of our kinetic model yielded different forms of material derivatives for the linear terms and higher-order derivative terms involving TKE from the second-order expansion.
To our knowledge, such closed-form Chapman–Enskog expressions for the TKE flux and its nonlinear corrections have not previously been reported for this class of turbulent kinetic models. To address wall-bounded turbulence, we developed two approaches within a unified kinetic framework. For applications outside the viscous sublayer, the non-equilibrium extrapolation boundary condition based on wall function approaches was proposed for the HR-BGK model, maintaining accurate logarithmic layer behaviour at lower computational expense. For direct viscous sublayer resolution, we constructed a low-Reynolds-number kinetic model incorporating a damping function and a viscous diffusion term, with the fully diffuse wall condition.
%\replaced{\color{red} enforcing }{that enforced}\highlight{no-slip} and zero TKE at the wall.
%It is worth noting that the present methods for wall-bounded flows are fundamentally different from the boundary conditions based on the wall function method of \citet{srinivasan1977turbulent}, devised for the kinetic model of \citet{lundgren1969model}. Our approach is not restricted to regions outside the viscous sublayer, enabling treatment of the complete flow domain from wall to bulk region. 

The present kinetic models were comprehensively validated against experimental data and DNS data for turbulent plane Couette flow, demonstrating accurate prediction of mean velocity profiles, skin friction coefficients, and Reynolds shear stress distributions. They can capture essential flow physics including logarithmic law behaviour and Reynolds stress anisotropy. Although not as pronounced as indicated by DNS, the present kinetic theoretic models are shown to predict the separations between the normal stress components. 
%In terms of predictive performance, the kinetic models achieve comparable accuracy to conventional nonlinear eddy viscosity models ~\citep{speziale1987nonlinear,nisizima1987turbulent,myong1990prediction,craft1996development} for mean velocity profiles, skin friction coefficients, and Reynolds shear stress components.
Nevertheless, accurate prediction of individual normal stress components in canonical wall-bounded flows remains a common challenge. A distinctive advantage of the kinetic approach lies in its robust theoretical foundation, requiring only a single relaxation time, whereas the conventional nonlinear eddy viscosity models necessitate multiple $ad \ hoc$ empirical coefficients for closure.
These findings suggest that future developments should explore higher-order information about the collision term $\mathcal{C}(F) \equiv -\nabla_{\boldsymbol{\xi}} \cdot \langle \boldsymbol{a}' f' \rangle$ to achieve more accurate modelling of stress tensor anisotropy. 
Along this direction, the nonlinear near-wall anisotropy mechanisms highlighted by~\cite{1997Prediction} and~\cite{wallin2000explicit} may guide extensions such as an ellipsoidal-statistical BGK (anisotropic Gaussian) equilibrium or a dual-relaxation-time formulation with a few extra parameters to regulate anisotropic relaxation.

Analysis of the reduced velocity distribution functions revealed significant departures from the local Gaussian equilibrium distribution, particularly for streamwise momentum and energy transport components. The non-equilibrium moments obtained from the kinetic model show excellent quantitative agreement with second-order Chapman-Enskog predictions, confirming the theoretical self-consistency of the approach. These non-equilibrium characteristics explain the ability of the kinetic model to capture non-Newtonian effects that linear eddy viscosity models cannot represent, highlighting the fundamental advantage of the kinetic approach over conventional turbulence modelling strategies.

Although this study focused on simple shear flows, the kinetic framework provides a foundation for modelling more complex turbulent phenomena. Future work should address three-dimensional flows, unsteady effects, and extension to compressible turbulence, fully exploiting the theoretical advantages of the kinetic approach. The kinetic theoretic based turbulence model represents a significant step towards describing the fundamental physics of turbulence, involving a wide range of scales without scale separations, it naturally captures higher order nonequilibrium turbulent flow properties as opposed to needing empirical parameters in traditional models.
%The kinetic turbulence model represents a significant step toward physics-based turbulence modelling, offering reduced dependence on empirical parameters while maintaining computational tractability for engineering applications.
We wish to emphasize that the present work is to propose a kinetic theory based turbulence model, in the form of a Boltzmann-BGK differential equation, for better and more naturally capturing certain fundamental underlying physics of the averaged dynamics of turbulence, instead of a kinetic theory based computational method for numerically solving any turbulence problems.

\section*{Acknowledgments} 
This work was supported by the National Natural Science Foundation of China (Grant No. 12472290) and the Interdisciplinary Research Program of HUST (2023JCYJ002). 
Z. Xin gratefully acknowledges the support of the China Scholarship Council (No. 202406160103).
The computation is completed in the HPC Platform of Huazhong University of Science and Technology.

\section*{Declaration of interests} 
The authors report no conflict of interest.

\appendix
\section{Derivations of the Reynolds stress and  turbulent kinetic energy flux via Chapman–Enskog expansion}\label{Appendix: CE}
In this Appendix, we provide a detailed derivation for the analytical expressions
of the Reynolds stress tensor and TKE flux (namely~\eqref{Eq:macro quantities}) via Chapman–Enskog expansion. The task involved in deriving a macroscopic representation of hydrodynamics is to express $\boldsymbol \sigma$ and $\boldsymbol Q$ in terms of the mean variables $\boldsymbol U$ and $K^{eq}$, as well as their (spatial and temporal) derivatives. If the system is at equilibrium, it is straightforward to show that
\begin{equation}
\begin{aligned}
    \sigma^{eq}_{\alpha \beta}&=-\int \left(\xi_\alpha - U_\alpha\right)\left(\xi_\beta - U_\beta\right)F^{eq}d\boldsymbol \xi = -\frac{2}{3} K^{eq} \delta_{\alpha \beta},\\
    Q^{eq}_\alpha &=\frac{1}{2}\int \left(\xi_\alpha - U_\alpha\right)\left(\xi_\beta - U_\beta\right)^2F^{eq}d\boldsymbol{\xi} = 0, 
\end{aligned}
\end{equation}
with 
\begin{equation}
    K^{eq} = \frac{1}{2}\int \left(\xi_\alpha - U_\alpha\right)^2 F^{eq} d \boldsymbol{\xi},
\end{equation}
where $\delta_{\alpha \beta}$ is the Kronecker delta function. On the other hand, when there exists a flow involving non-trivial (spatial or temporal) inhomogeneities, there will be additional contributions to the stress tensor due to the non-equilibrium part of the distribution function, $F^{neq}=F-F^{eq}$. The CE method is a systematic procedure to expand the velocity distribution function around its local equilibrium. This is possible if the ratio between the relaxation time $\tau$ and the representative advection time scale of the left hand side of~\eqref{Eq:kinetic equation} is small. Hence we may express the VDF in terms of a power series in $\mathscr{K}$:
\begin{equation}
    F = F^{(0)} + \mathscr{K} F^{(1)}+\mathscr{K}^2 F^{(2)}+ \cdots ,
\end{equation}
where $F^{(0)} = F^{eq}$, and the additional ($n>0$) terms represent deviations from equilibrium at various orders in $\mathscr{K}$. These non-equilibrium corrections do contribute to the fluxes. For instance, the TKE, Reynolds stress and TKE flux from the $n$th order are given by
\begin{equation}
\begin{aligned}
    K^{(n)}&=\frac{1}{2}\int \left(\xi_\alpha - U_\alpha\right)^2F^{(n)}d\boldsymbol \xi,\\
    \sigma^{(n)}_{\alpha \beta}&=-\int \left(\xi_\alpha - U_\alpha\right)\left(\xi_\beta - U_\beta\right)F^{(n)}d\boldsymbol \xi,\\
    Q_{\alpha}^{(n)}&=\frac{1}{2}\int \left(\xi_\alpha - U_\alpha\right)\left(\xi_\beta - U_\beta\right)^2F^{(n)}d\boldsymbol \xi.
\end{aligned}
\end{equation}

The CE expansion also requires expansion in time and space accordingly,
\begin{equation}
    \begin{aligned}
    \partial_t = \mathscr{K} \partial_{t_1} + \mathscr{K}^2\partial_{t_2}+ \cdots , \qquad
    \frac{\partial}{\partial x_\alpha} = \mathscr{K} \frac{\partial}{\partial x_{1\alpha}},
    \end{aligned}
\end{equation}
and the force and strain rate tensor are similarly expanded as
\begin{equation}
\begin{aligned}
    \bar a_\alpha &= \mathscr{K} \bar a^{(1)}_\alpha + \mathscr{K}^2 \bar a^{(2)}_\alpha, \quad \bar a^{(1)}_\alpha = - \frac{\partial \bar p}{\partial x_{1\alpha}}, \quad \bar a^{(2)}_\alpha = \nu_0 \nabla_1^2 U_\alpha,\\
    S_{\alpha\beta} &= \mathscr{K}S_{\alpha \beta}^{(1)}, \quad S_{\alpha\beta}^{(1)} = \frac{1}{2}\left( \frac{\partial U_\alpha}{\partial x_{1\beta}}+ \frac{\partial U_\beta}{\partial x_{1\alpha}}\right).
\end{aligned}
\end{equation}

Consequently, the kinetic equation~\eqref{Eq:kinetic equation} is turned into an infinite hierarchy of equations according to the order of $\mathscr{K}$,
\begin{equation}
    \sum_{k=1}^{n} \partial_{t_k}F^{(n-k)} + \xi_\alpha  \frac{\partial F^{(n-1)}}{\partial x_{1\alpha}}+\sum_{k=1}^{n}\bar a_\alpha^{(k)}  \frac{\partial F^{(n-k)}}{\partial \xi_\alpha} = -\frac{1}{\tau}F^{(n)}, \quad n=1,2, ..., \infty.
\end{equation}

In particular, for the first order ($n=1$), we have,
\begin{equation}\label{Eq:F(1)}
   \left(\partial_{t_1} + \xi_\alpha  \frac{\partial}{\partial x_{1\alpha}}+\bar a_\alpha ^{(1)} \frac{\partial }{\partial \xi_\alpha} \right) F^{eq} = -\frac{1}{\tau}F^{(1)},
\end{equation}
and for the second order ($n=2$),
\begin{equation}\label{Eq:F(2)}
     \left(\partial_{t_1} + \xi_\alpha  \frac{\partial}{\partial x_{1\alpha}}+\bar a_\alpha^{(1)}  \frac{\partial }{\partial \xi_\alpha} \right) F^{(1)} + \partial_{t_2}F^{eq}+ \bar a_\alpha^{(2)}  \frac{\partial F^{eq}}{\partial \xi_\alpha} =-\frac{1}{\tau}F^{(2)}.
\end{equation}

The fastest time derivative $\partial_{t_1}$ corresponding to isotropic turbulence is a result of the equilibrium VDF:
\begin{equation} \label{Eq:Euler Eq}
    \begin{aligned}
        \frac{\partial U_\alpha}{\partial x_{1\alpha}}&=0,\\
        \partial_{t_1} U_\alpha + U_\beta \frac{\partial U_\alpha}{\partial x_{1\beta}} &=-\frac{\partial \bar p}{\partial x_{1\alpha}}-\frac{\partial }{\partial x_{1\beta}}\left( \frac{2}{3} K^{eq} \delta_{\alpha \beta} \right),\\
         \partial_{t_1} K^{eq}+ U_\alpha \frac{\partial K^{eq}}{\partial x_{1\alpha}} &= -\frac{K^{(1)}}{\tau}.
    \end{aligned}
\end{equation}

Since the space–time dependence in $F^{eq}$ is only obtained through $U_\alpha$ and $K^{eq}$, we have
\begin{equation}
    \begin{aligned}
    \left(\partial_{t_1} + \xi_\alpha  \frac{\partial}{\partial x_{1\alpha}}+\bar a^{(1)}_\alpha  \frac{\partial }{\partial \xi_\alpha} \right) F^{eq} = &\frac{\partial F^{eq}}{\partial U_\beta}\left(\partial_{t_1} + \xi_\alpha  \frac{\partial}{\partial x_{1\alpha}} \right) U_\beta 
    \\&+\frac{\partial F^{eq}}{\partial K^{eq}}\left(\partial_{t_1} + \xi_\alpha  \frac{\partial}{\partial x_{1\alpha}}\right) K^{eq}+\bar a_\alpha^{(1)}  \frac{\partial F^{eq}}{\partial \xi_\alpha}.
    \end{aligned}
\end{equation}

Using the equilibrium VDF~\eqref{Eq: equilibrium VDF}, it can be shown directly that
\begin{equation}\label{Eq:partial Feq}
    \begin{aligned}
        \frac{\partial F^{eq}}{\partial U_\beta} &= \frac{3\left(\xi_\beta - U_\beta\right)}{2K^{eq}}F^{eq},\quad
        \frac{\partial F^{eq}}{\partial K^{eq}} = \frac{3}{2K^{eq}}\left( \frac{\left(\xi_\alpha - U_\alpha \right)^2}{2 K^{eq}} - 1\right)F^{eq},\\
        \frac{\partial F^{eq}}{\partial \xi_\alpha} &= -\frac{3\left(\xi_\alpha - U_\alpha \right)}{2K^{eq}}F^{eq},
    \end{aligned}
\end{equation}

and then combining~\eqref{Eq:F(1)}-\eqref{Eq:partial Feq}, we obtain
\begin{equation}\label{Eq:DFeq}
    \begin{aligned}
    F^{(1)}=&-\tau\left(\partial_{t_1} + \xi_\alpha  \frac{\partial}{\partial x_{1\alpha}}+\bar a_\alpha^{(1)}  \frac{\partial }{\partial \xi_\alpha} \right) F^{eq} 
    =- \frac{3\tau\left(\xi_\alpha-U_\alpha\right)}{2K^{eq}}\left(\frac{\left(\xi_\alpha - U_\alpha \right)^2}{2 K^{eq}} -\frac{5}{3} \right)\frac{\partial K^{eq}}{\partial x_{1\alpha}}F^{eq}\\ \quad \quad \quad
   & + \frac{3K^{(1)}}{2K^{eq}}\left(\frac{\left(\xi_\alpha - U_\alpha \right)^2}{2 K^{eq}} -1 \right)F^{eq}-\frac{3\tau\left(\xi_\alpha-U_\alpha\right)\left(\xi_\beta-U_\beta\right)}{2K^{eq}} \frac{\partial U_\alpha}{\partial x_{1\beta}}F^{eq}.
    \end{aligned}
\end{equation}
Substituting~\eqref{Eq:DFeq} into the stress and TKE flux moments, and after some straightforward algebra, it can be shown that
\begin{equation}
\begin{aligned}
    \sigma^{(1)}_{\alpha \beta}=-\frac{2}{3} K^{(1)}\delta_{\alpha \beta}+\frac{4\tau K^{eq}}{3}S_{\alpha \beta}^{(1)},\quad
    Q_{\alpha}^{(1)}=-\frac{10}{9}\tau K^{eq}\frac{\partial{K^{eq}}}{\partial x_{1\alpha}}.
\end{aligned}
\end{equation}

%Based on these results for the Chapman–Enskog expansion up to first order in $\tau$, we obtain the following relations
%\begin{equation}
%\begin{aligned}
%    \sigma_{\alpha \beta}&=\sigma^{(0)}_{\alpha \beta}+\tau\sigma^{(1)}_{\alpha \beta}=\frac{4\tau K^{eq}}{3}S_{\alpha \beta}-\frac{2}{3}\left( \tau \epsilon + K^{eq}\right)\delta_{\alpha \beta}=\frac{4\tau K^{eq}}{3}S_{\alpha \beta}-\frac{2K}{3}\delta_{\alpha \beta},\\
%    Q_{\alpha}&=Q_{\alpha}^{(0)}+ \tau Q_{\alpha}^{(1)}=-\frac{10}{9}\tau K^{eq}\frac{\partial{K^{eq}}}{\partial x_\alpha}.
%\end{aligned}
%\end{equation}

Furthermore, we present a derivation for the second-order CE expansion in the rest of this Appendix. Using the result of the first-order derivation above, we obtain the hydrodynamic time derivative at the first order. That is,
\begin{equation}
    \begin{aligned}
        \partial_{t_2} U_\alpha&=\frac{2}{3}\frac{\partial}{\partial x_{1\beta}} \left[ 2\tau K^{eq}S^{(1)}_{\alpha \beta}-K^{(1)} \delta_{\alpha \beta}\right]+\nu_0\nabla_1^2U_\alpha,\\
        \partial_{t_1} K^{(1)} +\partial_{t_2} K^{eq} &=\frac{\partial}{\partial x_{1\alpha}} \left[\frac{10}{9} \tau K^{eq}\frac{\partial K^{eq}}{\partial x_{1\alpha}}-U_\alpha K^{(1)} \right] + \frac{4\tau K^{eq}}{3}S^{(1)}_{\alpha \beta}S^{(1)}_{\alpha \beta}-\frac{K^{(2)}}{\tau}.
    \end{aligned}
\end{equation}
This is a direct result of the CE expansion up to the first order. Based on~\eqref{Eq:F(2)},~\eqref{Eq:partial Feq} and~\eqref{Eq:DFeq}, together with
\begin{equation}
    \partial_{t_2}F^{eq} = \frac{\partial F^{eq}}{\partial U_\beta}\partial_{t_2}U_\beta +  \frac{\partial F^{eq}}{\partial K^{eq}}\partial_{t_2} K^{eq},
\end{equation}
and it is easy to obtain
\begin{equation}\label{Eq: par t1 Feq}
\begin{aligned}
    \partial_{t_2}F^{eq}   =& \frac{3 F^{eq}}{2 K^{eq}}\Biggl\{ \frac{2(\xi_\alpha - U_\alpha)}{3} \left[\frac{\partial }{\partial x_{1\beta}}\left(2 \tau K^{eq} S_{\alpha\beta}^{(1)}\right)-\frac{\partial K^{(1)}}{\partial x_{1\alpha}}+\nu_0\nabla_1^2U_\alpha\right]
      + \left(\frac{(\xi_\alpha - U_\alpha)^2}{2 K^{eq}}- 1\right)\\
      &\times\left[
        \frac{\partial }{\partial x_{1\alpha}}
          \left( \frac{10}{9}\tau K^{eq}   \frac{\partial K^{eq} }{\partial x_{1\alpha}}
          \right)- U_\alpha \frac{\partial K^{(1)}}{\partial x_{1\alpha}}- \partial_{t_1}K^{(1)} - \frac{K^{(2)}}{\tau} + \frac{4\tau K^{eq}}{3}S^{(1)}_{\alpha \beta}S^{(1)}_{\alpha \beta}\right] \Biggr\}.
\end{aligned}
\end{equation}

Combining~\eqref{Eq:F(2)},~\eqref{Eq:Euler Eq},~\eqref{Eq:DFeq}, and~\eqref{Eq: par t1 Feq}, we have
\begin{equation}
    \begin{aligned}
        \sigma^{(2)}_{\alpha \beta} =&-\int \left(\xi_\alpha - U_\alpha\right)\left(\xi_\beta - U_\beta\right)F^{(2)}d\boldsymbol \xi\\
        =&-\frac{2 K^{(2)}}{3}\delta_{\alpha \beta}-\frac{4\tau}{3K^{eq}}\left(\partial_{t1}+U_\gamma \frac{\partial }{\partial x_\gamma}\right)\left[\tau (K^{eq})^2 S^{(1)}_{\alpha \beta}\right]
        \\&-\frac{8\tau^2K^{eq}}{3} \left(S^{(1)}_{\alpha \gamma}S^{(1)}_{\gamma \beta} - \frac{1}{3}\delta_{\alpha \beta} S^{(1)}_{\gamma \eta}S^{(1)}_{\gamma \eta}\right)
        +\frac{4\tau^2K^{eq}}{3}\left(S^{(1)}_{\alpha \gamma}\Omega^{(1)}_{\gamma \beta} +S^{(1)}_{\beta \gamma} \Omega^{(1)}_{\gamma \alpha} \right) 
        \\&-\frac{4\tau}{9}\left[\frac{\partial}{\partial x_{1\beta}}\left(\tau K^{eq}\frac{\partial K^{eq}}{\partial x_{1\alpha}}\right)+\frac{\partial}{\partial x_{1\alpha}}\left(\tau K^{eq}\frac{\partial K^{eq}}{\partial x_{1\beta}}\right)- \frac{2}{3}\delta_{\alpha \beta}\frac{\partial}{\partial x_{1\gamma}} \left( \tau K^{eq} \frac{\partial K^{eq}}{\partial x_{1\gamma}}\right)\right] ,
    \end{aligned}
\end{equation}
and 
\begin{equation}
    \begin{aligned}
        Q^{(2)}_{\alpha} =&\frac{1}{2}\int \left(\xi_\alpha - U_\alpha\right)\left(\xi_\beta - U_\beta\right)^2F^{(2)}d\boldsymbol \xi\\
        =&\frac{10\tau}{9K^{eq}}\left(\partial_{t1}+U_\gamma \frac{\partial }{\partial x_\gamma}\right)\left[\tau (K^{eq})^2 \frac{\partial K^{eq}}{\partial x_{1\alpha}}\right]
        +\frac{10\tau^2K^{eq}}{9}\frac{\partial K^{eq}}{\partial x_{1\beta}}\frac{\partial U_\beta}{\partial x_{1\alpha}}
        \\&-\frac{5\tau K^{eq}}{3}\frac{\partial}{\partial x_{1\beta}}\left(\frac{4\tau K^{eq}}{3}S^{(1)}_{\alpha \beta}\right) + \frac{5\tau K^{eq}}{3} a^{(1)}_{\alpha},
    \end{aligned}
\end{equation}
where $\Omega_{\alpha \beta}$ is the vorticity tensor of the mean velocity field, which is defined as
\begin{equation}
    \Omega_{\alpha \beta} \equiv \frac{1}{2}\left(\frac{\partial U_\alpha}{\partial x_\beta} -  \frac{\partial U_\beta}{\partial x_\alpha} \right)=\mathscr{K}\Omega^{(1)}_{\alpha \beta}, \quad  \Omega^{(1)}_{\alpha \beta} = \frac{1}{2}\left(\frac{\partial U_\alpha}{\partial x_{1\beta}} -  \frac{\partial U_\beta}{\partial x_{1\alpha}} \right).
\end{equation}

\section{The discretized velocity method of the kinetic model for (quasi-one-dimensional) turbulent  Couette flow}\label{sec: dvm}
Since only a quasi-one-dimensional turbulent Couette flow is considered in this study, four reduced VDFs are introduced to cast the three-dimensional velocity space $\left(\xi_x, \xi_y, \xi_z\right)$ into one dimension as
\begin{equation}
    \begin{aligned}
        \Phi_1\left(y, \xi_y,t\right) &= \int F\left(\boldsymbol{x},\boldsymbol{\xi},t\right) d\xi_x d\xi_z,\quad \quad
        \Phi_2\left(y, \xi_y,t\right) = \int \xi_x F\left(\boldsymbol{x},\boldsymbol{\xi},t\right) d\xi_x d\xi_z,\\
        \Phi_3\left(y, \xi_y,t\right) &=\int \xi_x^2F\left(\boldsymbol{x},\boldsymbol{\xi},t\right) d\xi_x d\xi_z, \quad
        \Phi_4\left(y, \xi_y,t\right) = \int \xi_z^2F\left(\boldsymbol{x},\boldsymbol{\xi},t\right) d\xi_x d\xi_z.
    \end{aligned}
\end{equation}
Then the mean variables can be computed by taking moments of the reduced VDFs,
\begin{equation}
    \begin{aligned}
        U_x\left(y,t\right) &= \int \Phi_2 d\xi_y, \qquad
        U_y\left(y,t\right) = \int \xi_y \Phi_1 d\xi_y,\\
        K\left(y,t\right) &= \frac{1}{2}\int \left(\left(\xi_y-U_y\right)^2\Phi_1+\Phi_3-U_x^2 \Phi_1+\Phi_4\right)d\xi_y- U_x^2,\\
    \end{aligned}
\end{equation}
and the high-order moments of interest are given by
\begin{equation}\label{Eq: sigma}
    \begin{aligned}
        \sigma_{xx}\left(y,t\right) &=- \int \left(\Phi_3-U_x^2 \Phi_1\right)d\xi_y + 2U_x^2,\quad
        \sigma_{xy}\left(y,t\right) = -\int \xi_y\Phi_2d\xi_y+U_xU_y,\\
        \sigma_{yy}\left(y,t\right) &= -\int\left(\xi_y-U_y\right)^2\Phi_1 d\xi_y,\qquad\qquad
        \sigma_{zz}\left(y,t\right) = -\int\Phi_4 d\xi_y.\\
    \end{aligned}
\end{equation}

Without loss of generality, we present the discretized velocity method for the LR-BGK model~\eqref{Eq:kinetic equation with source} and~\eqref{Eq: S term F}. The governing equations for the four reduced VDFs can be deduced as
\begin{equation}\label{Eq: phi eq}
    \partial_t \Phi_\alpha + \xi_y \frac{\partial \Phi_\alpha}{ \partial y} = \frac{\Phi^{eq}_\alpha-\Phi_\alpha}{\tau} + \mathcal{S}_\alpha, \qquad \alpha =1,2,3,4,
\end{equation}
where $\Phi^{eq}_\alpha$ are the reduced equilibrium VDFs, 
\begin{equation}
    \begin{aligned}
        \Phi_1^{eq}&=
        \left(\frac{4}{3}\pi K^{eq}\right)^{-\frac{1}{2}}\exp\left[-\frac{3\left(\xi_y-U_y\right)^2}{4K^{eq}}\right],\\
        \Phi_2^{eq}&= U_x \Phi_1^{eq},\qquad
        \Phi_3^{eq}= \left(\frac{2K^{eq}}{3}+U^2_x\right) \Phi_1^{eq},\qquad
        \Phi_4^{eq}= \frac{2K^{eq}}{3} \Phi_1^{eq},\qquad
    \end{aligned}
\end{equation}
and $\mathcal{S}_\alpha$ represent the corresponding source terms,
\begin{equation}
    \begin{aligned}
        \mathcal{S}_1&= -\bar a_y\frac{\partial \Phi_1}{\partial \xi_y}-\frac{\nu_0 \nabla^2K}{2K^{eq}}\left(\Phi_1+\left(\xi_y-U_y\right)\frac{\partial \Phi_1}{\partial \xi_y}\right),\\
        \mathcal{S}_2&= \bar a_x \Phi_1-\bar a_y\frac{\partial \Phi_2}{\partial \xi_y}-\frac{\nu_0 \nabla^2K}{2K^{eq}}\left(U_x\Phi_1+\left(\xi_y-U_y\right)\frac{\partial \Phi_2}{\partial \xi_y}\right),\\
        \mathcal{S}_3&= 2\bar a_x\Phi_2-\bar a_y\frac{\partial \Phi_3}{\partial \xi_y}-\frac{\nu_0 \nabla^2K}{2K^{eq}}\left(U_x\Phi_2-\Phi_3+\left(\xi_y-U_y\right)\frac{\partial \Phi_3}{\partial \xi_y}\right),\\
        \mathcal{S}_4&= -\bar a_y\frac{\partial \Phi_4}{\partial \xi_y}-\frac{\nu_0 \nabla^2K}{2K^{eq}}\left(-\Phi_4+\left(\xi_y-U_y\right)\frac{\partial \Phi_4}{\partial \xi_y}\right).
    \end{aligned}
\end{equation}
Within the finite-volume framework, the governing equation~\eqref{Eq: phi eq} for $\Phi_\alpha$ is discretized as
\begin{equation}\label{Eq: imp update f} 
    \Phi^{n+1}_{\alpha,j} = \Phi^n_{\alpha,j} - \frac{\xi_y \Delta t}{\Delta y}\left( \Phi^{n+1/2}_{\alpha,j+1/2}- \Phi^{n+1/2}_{\alpha,j-1/2}\right)+\frac{\Delta t}{\tau^{n+1}}\left(\Phi^{eq,n+1}_{\alpha,j}-\Phi^{n+1}_{\alpha,j} \right)+\Delta t \mathcal{S}^{n}_{\alpha,j},
\end{equation}
where $\Delta y = y_{j+1/2}-y_{j-1/2}$ is the length of the cell $j$, $\Phi_{j-1/2}^{n+1/2}$ and $\Phi_{j+1/2}^{n+1/2}$ are the left and right VDFs at the cell interface at time $t^{n+1/2}$, respectively. To eliminate the implicit collision term, we introduce the transformed VDFs     $\tilde{\Phi}_\alpha = \Phi_\alpha - \frac{\Delta t}{\tau}\left(\Phi^{eq}_\alpha -  \Phi_\alpha\right)$
%\begin{equation}\label{Eq:transform f}
%    \tilde{\Phi}_\alpha = \Phi_\alpha - \frac{\Delta t}{\tau}\left(\Phi^{eq}_\alpha -  \Phi_\alpha\right),
%\end{equation}
and the mean variables can be computed by taking moments of the transformed VDFs,
\begin{equation}\label{Eq:transform W}
    \begin{aligned}
        U_x &= \int \tilde \Phi_2 d\xi_y, \qquad
        U_y = \int \xi_y \tilde\Phi_1 d\xi_y,\\
        K&= \frac{1}{2}\int \left(\left(\xi_y-U_y\right)^2\tilde\Phi_1+\tilde\Phi_3-U_x^2 \tilde \Phi_1+\tilde\Phi_4\right)d\xi_y- U_x^2-\Delta t \epsilon.\\
    \end{aligned}
\end{equation}

Then equation~\eqref{Eq: imp update f} can be rewritten as
\begin{equation}\label{Eq:evol tilde f}
    \tilde \Phi^{n+1}_{\alpha,j} = \Phi^n_{\alpha,j} - \frac{\xi_y \Delta t}{\Delta y}\left( \Phi^{n+1/2}_{\alpha,j+1/2}- \Phi^{n+1/2}_{\alpha,j-1/2}\right)+\Delta t \mathcal{S}^{n}_{\alpha,j},
\end{equation}
where the mean pressure entering the source term is governed by the kinematic pressure–Poisson relation
$
\frac{\partial ^2\bar{p}^n}{\partial y^2} = \frac{\partial ^2\sigma_{yy}^n}{\partial y^2} 
$. 

As a result, the original VDFs and the mean variables at $t^{n+1}$ can be obtained from $\tilde \Phi^{n+1}_{\alpha,j}$,
\begin{equation}\label{Eq:evol f and W}
\begin{aligned}
    \Phi^{n+1}_{\alpha,j} &= \frac{\tau}{\tau +\Delta t} \tilde\Phi^{n+1}_{\alpha,j}+ \frac{\Delta t}{\tau +\Delta t} \Phi^{eq,n+1}_{\alpha,j},\\
    U_{x,j}^{n+1} &= \int \tilde \Phi_{2,j}^{n+1} d\xi_y, \\
    U_{y,j}^{n+1} &= \int \xi_y \tilde\Phi_{1,j}^{n+1} d\xi_y,\\
    K^{n+1}_j&= \frac{1}{2}\int \left(\left(\xi_y-U_{y,j}^{n+1}\right)^2\tilde\Phi_{1,j}^{n+1}+\tilde\Phi_{3,j}^{n+1}-U_x^2 \tilde \Phi_{1,j}^{n+1}+\tilde\Phi_{4,j}^{n+1}\right)d\xi_y- \left(U_{x,j}^{n+1}\right)^2-\Delta t \epsilon^{n+1}_j.\\
\end{aligned}
\end{equation}
and $\boldsymbol{\sigma}^{n+1}_j$ are calculated by the original VDFs $\Phi^{n+1}_{\alpha,j}$ according to~\eqref{Eq: sigma}.

The governing equation~\eqref{Eq: epsilon evol} for $\epsilon$ is also discretized using the finite-volume method, 
\begin{equation}\label{Eq:evol eps}
\begin{aligned}
     \epsilon^{n+1}_j &= \epsilon^n_j + \frac{\Delta t}{\Delta y}\left(\frac{\nu_T}{C_\epsilon} \left(\frac{\partial \epsilon}{\partial y}\right)^n_{j+1/2}-U^n_{j+1/2}\epsilon^n_{j+1/2}-\frac{\nu_T}{C_\epsilon} \left(\frac{\partial \epsilon}{\partial y}\right)^n_{j-1/2}+U^n_{j-1/2}\epsilon^n_{j-1/2}\right)\\
     &+C_{\epsilon1} \frac{\epsilon^n_j}{K_j^n} \left(\sigma_{xy,j}^n \left(\frac{\partial U_x}{\partial y}\right)^{n}_j+\sigma_{yy,j}^n \left(\frac{\partial U_y}{\partial y}\right)^{n}_j\right)-C_{\epsilon2}f_{\epsilon2}\frac{\left(\epsilon^n_j\right)^2}{K_j^n},
\end{aligned}
\end{equation}
where the velocity gradient is evaluated using a central finite difference.
Thus, the update rules for the VDFs and the macroscopic variables are given by equations~\eqref{Eq:evol tilde f},~\eqref{Eq:evol f and W} and~\eqref{Eq:evol eps}; the remaining task is to reconstruct the VDFs and macroscopic variables at interface.

First, the VDFs at cell interface are reconstructed from the local solution of the collisionless kinetic equations~\eqref{Eq: phi eq},
\begin{equation}
    \Phi_\alpha \left(y_{j+1/2},\xi_y,t^{n+1/2}\right) = \Phi_\alpha \left(y_{j+1/2}-\xi_y\Delta t/2,\xi_y,t^{n}\right).
\end{equation}
With the Taylor expansion around the cell interface $y_{j+1/2}$,  $\Phi_\alpha \left(y_{j+1/2},\xi_y,t^{n+1/2}\right)$ can be approximated by linear interpolations,
\begin{equation}\label{Eq: interface f}
    \Phi_\alpha \left(y_{j+1/2},\xi_y,t^{n+1/2}\right) =\Phi_\alpha \left(y_{j+1/2},\xi_y,t^{n}\right) - \frac{\xi_y \Delta t}{2} \frac{\Phi_\alpha \left(y_{j+1},\xi_y,t^{n}\right)-\Phi_\alpha \left(y_{j},\xi_y,t^{n}\right)}{y_{j+1}-y_{j}},
\end{equation}
and 
\begin{equation}\label{Eq: interface f1}
    \Phi_\alpha \left(y_{j+1/2},\xi_y,t^{n}\right) =\Phi_\alpha \left(y_{j},\xi_y,t^{n}\right) +\left(y_{j+1/2}-y_j\right)\frac{\Phi_\alpha \left(y_{j+1},\xi_y,t^{n}\right)-\Phi_\alpha \left(y_{j},\xi_y,t^{n}\right)}{y_{j+1}-y_{j}}.
\end{equation}

Then, $\epsilon$ at the cell interface are also obtained by linear interpolation,
\begin{equation}\label{Eq: interface W}
    \begin{aligned}
        %&\boldsymbol U_{j+1/2}^n=\boldsymbol U_{j}^n +\left(y_{j+1/2}-y_j\right) \frac{\boldsymbol U_{j+1}^n-\boldsymbol U_{j}^n}{y_{j+1}-y_{j}},\\
       % &K_{j+1/2}^n=K_{j}^n +\left(y_{j+1/2}-y_j\right) \frac{K_{j+1}^n-K_{j}^n}{y_{j+1}-y_{j}},\\
        \epsilon_{j+1/2}^n=\epsilon_{j}^n +\left(y_{j+1/2}-y_j\right) \frac{\epsilon_{j+1}^n-\epsilon_{j}^n}{y_{j+1}-y_{j}},\quad \left(\frac{\partial \epsilon}{\partial y}\right)^n_{j+1/2}=\left(y_{j+1/2}-y_j\right) \frac{\epsilon_{j+1}^n-\epsilon_{j}^n}{y_{j+1}-y_{j}}.
    \end{aligned}
\end{equation}

Additionally, this model still contains compressibility errors during time evolution. Here, we introduce a correction to the mean pressure defined by~\eqref{Eq: imp update f}. That is, the flow velocity is
\begin{equation}\label{Eq:u n+1}
\begin{aligned}
    U_{y,j}^{n*}&=U_{y,j}^{n}-\frac{\Delta t}{\Delta y} \left(\int \xi_y^2\Phi_{1,j+1/2}^{n+1/2}d\xi_y- \int \xi_y^2\Phi_{1,j-1/2}^{n+1/2}d\xi_y\right)+\Delta t \left(-\left(\frac{\partial \bar p}{\partial y}\right)^n_j + \left(\nu_0 \nabla^2 U\right)^n_{j}\right).
\end{aligned}
\end{equation}
Further, we define the pressure correction as the difference between the updated and the previous pressure field $\delta \bar p \equiv\bar p^{n+1}_j-\bar p^{n}_j$.
With this definition, the corrected velocity field is expressed as
\begin{equation}\label{eq:u_corr}
U^{n+1}_{y,j} = U^{n*}_{y,j} - \Delta t \frac{\partial (\delta \bar p)}{\partial y}.
\end{equation}
In order to enforce ${U}^{n+1}_{y,j}$ to be divergent free, $\delta p$ should satisfy the following Poisson equation,
\begin{equation}\label{Eq:phi}
\begin{aligned}
    \frac{\partial^2 (\delta \bar p)}{ \partial y^2}&=\frac{1}{\Delta t}\frac{\partial U_{y,j}^{n*}}{ \partial y}.
\end{aligned}
\end{equation}
After solving the above equation, one can obtain the fluid velocity ${U}^{n+1}_{y,j}$.

The above method is valid in a continuous velocity
space $\xi_y$. In practical computations, the velocity space is divided into a finite set of discrete velocities $\left\{\xi_{y,k} \mid k=1,2, \ldots, N\right\}$, where $N$ is the number of discrete velocities. With the discrete velocity space, the moments of a continuous distribution function have to be expressed as discrete moments,
\begin{equation}
    \int \phi(\xi_y) F(\xi_y) d \xi_y = \sum_k w_k\phi(\xi_{y,k}) F(\xi_{y,k}),
\end{equation}
where $\phi$ is a polynomial of the particle velocity $\xi_y$ and $\omega_k$ is the weight of the numerical quadrature at the discrete velocity $\xi_{y,k}$. It is important to discretize the velocity space based on certain quadrature rules, which are chosen according to the rarefaction degree.  The gradients of VDFs in the source term are approximated by central differences.
%\begin{equation}
%    \left(\frac{\partial \Phi_\alpha}{\partial \xi_y}\right)_k=\frac{\Phi_\alpha(\xi_{y,k+1})-\Phi_\alpha(\xi_{y,k-1})}{(\xi_{y,k+1}-\xi_{y,k-1})}.
%\end{equation}
%The discrete velocity method described above is similar to the well-known discrete unified gas kinetic scheme~\citep{guo2013discrete}, with the only difference being that the construction of the interface distribution function does not take into account collision effects.

For completeness, the general algorithmic framework consists of the following steps:
\begin{enumerate}
\item At the initial time ($n=0$), the variables ($\boldsymbol{U}^0$, $K^0$ and $\epsilon^0$) and the original VDFs $\Phi^0_\alpha$ are initialized at each cell center;
\item Evolve the VDFs $\Phi_\alpha$ at each cell interface at time $t^{n+1/2}$ according to~\eqref{Eq: interface f} and~\eqref{Eq: interface f1}, and then calculate $\epsilon$ at each cell interface according to~\eqref{Eq: interface W};
\item Update $\epsilon^{n+1}$ at each cell interface according to~\eqref{Eq:evol eps};
\item Update the transformed VDFs $\tilde \Phi_\alpha^{n+1}$, and then update VDFs $\Phi_\alpha^{n+1}$ and mean variables $\boldsymbol{U}^{n+1}$ and $K^{n+1}$ using~\eqref{Eq:evol f and W};
\item Correct the mean velocity ${U}^{n+1}_{y,j}$ according to~\eqref{Eq:u n+1}, ~\eqref{eq:u_corr}, and~\eqref{Eq:phi};
\item Advance the time step $(n \to n + 1)$ and repeat step (ii) to (v) until the end.
\end{enumerate}

\bibliographystyle{jfm}
\bibliography{Literature}

@article{pope2001turbulent,
  title={Turbulent flows},
  author={Pope, Stephen B},
  journal={Measurement Science and Technology},
  volume={12},
  number={11},
  pages={2020--2021},
  year={2001}
}

@article{moin1998direct,
  title={Direct numerical simulation: a tool in turbulence research},
  author={Moin, Parviz and Mahesh, Krishnan},
  journal={Annual Review of Fluid Mechanics},
  volume={30},
  number={1},
  pages={539--578},
  year={1998},
  publisher={Annual Reviews 4139 El Camino Way, PO Box 10139, Palo Alto, CA 94303-0139, USA}
}

@article{reynolds1895iv,
  title={{IV}. {On} the dynamical theory of incompressible viscous fluids and the determination of the criterion},
  author={Reynolds, Osborne},
  journal={Philosophical Transactions of the Royal Society of London.(a.)},
  number={186},
  pages={123--164},
  year={1895},
  publisher={The Royal Society London}
}

@article{Prandtl1925,
  author    = {Prandtl, Ludwig},
  title     = {\"Uber die ausgebildete Turbulenz},
  journal   = {Zeitschrift f{\"u}r angewandte Mathematik und Mechanik},
  volume    = {5},
  year      = {1925},
  pages     = {136--139},
  language  = {German},
  note      = {In German}
}

@article{durbin2018some,
  title={Some recent developments in turbulence closure modeling},
  author={Durbin, Paul A},
  journal={Annual Review of Fluid Mechanics},
  volume={50},
  number={1},
  pages={77--103},
  year={2018},
  publisher={Annual Reviews}
}

@article{launder1975progress,
  title={Progress in the development of a {Reynolds}-stress turbulence closure},
  author={Launder, Brian Edward and Reece, G Jr and Rodi, W},
  journal={Journal of Fluid Mechanics},
  volume={68},
  number={3},
  pages={537--566},
  year={1975},
  publisher={Cambridge University Press}
}

@book{monin2013statistical,
  title={Statistical {Fluid} {Mechanics}, {Volume} II: {Mechanics} of {Turbulence}},
  author={Monin, Andre Sergeevich and Yaglom, Akiva M},
  year={2013},
  publisher={Courier Corporation}
}

@article{hamlington2008reynolds,
  title={{Reynolds} stress closure for nonequilibrium effects in turbulent flows},
  author={Hamlington, Peter E and Dahm, Werner JA},
  journal={Physics of Fluids},
  volume={20},
  number={11},
  pages = {115101},
  year={2008},
  publisher={AIP Publishing}
}

@article{chen2023average,
  title={Average turbulence dynamics from a one-parameter kinetic theory},
  author={Chen, Hudong and Staroselsky, Ilya and Sreenivasan, Katepalli R and Yakhot, Victor},
  journal={Atmosphere},
  volume={14},
  number={7},
  pages={1109},
  year={2023},
  publisher={MDPI}
}

@article{chen2024average,
  title={Average turbulence dynamics from a one-parameter kinetic theory: Estimation of the relaxation time},
  author={Chen, Hudong and Staroselsky, Ilya and Sreenivasan, Katepalli R and Yakhot, Victor},
  journal={Physics of Fluids},
  volume={36},
  number={3},
  pages={035156},
  year={2024},
  publisher={AIP Publishing}
}

@incollection{launder1983numerical,
  title={The numerical computation of turbulent flows},
  author={Launder, Brian Edward and Spalding, Dudley Brian},
  booktitle={Numerical {Prediction} of {Flow}, {Heat} {Transfer}, {Turbulence} and {Combustion}},
  pages={96--116},
  year={1983},
  publisher={Elsevier}
}

@article{klimontovich1969statistical,
  title={The statistical theory of non-equilibrium processes in a plasma},
  author={Klimontovich, Yu L},
  journal={Journal of Plasma Physics},
  volume={3},
  pages={148},
  year={1969}
}

@book{chapman1990mathematical,
  title={The {Mathematical} {Theory} of {Non-Uniform} {Gases}},
  author={Chapman, Sydney and Cowling, Thomas George},
  year={1990},
  publisher={Cambridge University Press}
}

@article{sreenivasan2021dynamics,
  title={Dynamics of three-dimensional turbulence from {Navier}-{Stokes} equations},
  author={Sreenivasan, Katepalli R and Yakhot, Victor},
  journal={Physical Review Fluids},
  volume={6},
  number={10},
  pages={104604},
  year={2021},
  publisher={APS}
}

@article{sreenivasan2024saturation,
  title={Saturation of exponents and the asymptotic fourth state of turbulence},
  author={Sreenivasan, Katepalli R and Yakhot, Victor and Staroselsky, Ilya and Chen, Hudong},
  journal={Physical Review Research},
  volume={6},
  number={3},
  pages={033087},
  year={2024},
  publisher={APS}
}

@inproceedings{spalart1992one,
  title={A one-equation turbulence model for aerodynamic flows},
  author={Spalart, Philippe and Allmaras, Steven},
  booktitle={30th {Aerospace} {Sciences} {Meeting} and {Exhibit}},
  pages={439},
  year={1992}
}

@article{menter1997eddy,
  title={Eddy viscosity transport equations and their relation to the $k$-$\varepsilon$ model},
  author={Menter, Florian R},
  journal={ASME Journal of Fluids Engineering},
  volume={119},
  number={4},
  pages={876--884},
  year={1997},
}

@article{wilcox2008formulation,
  title={Formulation of the $k$-$\omega$ turbulence model revisited},
  author={Wilcox, David C},
  journal={AIAA Journal},
  volume={46},
  number={11},
  pages={2823--2838},
  year={2008}
}

@article{pope1975more,
  title={A more general effective-viscosity hypothesis},
  author={Pope, Stephen B},
  journal={Journal of Fluid Mechanics},
  volume={72},
  number={2},
  pages={331--340},
  year={1975},
  publisher={Cambridge University Press}
}

@article{speziale1987nonlinear,
  title={On nonlinear $K$-$l$ and $K$-$\varepsilon$ models of turbulence},
  author={Speziale, Charles G},
  journal={Journal of Fluid Mechanics},
  volume={178},
  pages={459--475},
  year={1987},
  publisher={Cambridge University Press}
}

@article{gatski1993explicit,
  title={On explicit algebraic stress models for complex turbulent flows},
  author={Gatski, Thomas B and Speziale, Charles G},
  journal={Journal of Fluid Mechanics},
  volume={254},
  pages={59--78},
  year={1993},
  publisher={Cambridge University Press}
}

@article{craft1996development,
  title={Development and application of a cubic eddy-viscosity model of turbulence},
  author={Craft, TJ and Launder, BE and Suga, K},
  journal={International Journal of Heat and Fluid Flow},
  volume={17},
  number={2},
  pages={108--115},
  year={1996},
  publisher={Elsevier}
}

@article{wallin2000explicit,
  title={An explicit algebraic {Reynolds} stress model for incompressible and compressible turbulent flows},
  author={Wallin, Stefan and Johansson, Arne V},
  journal={Journal of Fluid Mechanics},
  volume={403},
  pages={89--132},
  year={2000},
  publisher={Cambridge University Press}
}

@book{wilcox1998turbulence,
  title={Turbulence {Modeling} for CFD},
  author={Wilcox, David C},
  year={1998},
  publisher={DCW Industries}
}

@article{venugopal2019non,
  title={Non-equilibrium thermal transport and entropy analyses in rarefied cavity flows},
  author={Venugopal, Vishnu and Praturi, Divya Sri and Girimaji, Sharath S},
  journal={Journal of Fluid Mechanics},
  volume={864},
  pages={995--1025},
  year={2019},
  publisher={Cambridge University Press}
}

@article{grotzbach1982direct,
  title={Direct numerical simulation of the turbulent momentum and heat transfer in an internally heated fluid layer},
  author={Gr{\"o}tzbach, G},
  journal={Heat Transfer},
  volume={2},
  pages={141--146},
  year={1982},
  publisher={Hemisphere Publishing Washington, DC}
}

@article{moeng1989evaluation,
  title={Evaluation of turbulent transport and dissipation closures in second-order modeling},
  author={Moeng, Chin-Hoh and Wyngaard, John C},
  journal={Journal of Atmospheric Sciences},
  volume={46},
  number={14},
  pages={2311--2330},
  year={1989}
}

@article{chandra2007analysis,
  title={Analysis and modeling of the turbulent diffusion of turbulent kinetic energy in natural convection},
  author={Chandra, Laltu and Gr{\"o}tzbach, G{\"u}nther},
  journal={Flow, Turbulence and Combustion},
  volume={79},
  pages={133--154},
  year={2007},
  publisher={Springer}
}

@article{chen2004expanded,
  title={Expanded analogy between {Boltzmann} kinetic theory of fluids and turbulence},
  author={Chen, Hudong and Orszag, Steven A and Staroselsky, Ilya and Succi, Sauro},
  journal={Journal of Fluid Mechanics},
  volume={519},
  pages={301--314},
  year={2004},
  publisher={Cambridge University Press}
}

@article{heinz2007unified,
  title={Unified turbulence models for {LES} and {RANS}, {FDF} and {PDF} simulations},
  author={Heinz, Stefan},
  journal={Theoretical and Computational Fluid Dynamics},
  volume={21},
  pages={99--118},
  year={2007},
  publisher={Springer}
}

@article{luan2025constructing,
  title={Constructing turbulence models using the kinetic {Fokker}--{Planck} equation},
  author={Luan, Peng and Zhang, Haoyuan and Zhang, Jun},
  journal={Journal of Fluid Mechanics},
  volume={1011},
  pages={A44},
  year={2025},
  publisher={Cambridge University Press}
}

@article{yakhot1992development,
  title={Development of turbulence models for shear flows by a double expansion technique},
  author={Yakhot, VSASTBCG and Orszag, Steven A and Thangam, Siva and Gatski, TB and Speziale, CG},
  journal={Physics of Fluids A},
  volume={4},
  number={7},
  pages={1510--1520},
  year={1992},
  publisher={American Institute of Physics}
}

@article{nisizima1987turbulent,
  title={Turbulent channel and {Couette} flows using an anisotropic $k$-$\varepsilon$ model},
  author={Nisizima, Shoiti and Yoshizawa, Akira},
  journal={AIAA Journal},
  volume={25},
  number={3},
  pages={414--420},
  year={1987}
}

@article{rubinstein1990nonlinear,
  title={Nonlinear {Reynolds} stress models and the renormalization group},
  author={Rubinstein, Robert and Barton, J Michael},
  journal={Physics of Fluids A},
  volume={2},
  number={8},
  pages={1472--1476},
  year={1990},
  publisher={American Institute of Physics}
}

@article{myong1990prediction,
  title={Prediction of anisotropy of the near-wall turbulence with an anisotropic low-{Reynolds}-number $k$-$\varepsilon$ turbulence model},
  author={Myong, Hyon Kook and Kasagi, Nobuhide},
  journal={ASME Journal of Fluids Engineering},
  volume={112},
  pages={521--524},
  year={1990}
}

@article{jones1972prediction,
  title={The prediction of laminarization with a two-equation model of turbulence},
  author={Jones, W Peter and Launder, Brian Edward},
  journal={International Journal of Heat and Mass Transfer},
  volume={15},
  number={2},
  pages={301--314},
  year={1972},
  publisher={Elsevier}
}

@article{myong1990new,
  title={A new approach to the improvement of $k$-$\varepsilon$ turbulence model for wall-bounded shear flows},
  author={Myong, Hyon Kook and Kasagi, Nobuhide},
  journal={JSME International Journal Ser. 2},
  volume={33},
  number={1},
  pages={63--72},
  year={1990},
  publisher={The Japan Society of Mechanical Engineers}
}

@article{nagano1990improved,
  title={An improved $k$-$\varepsilon$ model for boundary layer flows},
  author={Nagano, Y and Tagawa, M},
  journal={ASME Journal of Fluids Engineering},
  volume={112},
  number={1},
  pages={33-39},
  year={1990}
}

@article{lundgren1969model,
  title={Model equation for nonhomogeneous turbulence},
  author={Lundgren, TS},
  journal={Physics of Fluids},
  volume={12},
  number={3},
  pages={485--497},
  year={1969}
}

@article{maxwell1879vii,
  title={{VII}. {On} stresses in rarified gases arising from inequalities of temperature},
  author={Maxwell, James Clerk},
  journal={Philosophical Transactions of the Royal Society of London},
  number={170},
  pages={231--256},
  year={1879},
  publisher={The Royal Society London}
}

@article{zhao2002non,
  title={Non-equilibrium extrapolation method for velocity and pressure boundary conditions in the lattice {Boltzmann} method},
  author={Guo, Zhaoli and Zheng, Chuguang and Shi, Baochang},
  journal={Chinese Physics},
  volume={11},
  number={4},
  pages={366},
  year={2002},
  publisher={IOP Publishing}
}

@article{cercignani1971kinetic,
  title={Kinetic models for gas-surface interactions},
  author={Cercignani, Carlo and Lampis, Maria},
  journal={Transport Theory and Statistical Physics},
  volume={1},
  number={2},
  pages={101--114},
  year={1971},
  publisher={Taylor \& Francis}
}

@article{cercignani1972scattering,
  title={Scattering kernels for gas-surface interactions},
  author={Cercignani, Carlo},
  journal={Transport Theory and Statistical Physics},
  volume={2},
  number={1},
  pages={27--53},
  year={1972},
  publisher={Taylor \& Francis}
}

@article{lord1991some,
  title={Some extensions to the {Cercignani}-{Lampis} gas-surface scattering kernel},
  author={Lord, RG},
  journal={Physics of Fluids A},
  volume={3},
  number={4},
  pages={706--710},
  year={1991}
}

@article{aoki2022boundary,
  title={Boundary conditions for the {Boltzmann} equation from gas-surface interaction kinetic models},
  author={Aoki, Kazuo and Giovangigli, Vincent and Kosuge, Shingo},
  journal={Physical Review E},
  volume={106},
  number={3},
  pages={035306},
  year={2022},
  publisher={APS}
}

@article{kosuge2025applications,
  title={Applications of new boundary conditions for the {Boltzmann} equation derived from a kinetic model of gas-surface interaction},
  author={Kosuge, Shingo and Aoki, Kazuo and Giovangigli, Vincent and Golse, Fran{\c{c}}ois},
  journal={Physical Review Fluids},
  volume={10},
  number={5},
  pages={053401},
  year={2025},
  publisher={APS}
}

@article{chen2003extended,
  title={Extended {Boltzmann} kinetic equation for turbulent flows},
  author={Chen, Hudong and Kandasamy, Satheesh and Orszag, Steven and Shock, Rick and Succi, Sauro and Yakhot, Victor},
  journal={Science},
  volume={301},
  number={5633},
  pages={633--636},
  year={2003},
  publisher={American Association for the Advancement of Science}
}

@article{basu2021turbulent,
  title={Turbulent {Prandtl} number and characteristic length scales in stably stratified flows: steady-state analytical solutions},
  author={Basu, Sukanta and Holtslag, Albert AM},
  journal={Environmental Fluid Mechanics},
  volume={21},
  number={6},
  pages={1273--1302},
  year={2021},
  publisher={Springer}
}

@article{kays1994turbulent,
  title={Turbulent {Prandtl} number. Where are we?},
  author={Kays, William M},
  journal={ASME Journal of Heat Transfer},
  volume={116},
  number={2},
  pages={284--295},
  year={1994}
}

@book{tennekes1972first,
  title={A {First} {Course} in {Turbulence}},
  author={Tennekes, Hendrik and Lumley, John Leask},
  year={1972},
  publisher={MIT Press}
}

@book{sutton2020atmospheric,
  title={Atmospheric {Turbulence}},
  author={Sutton, Oliver Graham},
  year={2020},
  publisher={Routledge}
}

@article{henry1984analytical,
  title={Analytical solution of two gradient-diffusion models applied to turbulent {Couette} flow},
  author={Henry, FS and Reynolds, AJ},
  journal = {ASME Journal of Fluids Engineering},
  volume={106},
  number={2},
  pages={211--216},
  year={1984}
}

@article{schneider1989reynolds,
  title={On {Reynolds} stress transport in turbulent {Couette} flow},
  author={Schneider, W},
  journal={Zeitschrift fur Flugwissenschaften und Weltraumforschung},
  volume={13},
  pages={315--319},
  year={1989}
}

@article{andersson1994modeling,
  title={Modeling plane turbulent {Couette} flow},
  author={Andersson, HI and Pettersson, BA},
  journal={International Journal of Heat and Fluid Flow},
  volume={15},
  number={6},
  pages={447--455},
  year={1994},
  publisher={Elsevier}
}

@article{tavoularis1981experiments,
  title={Experiments in nearly homogenous turbulent shear flow with a uniform mean temperature gradient. {Part} 1},
  author={Tavoularis, Stavros and Corrsin, Stanley},
  journal={Journal of Fluid Mechanics},
  volume={104},
  pages={311--347},
  year={1981},
  publisher={Cambridge University Press}
}

@book{frisch1996turbulence,
  title={Turbulence},
  author={Frisch, Uriel},
  year={1996},
  publisher={Cambridge University Press}
}

@article{hanjalic1976contribution,
  title={Contribution towards a {Reynolds}-stress closure for low-{Reynolds}-number turbulence},
  author={Hanjali{\'c}, K and Launder, Brian Edward},
  journal={Journal of Fluid Mechanics},
  volume={74},
  number={4},
  pages={593--610},
  year={1976},
  publisher={Cambridge University Press}
}

@article{wu2014solving,
  title={Solving the {Boltzmann} equation deterministically by the fast spectral method: application to gas microflows},
  author={Wu, Lei and Reese, Jason M and Zhang, Yonghao},
  journal={Journal of Fluid Mechanics},
  volume={746},
  pages={53--84},
  year={2014},
  publisher={Cambridge University Press}
}

@article{bech1995investigation,
  title={An investigation of turbulent plane {Couette} flow at low {Reynolds} numbers},
  author={Bech, Knut H and Tillmark, Nils and Alfredsson, P Henrik and Andersson, Helge I},
  journal={Journal of Fluid Mechanics},
  volume={286},
  pages={291--325},
  year={1995},
  publisher={Cambridge University Press}
}

@book{robertson1959study,
  title={A {Study} of {Turbulent} {Plane} {Couette} {Flow}},
  author={Robertson, James M},
  year={1959},
  publisher={University of Illinois, Department of Theoretical and Applied Mechanics}
}

@article{pirozzoli2014turbulence,
  title={Turbulence statistics in {Couette} flow at high {Reynolds} number},
  author={Pirozzoli, Sergio and Bernardini, Matteo and Orlandi, Paolo},
  journal={Journal of Fluid Mechanics},
  volume={758},
  pages={327--343},
  year={2014},
  publisher={Cambridge University Press}
}

@article{kitoh2005experimental,
  title={Experimental study on mean velocity and turbulence characteristics of plane {Couette} flow: low-{Reynolds}-number effects and large longitudinal vortical structure},
  author={Kitoh, Osami and Nakabyashi, Koichi and Nishimura, Futoshi},
  journal={Journal of Fluid Mechanics},
  volume={539},
  pages={199--227},
  year={2005},
  publisher={Cambridge University Press}
}

@article{el1980velocity,
  title={Velocity distributions in plane turbulent channel flows},
  author={El Telbany, MMM and Reynolds, AJ},
  journal={Journal of Fluid Mechanics},
  volume={100},
  number={1},
  pages={1--29},
  year={1980},
  publisher={Cambridge University Press}
}

@inproceedings{robertson1959turbulent,
  title={On turbulent plane {Couette} flow},
  author={Robertson, James M},
  booktitle={Proc. 6th Midwestern Conf. on Fluid Mech., University of Texas, Austin},
  pages={169--182},
  year={1959}
}

@article{reichardt1956geschwindigkeitsverteilung,
  title={{\"U}ber die Geschwindigkeitsverteilung in einer geradlinigen turbulenten Couettestr{\"o}mung},
  author={Reichardt, H},
  journal={ZAMM-Journal of Applied Mathematics and Mechanics/Zeitschrift f{\"u}r Angewandte Mathematik und Mechanik},
  volume={36},
  number={S1},
  pages={S26--S29},
  year={1956},
  publisher={Wiley Online Library}
}

@article{tsukahara2006dns,
  title={{DNS} of turbulent {Couette} flow with emphasis on the large-scale structure in the core region},
  author={Tsukahara, Takahiro and Kawamura, Hiroshi and Shingai, Kenji},
  journal={Journal of Turbulence},
  number={7},
  pages={N19},
  year={2006},
  publisher={Taylor \& Francis}
}

@article{launder1974application,
  title={Application of the energy-dissipation model of turbulence to the calculation of flow near a spinning disc},
  author={Launder, Brian Edward and Sharma, Bahrat I},
  journal={Letters in Heat and Mass Transfer},
  volume={1},
  number={2},
  pages={131--137},
  year={1974},
  publisher={Pergamon}
}

@article{xu2007multiple,
  title={Multiple-temperature kinetic model for continuum and near continuum flows},
  author={Xu, Kun and Liu, Hongwei and Jiang, Jianzheng},
  journal={Physics of Fluids},
  volume={19},
  number={1},
  pages={016101},
  year={2007},
  publisher={AIP Publishing}
}

@incollection{cercignani1988boltzmann,
  title={The {Boltzmann} equation},
  author={Cercignani, Carlo},
  booktitle={The {Boltzmann} {Equation} and {Its} {Applications}},
  pages={40--103},
  year={1988},
  publisher={Springer}
}

@article{chen1999analysis,
  title={Analysis of subgrid scale turbulence using the {Boltzmann} {Bhatnagar}-{Gross}-{Krook} kinetic equation},
  author={Chen, Hudong and Succi, Sauro and Orszag, Steven},
  journal={Physical Review E},
  volume={59},
  number={3},
  pages={R2527},
  year={1999},
  publisher={APS}
}

@article{bhatnagar1954model,
  title={A model for collision processes in gases. I. Small amplitude processes in charged and neutral one-component systems},
  author={Bhatnagar, Prabhu Lal and Gross, Eugene P and Krook, Max},
  journal={Physical Review},
  volume={94},
  number={3},
  pages={511},
  year={1954},
  publisher={APS}
}

@article{ansumali2004kinetic,
  title={Kinetic theory of turbulence modeling: smallness parameter, scaling and microscopic derivation of {Smagorinsky} model},
  author={Ansumali, Santosh and Karlin, Iliya V and Succi, Sauro},
  journal={Physica A},
  volume={338},
  number={3-4},
  pages={379--394},
  year={2004},
  publisher={Elsevier}
}

@article{girimaji2007boltzmann,
  title={{Boltzmann} kinetic equation for filtered fluid turbulence},
  author={Girimaji, Sharath S},
  journal={Physical Review Letters},
  volume={99},
  number={3},
  pages={034501},
  year={2007},
  publisher={APS}
}

@article{lundgren1967distribution,
  title={Distribution functions in the statistical theory of turbulence},
  author={Lundgren, TS},
  journal={Physics of Fluids},
  volume={10},
  number={5},
  pages={969--975},
  year={1967}
}

@article{cohen1962fundamental,
  title={Fundamental problems in statistical mechanics},
  author={Cohen, Ezechiel Godert David and Balazs, Nandor L},
  journal={Physics Today},
  volume={15},
  number={12},
  pages={64--66},
  year={1962},
  publisher={American Institute of Physics}
}

@article{srinivasan1977turbulent,
  title={Turbulent plane {Couette} flow using probability distribution functions},
  author={Srinivasan, R and Giddens, DP and Bangert, LH and Wu, JC},
  journal={The Physics of Fluids},
  volume={20},
  number={4},
  pages={557--567},
  year={1977},
  publisher={American Institute of Physics}
}

@article{yoshizawa1993nonequilibrium,
  title={A nonequilibrium representation of the turbulent viscosity based on a two-scale turbulence theory},
  author={Yoshizawa, Akira and Nisizima, Shoiti},
  journal={Physics of Fluids A},
  volume={5},
  number={12},
  pages={3302--3304},
  year={1993},
  publisher={American Institute of Physics}
}

@article{yang2025wave,
  title={Wave-Particle Based Multiscale Modeling and Simulation of Non-equilibrium Turbulent Flows},
  author={Yang, Xiaojian and Xu, Kun},
  journal={arXiv preprint arXiv:2503.07207},
  year={2025}
}

@article{pope1983lagrangian,
  title={A {Lagrangian} two-time probability density function equation for inhomogeneous turbulent flows},
  author={Pope, Stephen B},
  journal={Physics of Fluids},
  volume={26},
  number={12},
  pages={3448--3450},
  year={1983},
  publisher={AIP Publishing}
}

@article{mansour1988reynolds,
  title={Reynolds-stress and dissipation-rate budgets in a turbulent channel flow},
  author={Mansour, N Nd and Kim, John and Moin, Parviz},
  journal={Journal of Fluid Mechanics},
  volume={194},
  pages={15--44},
  year={1988},
  publisher={Cambridge University Press}
}

@article{1997Prediction,
  title={Prediction of turbulent transitional phenomena with a nonlinear eddy-viscosity model},
  author={ Craft, T. J.  and  Launder, B. E.  and  Suga, K. },
  journal={International Journal of Heat and Fluid Flow},
  volume={18},
  number={1},
  pages={15-28},
  year={1997},
}

\end{document}